\documentclass[10pt,twocolumn,twoside]{IEEEtran}
\usepackage[numbers]{natbib}
\ifCLASSINFOpdf
\else
\fi
\usepackage{graphics}
\usepackage{graphicx}
\usepackage{subcaption}
\usepackage{multirow}

\usepackage{amsmath,,amssymb}
\usepackage{amsthm}

\usepackage{pstricks, pst-node}
\usepackage{tikz}
\usepackage{pgfplots}
\usepackage{algorithmic}
\usepackage{algorithm}

\usepackage{url}

\newtheorem{theorem}{Theorem}

\newtheorem{proposition}{Proposition}

\def\II{{\mathbb I}}
\def\sign{{\rm sign}}
\def\Kcal{\mathcal K}
\def\RR{\mathbb{R}}
\def\SS{\mathbb{S}}
\def\PP{\mathbb{P}}

\def\EE{\mathbb{E}}

\def\Fcal{\mathcal F}
\def\bK{\mathbf K}

\begin{document}
%
\title{Spatially Adaptive Colocalization Analysis in Dual-Color Fluorescence Microscopy}
%
%
%

\author{Shulei~Wang$^{\ast,\dag,\natural}$,
        Ellen~T.~Arena$^\dag$,
        Jordan~T.~Becker$^\dag$,
        William~M.~Bement$^\dag$,\\
        Nathan~M.~Sherer$^\dag$,
        Kevin~W.~Eliceiri$^\dag$,
        and~Ming~Yuan$^{\ast,\dag,\natural}$ 
\thanks{$^\ast$Columbia University}
\thanks{$^\dag$University of Wisconsin-Madison}
\thanks{$^\natural$Address for Correspondence: Department of Statistics, Columbia University, 1255 Amsterdam Avenue, New York, NY 10027.}}

%
%

\markboth{IEEE TRANSACTIONS ON IMAGE PROCESSING,~Vol.~XX, No.~XX, XXXX~XXXX}%
{Wang \MakeLowercase{\textit{et al.}}: Spatially Adaptive Colocalization Analysis in Dual-Color Fluorescence Microscopy}
%



\maketitle

\begin{abstract}
Colocalization analysis aims to study complex spatial associations between bio-molecules via optical imaging techniques. However, existing colocalization analysis workflows only assess an average degree of colocalization within a certain region of interest and ignore the unique and valuable spatial information offered by microscopy. In the current work, we introduce a new framework for colocalization analysis that allows us to quantify colocalization levels at each individual location and automatically identify pixels or regions where colocalization occurs. The framework, referred to as spatially adaptive colocalization analysis (SACA), integrates a pixel-wise local kernel model for colocalization quantification and a multi-scale adaptive propagation-separation strategy for utilizing spatial information to detect colocalization in a spatially adaptive fashion. Applications to simulated and real biological datasets demonstrate the practical merits of SACA in what we hope to be an easily applicable and robust colocalization analysis method. In addition, theoretical properties of SACA are investigated to provide rigorous statistical justification.
\end{abstract}


%
\IEEEpeerreviewmaketitle

\section{Introduction}
%
%
%
%

\IEEEPARstart{C}{olocalization} analysis is a powerful tool to study the spatial relationships between macromolecules. The primary goal of colocalization analysis is to better understand the underlying associations between fluorescently-labeled molecules by quantifying the co-occurrence and correlation between them. Arguably, one of the most widely used quantitative colocalization analysis workflows can be described by the following procedure \citep[see][]{dunn2011}, called the ``3-step procedure" hereafter: 1) an appropriate region of interest (ROI) is selected by either a manual or automatic image segmentation method; 2) the level of colocalization is measured by making a scatter plot of pixel data within the ROI and calculating the colocalization quantification index accordingly \citep[see][]{manders1992dynamics,manders1993measurement,Li04,Khanna06,Helmuth2010,lagache2013statistical}; 3) statistical significance is evaluated under a hypothesis test framework \citep[see][]{costes2004}. 

\begin{figure}[h!]
    \centering
     \begin{subfigure}[b]{0.24\textwidth}
     \centering
        \begin{tikzpicture}[scale=1]
  \node[anchor=south west,inner sep=0] at (0,0) {\includegraphics[width=0.97\textwidth]{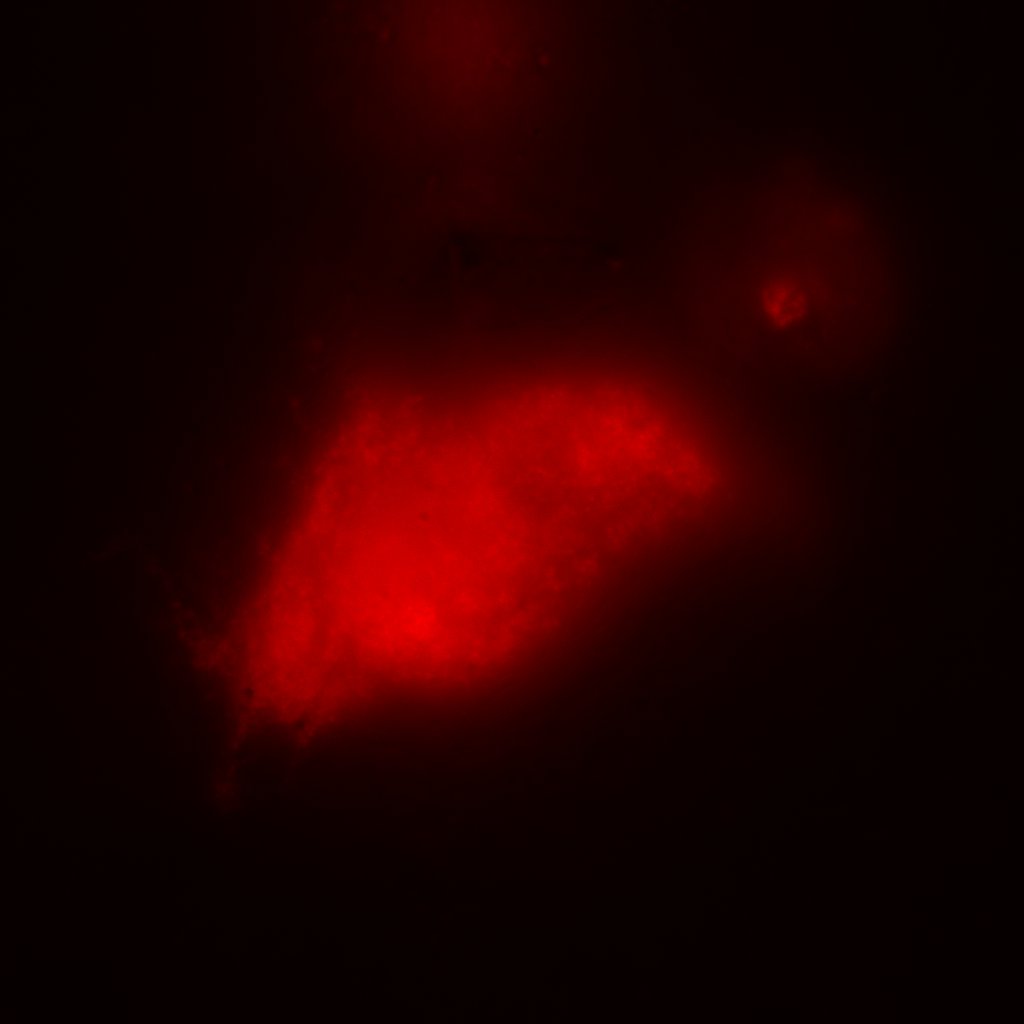}};
    \end{tikzpicture}
        \caption{MS2-YFP}
        \label{fg:toyred}
    \end{subfigure}
    \begin{subfigure}[b]{0.24\textwidth}
    \centering
        \begin{tikzpicture}[scale=1]
  \node[anchor=south west,inner sep=0] at (0,0) {\includegraphics[width=0.97\textwidth]{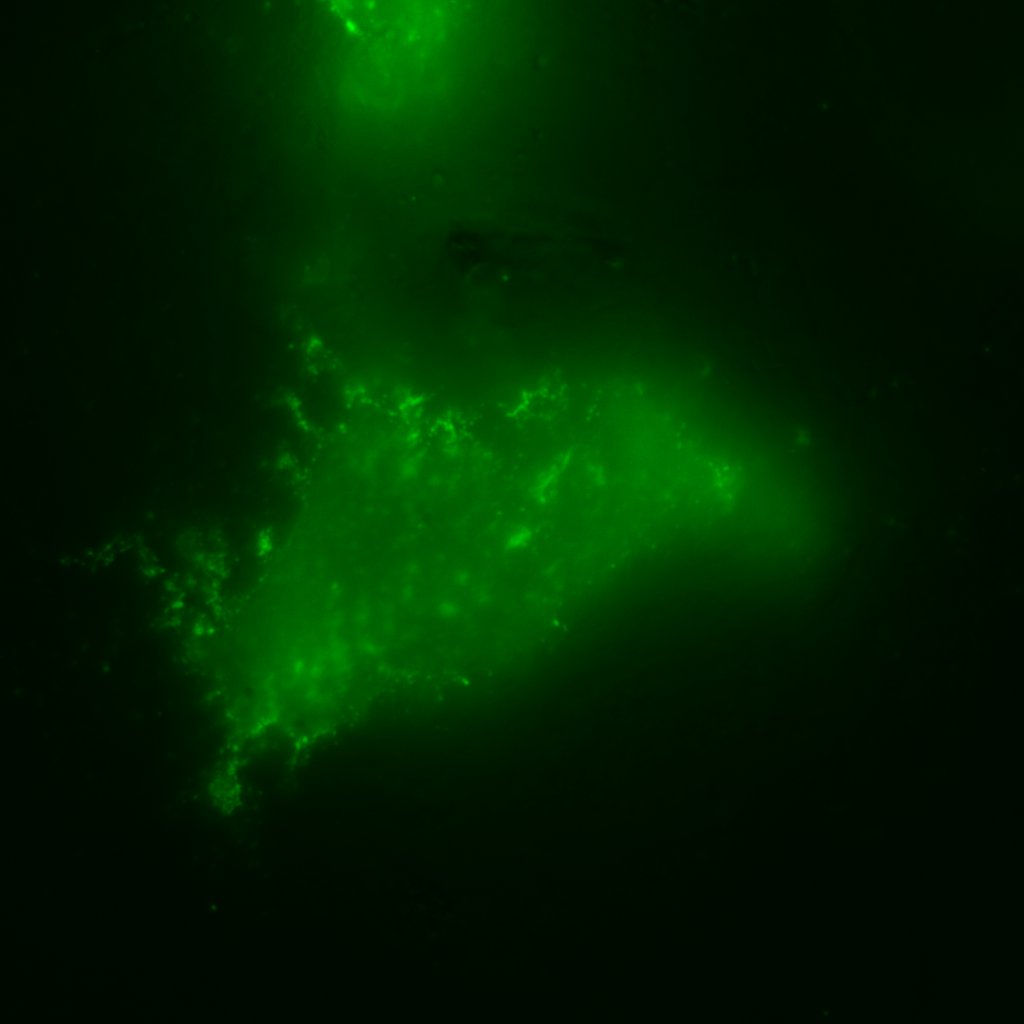}};
    \end{tikzpicture}
        \caption{Gag-CFP}
        \label{fg:toygreen}
    \end{subfigure}
    \begin{subfigure}[b]{0.24\textwidth}
        \centering
        \begin{tikzpicture}[scale=1]
\node[anchor=south west,inner sep=0] at (0,0) {\includegraphics[width=0.97\textwidth]{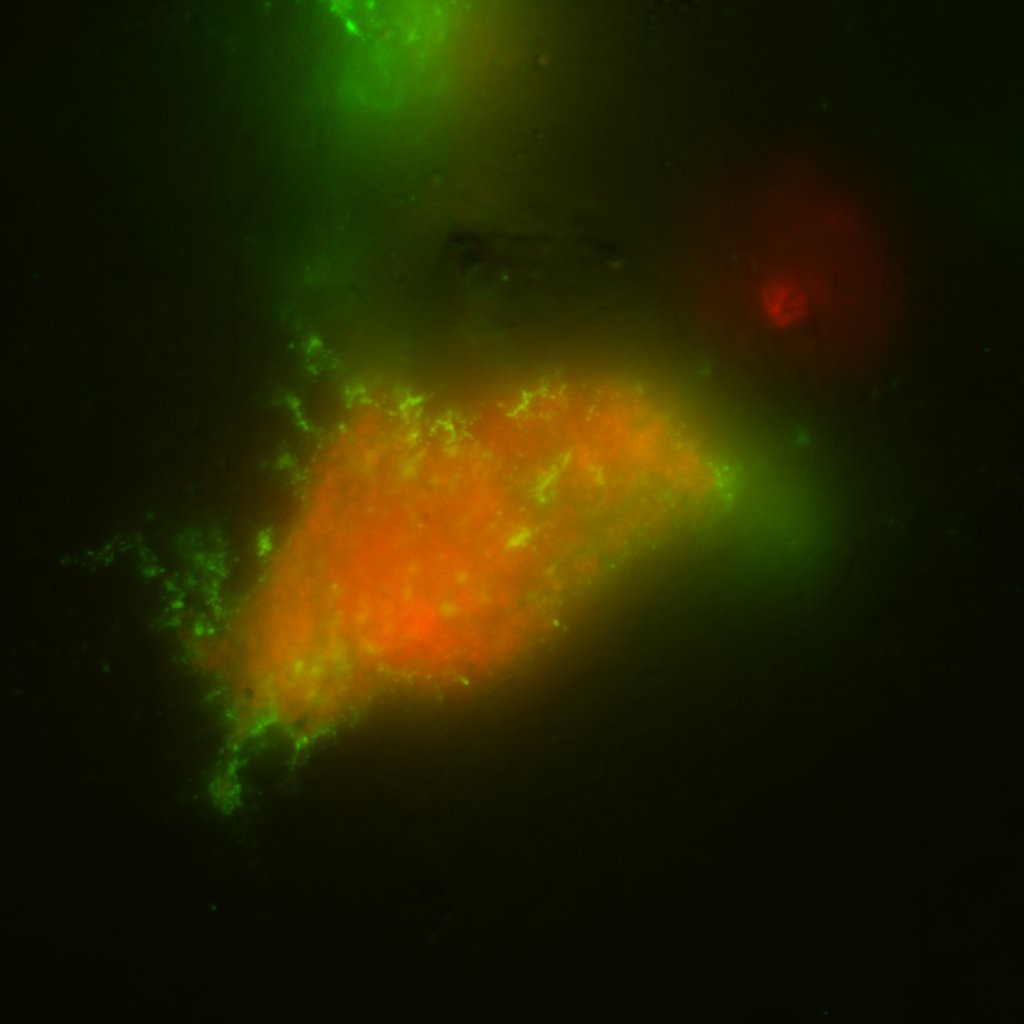}};
    \end{tikzpicture}
        \caption{Merged Image}
        \label{fg:toymerge}
    \end{subfigure}
         \begin{subfigure}[b]{0.24\textwidth}
     \centering
        \begin{tikzpicture}[scale=1]
  \node[anchor=south west,inner sep=0] at (0,0) {\includegraphics[width=0.97\textwidth]{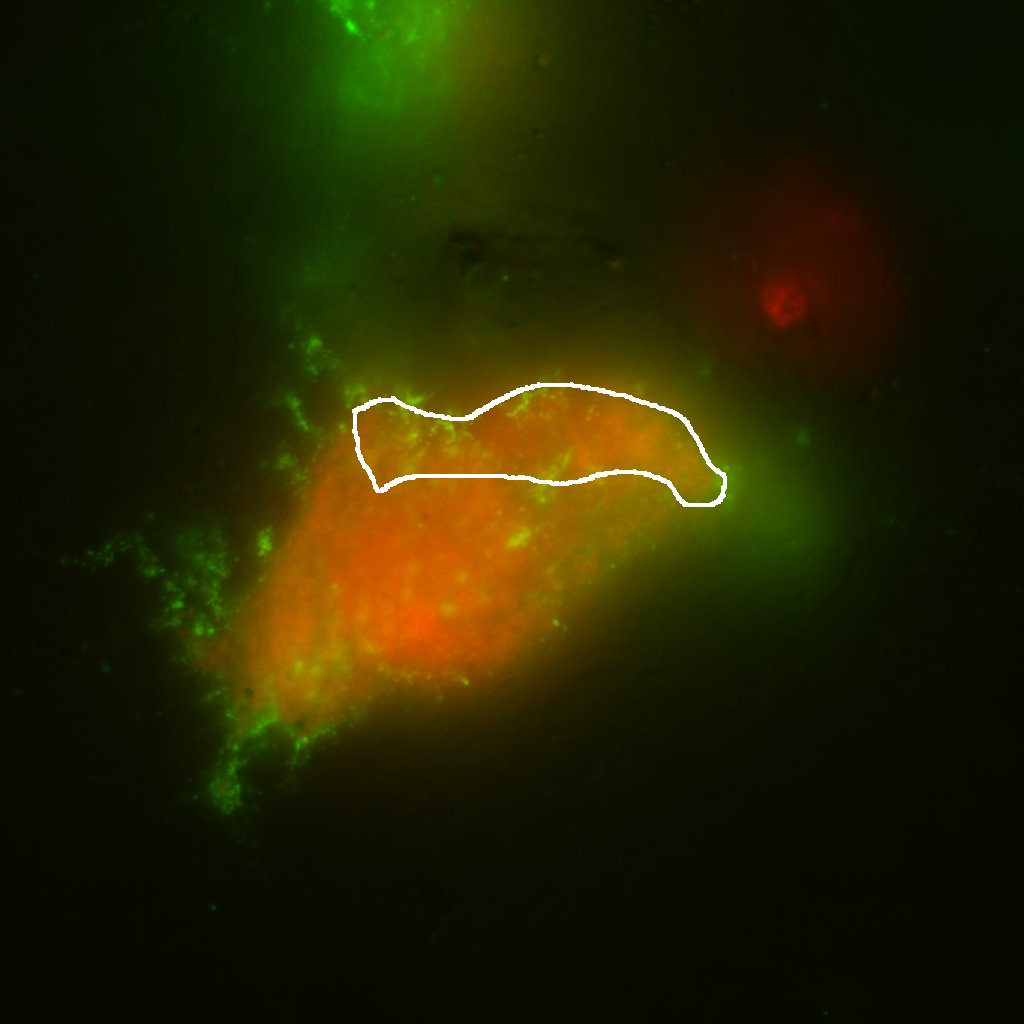}};
    \end{tikzpicture}
        \caption{Region of Interest}
        \label{fg:toyroi}
    \end{subfigure}
    \hspace{0.05\textwidth}
        \begin{subfigure}[b]{0.24\textwidth}
     \centering
        \begin{tikzpicture}[scale=1]
  \node[anchor=south west,inner sep=0] at (0,0) {\includegraphics[width=0.97\textwidth]{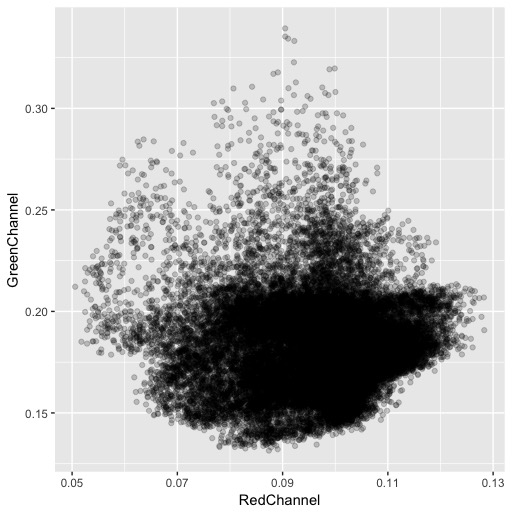}};
    \end{tikzpicture}
        \caption{Scatter Plot within ROI}
        \label{fg:toyplot}
    \end{subfigure}
    \caption{Colocalization analysis on a typical example, an HIV model of viral particle assembly.}\label{fg:realdatatoy}
\end{figure}

To illustrate how this 3-step procedure works, we applied it to microscopic images of HeLa cells (see Figure~\ref{fg:toyred}, \ref{fg:toygreen}, and \ref{fg:toymerge}) expressing HIV-1-Gag, a structural protein of human immunodeficiency virus type 1 (HIV-1) fused to cyan fluorescence protein (CFP) (green channel) and MS2 protein fused to yellow fluorescence protein (YFP) (red channel).  Gag-CFP was expressed from an mRNA engineered to contain multiple copies of an RNA stem loop that binds MS2-YFP with high specificity; significantly higher colocalization levels between Gag-CFP and MS2-YFP are expected at the edge of cells where Gag and viral RNAs are converging at sites of virus particle assembly. A ROI was selected manually in Figure~\ref{fg:toyroi} with the goal to meaure the level of colocalization within that region, one that is expected to contain higher levels of colocalization. Following step 2 and 3, we made the scatter plot in Figure~\ref{fg:toyplot} and calculated the colocalization quantification index  and corresponding $p$-value within the selected ROI by pixel shuffling (Pearson correlation $r=-0.016$ ($p=99.4\%$) and Mander's collocalization coefficients $M_1=0.839$ ($p=0.1\%$) and $M_2=0.196$ ($p=0.1\%$)). A conclusion is then made based on the scatter plot and the $p$-value of colocalization quantification indices.

The 3-step procedure above, however, neglects the most valuable information offered by microscopy, the spatial information of pixels. In most, if not all, applications of microscopy, high-resolution spatial information of targeted macromolecules is extracted from images to precisely locate the biological event of interest. This is also very much needed for colocalization analysis, where the accurate location of association between macromolecules is of notable interest. Despite this demand, the existing 3-step procedure is only able to identify the location of colocalization within the pre-selected ROI, as all spatial resolution is lost thereafter. An ideal colocalization analysis method should be able to identify the location of association at a finer resolution, at the pixel level, by taking advantage of neighboring spatial information. 

Another fact overlooked by the 3-step procedure above is that microscopic images are rarely spatially homogeneous, as the distribution of fluorescent signal varies within cells dramatically. This spatial heterogeneity in signal is ubiquitous in microscopic images, bringing difficulties in all aspects of image processing, such as segmentation, noise reduction, cell tracking among many others \citep[see][]{dunn2011,mueller2013,buggenthin2013}. Most existing colocalization analyses also cannot escape the fate of image heterogeneity; the 3-step procedure above only obtained an average degree of colocalization measured within the selected ROI. One strategy already in practice to alleviate this issue of heterogeneity is to select a ROI in a homogeneous region of a microscopic image. However, this non-reproducible method can introduce selection bias and neglects the phenomenon that colocalization may also occur at the boundary of two homogeneous regions. Thus, there is a clear need for a colocalization analysis method that automatically accounts for the spatial heterogeneity in microscopic images and reports colocalized regions at the pixel level by taking advantage of all available spatial information. The goal of this paper is to address this challenge and develop a principled and effective local colocalization analysis framework to fill this need. 

More specifically, we introduce a general Spatially Adaptive Colocalization Analysis framework (SACA) to infer the degree of colocalization pixel-by-pixel. Instead of evaluating the average colocalization level within a ROI, SACA allows quantification of colocalization at each pixel across the entire image. In doing so, spatial heterogeneity is taken into full consideration, and colocalization can be identified at the pixel level. The framework adopts the multiscale strategy of the propagation and separation (PS) approach \citep[see][]{Polzehl2000,Polzehl2006} to expand the neighborhood adaptively so that one could obtain a more precise estimate of the colocalization level at each pixel, including more detection power for more subtle levels of colocalization. 

\begin{figure}[h!]
    \centering
    \begin{subfigure}[b]{0.24\textwidth}
    \centering
        \begin{tikzpicture}[scale=1]
  \node[anchor=south west,inner sep=0] at (0,0) {\includegraphics[width=0.97\textwidth]{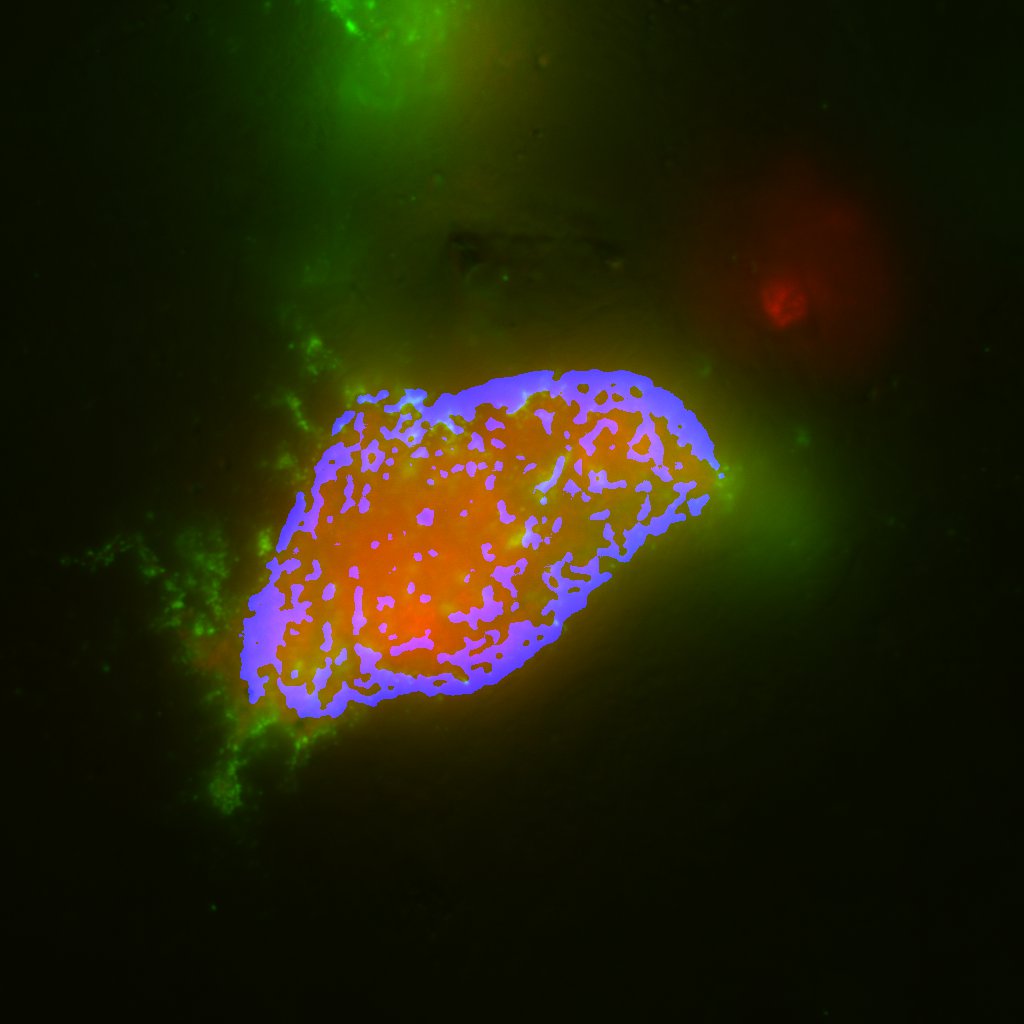}};
    \end{tikzpicture}
        \caption{Colocalized region, as indicated by blue }
        \label{fg:toyregion}
    \end{subfigure}
    \begin{subfigure}[b]{0.24\textwidth}
     \centering
        \begin{tikzpicture}[scale=1]
  \node[anchor=south west,inner sep=0] at (0,0) {\includegraphics[width=0.97\textwidth]{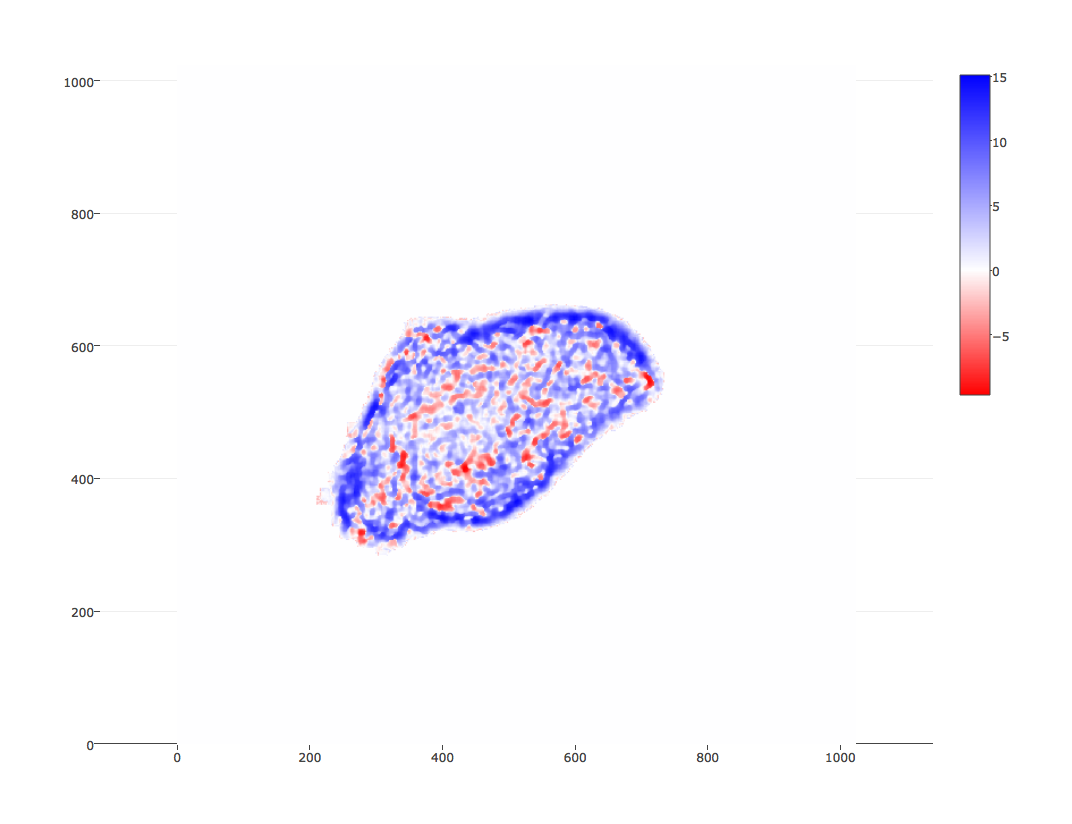}}; 
    \end{tikzpicture}
        \caption{Pixel-level colocalization}
        \label{fg:toylaca}
    \end{subfigure}
    \caption{SACA on a typical colocalization analysis example.}\label{fg:realdatatoylaca}
\end{figure}

To demonstrate the practical merits of SACA, we applied it to the same microscopic image used in Figure~\ref{fg:realdatatoy}.  Compared with the classical 3-step procedure, SACA does not call for a predetermined ROI and can quantify the degree of colocalization at each pixel across the entire image (Figure~\ref{fg:realdatatoylaca}). Moreover, the colocalized region in Figure~\ref{fg:toyregion} can be identified automatically by multiple comparison corrections. SACA analysis in Figure~\ref{fg:realdatatoylaca} reveals that pixels with a high degree of colocalization concentrate on the edge of cell, and there are few colocalized regions inside the cell; this is due to the fact that the two particles labelled by fluorescence in this biological case are in fact assembled along the plasma membrane. Through this real, biological example, we can conclude that SACA is not only able to detect the existence of colocalization within microscopic images, but, more importantly, can also identify regions with high colocalization levels in a reproducible and objective way. To our best knowledge, this is the first method that can provide such a pixel-level spatial inference on colocalization. 

\subsection{Related Work}

In literature, there are methods which also incorporate spatial information of microscopy images into colocalization analysis. In particular, object-based colocalization methods are developed to take advantage of spatial information \citep[see][]{Helmuth2010,zhang2008statistical,lagache2013statistical}. Object based methods identify the objects first and then conduct spatial analysis among these detected objects \citep[see][]{worz20103d,ruusuvuori2014quantitative,Lagache2015,moser2017fluorescence}. The spatial association between spots can be exploited under a marked point process framework \citep[see][]{illian2008statistical,Helmuth2010,lagache2013statistical,Lagache2015,lagache2018mapping}. Under this framework, second-order statistics, such as the nearest neighbor or Ripley's K function, is used to quantify colocalization events. Besides employing marked point process, there are methods that measure point-pattern matching on a nearest neighbor graph and then quantify spatial association through point-pattern matching \cite{ruusuvuori2014quantitative}. Moreover, other work shows the superiority of object based colocalization method to intensity based colocalization methods \cite{Lagache2015}. However, current object based methods adopt spatial homogeneity assumptions implicitly and thus can only provide estimations of average colocalization levels within ROIs, but not quantification of colocalization at each location. Furthermore, the success of object based methods heavily relies on high quality of object identification \citep[see][]{lagache2018mapping}. It is not always easy to distinguish such regions from background or local environment signals in some biological applications \citep[see][and Figure~\ref{fg:realdatatoy}]{pike2017quantifying}. Compared with object based methods, SACA does not rely on object detection and is able to provide colocalization quantification at each pixel. 

The rest of the paper is organized as follows. In the next section, we introduce the general statistical framework for SACA. Further details on implementation and computational considerations are given in Section 3. Numerical experiments, both simulated and real biological test cases, are presented in Section 4 to demonstrate the merits of SACA. Theoretical properties of SACA are investigated in Section 5, and we finish with concluding remarks in Section 6.

\section{Spatially Adaptive Colocalization Analysis}

Given a ROI, colocalization analysis can naturally be cast as a hypothesis testing problem. Let $\II$ be the index set of all pixels in the ROI and $(X_k, Y_k)$ be the intensity of the two channels measured at pixel $k\in\II$. If each $(X_k,Y_k)$ is from bivariate distribution $F$, then colocalization can be formulated as testing the following hypotheses:
\begin{equation}
\label{eq:global}
H_{0}: F\in \Fcal_0\qquad {\rm v.s.}\qquad H_{1}: F\in \Fcal_1,
\end{equation}
where $\Fcal_0$ and $\Fcal_1$ are the families of distributions that characterize non-colocalization and colocalization, respectively. As mentioned previously, this formulation does not take the spatial heterogeneity of microscopic images into account and thus is not able to reveal any spatial information regarding colocalization. To address this deficiency, we shall consider a general framework to quantify the degree of colocalization at the pixel level.

\subsection{Quantifying Colocalization Locally}
\label{sc:localquan}

To infer the level of colocalization at each pixel, we may consider a pixel-wise version of the hypotheses given in \eqref{eq:global}:
\begin{equation}
\label{eq:multipletest}
H_{0,k}: F_k\in \Fcal_0\qquad {\rm v.s.}\qquad H_{1,k}: F_k\in \Fcal_1,\qquad k\in\II.
\end{equation}
Here $F_k$ is a bivariate distribution of measurement $(X_k,Y_k)$ at each individual pixel $k$. Although it fully accounts for the possible heterogeneity among pixels, the main difficulty of this framework is that we have only one sample $(X_k,Y_k)$ available, insufficient for assessing association between the two channels. To address this challenge, we appeal to the fact that neighboring pixels are more likely to share a similar level of colocalization and consider testing $H_{0,k}$ against $H_{1,k}$ by utilizing information from neighboring pixels.

Our general framework can be applied with a generic colocalization index, such as Pearson's correlation or Mander's split coefficient. To be specific, we shall characterize colocalization using a general and robust criterion recently introduced by \cite{wang2017}:
\begin{equation}
\label{eq:colocaldef}
Q(F):=\EE\left(\sign(X-\tilde{X})(Y-\tilde{Y})\middle|X,\tilde{X}>t_X;Y,\tilde{Y}>t_Y\right)
\end{equation}
for some pre-specified signal strength thresholds $t_X$ and $t_Y$. Here $(X,Y)$ and $(\tilde{X},\tilde{Y})$ are independent, identical copies following distribution $F$. Assuming that for all pixels $i$ in a ROI, $(X_i,Y_i)\sim F$, the level of colocalization, as measured by \eqref{eq:colocaldef}, in the ROI with index set $\II$ can be conveniently measured by Kendall's tau coefficient over all pixels in $\II$
\begin{equation}
\label{eq:tructkendalltau}
\tau(t_X,t_Y)={\sum_{i,j\in\Kcal(t_X,t_Y):i\ne j}\sign(X_i-X_j)\sign(Y_i-Y_j)\over n_{t_X,t_Y}(n_{t_X,t_Y}-1)},
\end{equation}
where $\Kcal(t_X,t_Y)=\{i\in\II:X_i>t_X,Y_i>t_Y\}$ and $n_{t_X,t_Y}=|\Kcal(t_X,t_Y)|$. See \cite{wang2017} for further details.

In the presence of heterogeneity, however, $\tau(t_X,t_Y)$ can only be viewed as an index for the ``average'' level of colocalization. To define a pixel specific index for colocalization, i.e., for testing \eqref{eq:multipletest}, we borrow a classical idea in nonparametric statistics by assigning weights to each pixel determined by their ``closeness'' to pixel $k$ and evaluate a weighted Kendall's tau coefficient \citep{Shieh1998}. More specifically, let $w_i\ge 0$ be the weight of pixel $i\in\II$, then weighted Kendall's tau coefficient is defined as
\begin{equation}
\label{eq:wtkendalltau}
\tau_w:={1 \over \sum_{i\ne j}w_iw_j}\sum_{i\ne j}w_iw_j\sign(X_i-X_j)\sign(Y_i-Y_j).
\end{equation}
We shall also adopt the convention that $\tau_w=0$, if all weights $w_i=0$. In particular, if we take
$$
w_i=\mathbf{I}_{(i\in\II, X_i>t_X,Y_i>t_Y)},
$$
where $\mathbf{I}$ is the indicator function, then \eqref{eq:wtkendalltau} becomes the $\tau(t_X,t_Y)$, as given in \eqref{eq:tructkendalltau}.

The power of the weights, however, lies in their flexibility to incorporate neighboring pixels in evaluating colocalization at pixel $k$. More specifically, we want to choose weights that incorporate two types of information: 1) how far a pixel is from pixel $k$; 2) whether or not a pixel is background or signal. To this end, we consider weights of the form:
\begin{equation}
\label{eq:nonadpwt}
w_i(k;r)=\bK_l\left(d(i,k)\over r\right) \bK_b\left(X_i,Y_i\right),
\end{equation}
where $\bK_l: \RR^+\to\RR$ is a non-negative and non-increasing kernel function with compact support, $d(i,k)$ is the distance between pixels $i$ and $k$, and $\bK_b\left(X_i,Y_i\right)=\mathbf{I}_{(X_i>t_X,Y_i>t_Y)}$. Here the tuning parameter $r$ represents the radius of the neighborhood around $k$. We shall view it as known in this subsection and discuss an automated and spatially adaptive strategy to choose it in the next section. The simplest example of $\bK_l$ is $\mathbf{I}_{i\in B(k,r)}$, where $B(k,r)=\{i: d(i,k)\le r\}$ is the neighborhood of $k$ of size $r$. With the weights defined above, we now consider the following weighted Kendall's tau coefficient:
$$
\tau_w(k;r)={\sum_{i\ne j}w_i(k;r)w_j(k;r)\sign(X_i-X_j)\sign(Y_i-Y_j)\over \sum_{i\neq j}w_i(k;r)w_j(k;r)}.
$$

We are now in position to test the hypotheses in (\ref{eq:multipletest}). A natural pixel-wise test statistic for the hypotheses in (\ref{eq:multipletest}) is the pixel-wise standardized weighted Kendall's tau coefficient:
$$
Z(k;r):={3\over 2}\sqrt{\tilde{N}_k} \cdot \tau_w(k;r),\qquad k\in\II,
$$
where $\tilde{N}_k$ is ``effective sample size" of\ $\tau_w(k;r)$: 
\begin{equation}
\label{eq:sampsize}
\tilde{N}_k= \left(\sum_{i}w_i(k;r)\right)^2\bigg/\sum_{i}w^2_i(k;r).
\end{equation}
See Section~\ref{sc:theprop} for detailed discussion. As shown in \cite{Shieh1998}, when the weights are independent from data $X_i$ and $Y_i$,
$$
Z(k;r)\to_{d} N(0,1),
$$
where $\to_d$ refers to convergence in distribution, and $N(0,1)$ is the standard normal distribution. Thus, $Z(k;r)$ can be viewed as a $z$-score for colocalization at pixel $k$.

Obviously, we want to compute $Z(k;r)$ for all pixels $k$. How to translate them into $p$-values requires caution because of its multiple comparison nature. There are various ways we could adjust for multiple comparisons. The list includes Bonferroni correction \citep[see][]{dunn1961} or false discovery rate control \citep[see][]{Benjamini1995,Efron2001} among many others.

\subsection{Adaptive Colocalization Analysis}
\label{sc:adpwt}

The choice of $r$ is of clear importance to the success of our approach. A large $r$ could result in over-smoothing and blur the sharp edges of colocalized regions; on the other hand, the signal-to-noise ratio could be inflated if $r$ is small (see also numerical experiments in section~\ref{sc:numerical}). In practice, it is often hard to strike a good balance and choose an $r$ to ensure that the pixel-wise standardized weighted Kendall's tau coefficient $Z(k;r)$ is the most efficient. To overcome this problem, we adopt a propagation and separation (PS) approach to select $r$ and neighborhood shape simultaneously. The PS approach was introduced by  \cite{Polzehl2000,Polzehl2006} and is widely used in neuroimaging and image restoration \citep[see][]{Li2011,Polzehl2000}. Instead of determining the size of neighborhood $r$ and weights $w_i(k;r)$ in advance, the PS approach updates them iteratively in a sequence of nested neighborhoods.  

\begin{figure}[h!]
\begin{center}
\begin{tikzpicture}[scale=0.7]
\fill[fill=yellow!50!white] plot[smooth,samples=100,domain=-0:3.5] (\x,{0.5-2/(\x+0.5)}) -- plot [smooth,tension=1] coordinates {(3.5,0) (2.4,2.4) (0,3.5) } -- plot[smooth,samples=100,domain=0:-3.5] (\x,{-0.5+2/(-\x+0.5)}) -- plot [smooth,tension=1] coordinates {(-3.5,0) (-2.4,-2.4) (0,-3.5) }-- cycle;
\draw[line width=1.5pt] plot[smooth,samples=100,domain=-0:3.5] (\x,{0.5-2/(\x+0.5)});
\draw[line width=1.5pt] plot[smooth,samples=100,domain=0:-3.5] (\x,{-0.5+2/(-\x+0.5)});
\draw[black!80!white,line width=1.5pt] (0,0) -- ({sin(-80)},{cos(-80)});
\draw[thick,black](-.5,0.25)node{$r_0$};
\draw[black!80!white,line width=1.5pt] (0,0) -- ({1.8*sin(-110)},{1.8*cos(-110)});
\draw[thick,black](-1.4,-0.25)node{$r_1$};
\draw[black!80!white,line width=1.5pt] (0,0) -- ({3*sin(-140)},{3*cos(-140)});
\draw[thick,black](-1.7,-1.6)node{$r_2$};
\fill[red,line width=1.5pt] (0,0) ellipse (0.05 and 0.05);
\draw[orange!80!white,line width=1.5pt] (0,0) ellipse (1 and 1);
\draw[orange!80!white,line width=1.5pt] (0,0) ellipse (1.8 and 1.8);
\draw[orange!80!white,line width=1.5pt] (0,0) ellipse (3 and 3);
\draw[thick,black](0.2,-0.3)node{$(\mathbf{A})$};

\fill[red,line width=1.5pt] (-2.05,2.05) ellipse (0.05 and 0.05);
\draw[red!80!white,dashed,line width=1.5pt] (-2.05,2.05) ellipse (1 and 1);
\draw[red!80!white,dashed,line width=1.5pt] (-2.05,2.05) ellipse (1.8 and 1.8); 
\draw[thick,black](-1.5,2)node{$(\mathbf{B})$};

\fill[red,line width=1.5pt] (1.4,1.5) ellipse (0.05 and 0.05);
\draw[red!80!white,dashed,line width=1.5pt] (1.4,1.5) ellipse (1 and 1);
\draw[red!80!white,dashed,line width=1.5pt] (1.4,1.5) ellipse (1.8 and 1.8); 
\draw[thick,black](1.8,1.7)node{$(\mathbf{C})$};

\draw[thick,black](4.2,0.5)node{Colocalized};
\draw[thick,black](4.5,-0.5)node{Non-colocalized};
\end{tikzpicture}
\caption{PS-based procedure can separate regions with different colocalization levels, and the adaptive neighborhood in PS can extend to homogeneous regions freely. After two iterations, both pixels $B$ and $C$ start to enter into the neighborhood of $A$, i.e. the ball with center at $A$ and with radius $r_2$. The weights are given adaptively so that it is small between $B$ and $A$, but large between $C$ and $A$, so that only $C$ is incorporated into the estimation process of the colocalization level at $A$. }
\label{fg:psillu}
\end{center}
\end{figure}
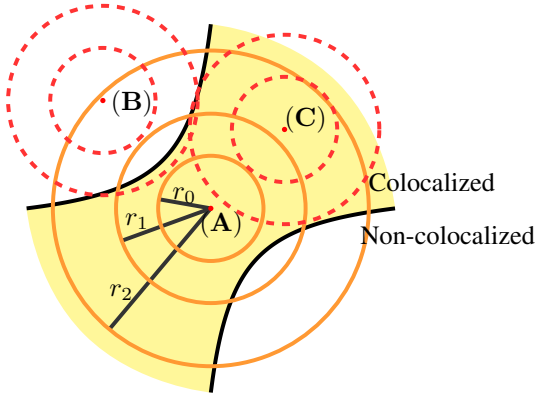

More specifically, let $r_0<r_1<\cdots<r_t<\cdots$ be a sequence of nonnegative radii and $B(k,r_t)$, $t\ge 0$, be a sequence of corresponding nested neighborhoods of pixel $k$ with increasing radii. At iteration $t$, the weights $w_i(k;r_t)$ not only include the spatial distance and background identification as we describe in Section~\ref{sc:localquan}, but also reflect if there is a significant difference in the values of weighted Kendall's tau coefficient $\tau_w$ estimated in the previous iteration, i.e., weighted Kendall's tau coefficient at pixel $i$, $\tau_w(i;r_{t-1})$, and at pixel $k$, $\tau_w(k;r_{t-1})$. Thus the weights $w_i(k;r_t)$ in $\tau_w(k;r_t)$ at iteration $t$ are given by 
\begin{equation}
\label{eq:adpwt}
w_i(k;r_t)=\bK_l\left(d(i,k)\over r_t\right) \bK_b(X_i,Y_i)\bK_s\left(D_i(k;r_{t-1}) \over D_n \right) ,
\end{equation}
where $D_i(k;r_{t})$ is the normalized distance between $\tau_w(i;r_{t})$ and $\tau_w(k;r_{t})$:
$$
D_i(k;r_{t})=\sqrt{\tilde{N}_k^{(t)}}|\tau_w(i;r_{t})-\tau_w(k;r_{t})|, 
$$
and $\tilde{N}^{(t)}_k$ is the ``effective sample size" as defined in (\ref{eq:sampsize}):
$$
\tilde{N}^{(t)}_k= \left(\sum_{i}w_i(k;r_t)\right)^2\bigg/\sum_{i}w^2_i(k;r_t).
$$
We follow the convention that the distance $D_i(k;r_{t-1})=0$ when $t=0$, and $\tilde{N}^{(t)}_k=0$ when $w_i(k;r_t)=0$ for all $i\in\II$. Here, kernel $\bK_s$ is some nonnegative kernel which downweights pixel $i$ if weighted Kendall's tau coefficient at pixel $i$ ($\tau_w(i;r_{t-1})$) and at pixel $k$ ($\tau_w(k;r_{t-1})$) are very different. In this way, new pixel $i$ is included in the estimation of $\tau_w(k;r_t)$ at iteration $t$, only provided that $\tau_w(i;r_{t-1})$ and $\tau_w(k;r_{t-1})$ are similar. After the weights $w_i(k;r_t)$ are updated, weighted Kendall's tau coefficient $\tau_w(k;r_t)$ can be calculated accordingly. The iterative procedure can be schematically described by the following diagram.
\begin{center}
\begin{tikzpicture}[scale=0.7]
\draw[thick,black](0,0)node{$w_i(k;r_0)$};
\draw[-latex,thick,black](0.4,-0.3)node[left]{} to[out=-45,in=135] (1.1,-1);
\draw[thick,black](1.5,-1.3)node{$\tau_w(k;r_0)$};

\draw[-latex,thick,black](1.9,-1)node[left]{} to[out=45,in=225] (2.6,-0.3);
\draw[thick,black](3,0)node{$w_i(k;r_1)$};
\draw[thick,black](4.3,0)node{$\ldots$};
\draw[thick,black](4.3,-0.65)node{$\ldots$};
\draw[thick,black](4.3,-1.3)node{$\ldots$};
\draw[thick,black](5.8,-1.3)node{$\tau_w(k;r_{t-1})$};

\draw[-latex,thick,black](6.2,-1)node[left]{} to[out=45,in=225] (6.9,-0.3);
\draw[thick,black](7.3,0)node{$w_i(k;r_t)$};
\draw[-latex,thick,black](7.7,-0.3)node[left]{} to[out=-45,in=135] (8.4,-1);
\draw[thick,black](8.8,-1.3)node{$\tau_w(k;r_t)$};

\draw[thick,black](10.1,0)node{$\ldots$};
\draw[thick,black](10.1,-0.65)node{$\ldots$};
\draw[thick,black](10.1,-1.3)node{$\ldots$};

\end{tikzpicture}
\end{center}
There are two key benefits of using adaptive weights $w_i(k;r_t)$: ({\it propagation}) the pixels in the homogeneous region of the neighborhood, which does not need to be the same shape as $B(k,r)$ and can be an arbitrary shape, are involved in estimation of $\tau_w(k;r_t)$; ({\it separation}) $\tau_w(k;r_t)$ avoids the disturbance from regions of different colocalization levels. Figure~\ref{fg:psillu} illustrates an example to show how this procedure works, where yellow regions are colocalized and outside regions are non-colocalized.  

\begin{figure}[h!]
\begin{center}
\begin{tikzpicture}[scale=0.6]
\draw[-latex,line width=1.5pt] (0,0) -- (12,0);
\draw[thick,black](12.2,-0.3)node{$t$};
\draw[line width=1.5pt] (1,-0.1) -- (1,0.1);
\draw[thick,black](1,-0.5)node{$0$};
\draw[line width=1.5pt] (4,-0.1) -- (4,0.1);
\draw[thick,black](4,-0.5)node{$T^L$};
\draw[line width=1.5pt] (9,-0.1) -- (9,0.1);
\draw[thick,black](9,-0.5)node{$T^U$};

\draw [decorate,decoration={brace,amplitude=10pt},xshift=0pt,yshift=4pt]
(1,0) -- (4,0) node [black,midway,xshift=0cm,yshift=18pt] 
{No stop};
\draw [decorate,decoration={brace,amplitude=10pt},xshift=0pt,yshift=4pt]
(9,0) -- (12,0) node [black,midway,xshift=0cm,yshift=18pt] 
{Stop};

\draw [decorate,decoration={brace,amplitude=10pt},xshift=0pt,yshift=4pt]
(4,0) -- (9,0) node [black,midway,xshift=0cm,yshift=18pt] 
{If $\Delta \tau_k^{(t)}>\Lambda$, stop};
\end{tikzpicture}
\caption{Stopping criteria for multiscale adaptive estimation.}
\label{fg:stopct}
\end{center}
\end{figure}
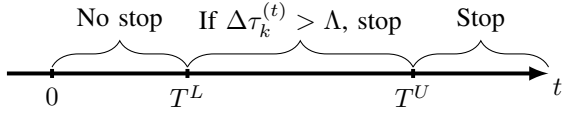

The above iterative procedure ends when the stopping criteria are satisfied. The stopping criteria are based on the number of iterations $t$ and the estimated weighted Kendall's tau coefficient $\tau_w(k;r_t)$ at iteration $t$. Let $T^L$ and $T^U$ be two positive integers satisfying $0<T^L<T^U$. We shall compare $t$ with $T^L$ and $T^U$. If $t$ is smaller than lower bound $T^L$, then iterations never stop, as the estimates in the first several iterations are not stable. If $t\ge T^U$, iterative updates stop. When $T^L\le t<T^U$, we compare the threshold $\Lambda$ and normalized difference $\Delta \tau_k^{(t)}$ between $\tau_w(k;r_t)$ and estimation at step $T_L$ 
$$
\Delta \tau_k^{(t)}:=\sqrt{\tilde{N}_k^{(T^L)}}|\tau_w(k;r_t)-\tau_w(k;r_{T^L})|.
$$ 
To avoid $\tau_w(k;r_t)$ deviating from benchmark $\tau_w(k;r_{T^L})$ too much, we stop updating it if $\Delta \tau_k^{(t)}>\Lambda$ and otherwise continue the iterative updates. The stopping criteria are summarized in Figure~\ref{fg:stopct}. After the iterative procedure stops, we denote the Kendall tau's correlation estimations by $\tau_w(k;r_T)$ and weights by $w_i(k;r_T)$.

After $\tau_w(k;r_T)$s are calculated, we follow the same approach as before to obtain pixel-wise test statistics $Z(k;r_T)$ ($z$-score)
$$
Z(k;r_T):={3\over 2}\sqrt{\tilde{N}^{(T)}_k} \cdot \tau_w(k;r_T),\qquad k\in\II,
$$
where $\tilde{N}^{(T)}_k$ is ``effective sample size", transforming them into $p$-values and doing inference by multiple comparison corrections. The new adaptive version local colocalization analysis (SACA) is summarized in Algorithm~\ref{ag:laca}. The pixel-wise $z$-score output from SACA not only reflects colocalization or anti-colocalization at each pixel by its sign, but also shows the degree of colocalization or anti-colocalization through its absolute values. 

\begin{algorithm}[h!]
\caption{Spatially Adaptive Colocalization Analysis (SACA)}
\label{ag:laca}
\begin{algorithmic}
\REQUIRE Image data with two channels $(X_k,Y_k)$, $k\in \II$.
\ENSURE Pixel-wise $z$ score, $p$-values, and significant region $R$
\STATE Evaluate  $\tau_w(k;r_0)$ by data in neighborhood $B(k,r_0)$
\FOR{$t=1,\ldots,T^U$}
\STATE Update adaptive weights $w_i(k;r_{t-1})$ for all $i\in B(k,r_t)$
\STATE Evaluate $\tau_w(k;r_t)$ by data in neighborhood $B(k,r_t)$
\STATE Check the stopping criteria 
\ENDFOR
\STATE Evaluate pixel-wise $z$ score $Z(k;r_T)$
\STATE Transform $z$ score $Z(k;r_T)$ to pixel-wise $p$-value $P_k$
\STATE Do inference by correcting multiple comparisons and get significant region $R$

\RETURN $Z(k;r_T)$, $P_k$ and $R$
\end{algorithmic}
\end{algorithm}

\section{Practical Considerations}

When implementing the framework described in the previous section, several practical issues need to be more carefully examined; one is the potential computational cost, which we shall address in Section \ref{sc:fastalgm}, and the other is the choice of various tuning parameters, which we shall discuss in detail in Section \ref{sc:paramter}.

\subsection{Fast Algorithm for Calculating $\tau_w$}
\label{sc:fastalgm}

In this section, we mainly discuss the fast algorithm for calculating weighted Kendall's tau coefficient $\tau_w$ in (\ref{eq:wtkendalltau}). Naive calculation of $\tau_w$ involves $n(n-1)/2$ terms, and thus its computation complexity is $O(n^2)$. This could be quite expensive to compute for pixel-wise local models, and particularly so as we need to repeat calculating $\tau_w$ many times at each pixel in the SACA algorithm.

Naive calculation for the unweighted Kendall's tau coefficient has a complexity $O(n^2)$. A much more efficient algorithm was introduced by \cite{knight1966}. It computes the unweighted Kendall's tau coefficient based on merging sorting algorithms and only requires $O(n\log n)$ floating-point operations. Knight's algorithm relies on a key observation that $\tau$ can be calculated from the number of discordant pairs $s$ if no ties exists in $X_i$ and $Y_i$,
$$
\tau=1-{4s\over n(n-1)}.
$$
The number of discordant pairs $s$ can be calculated as exchange count, which is the number of exchanges made in sorting the pairs of $(X_i,Y_i)$ with respect to $Y_i$, when they are originally ordered with respect to $X_i$. To obtain $s$, the merging sort algorithm is used to compute the number of exchanges.

The same strategy can also be extended to compute the weight Kendall's tau coefficient $\tau_w$. To the end, write $\tau_w$ as
$$
\tau_w=1-{4s_w\over \sum_{i\ne j}w_iw_j},
$$
where the weighted concordance $s_w$ is given by 
$$
s_w=\sum_{i>j}w_iw_j \mathbf{I}_{(X_i-X_j)(Y_i-Y_j)<0}.
$$
Here, we only discuss the case where there is no tie in data $(X_i, Y_i)$, and ties can be broken by a random perturbation on $X_i$ or $Y_i$. If $(X_i,Y_i)$ is ordered with respect to $X_i$ increasingly, i.e. $X_i>X_j$ for $i>j$, then $s_w$ can be written as
$$
s_w=\sum_{i>j}w_iw_j \mathbf{I}_{Y_i<Y_j}.
$$
To compute for $s_w$ above, we adopt a weighted version of the merging sort algorithm in Algorithm~\ref{ag:fasttau}. As with Knight's algorithm, the complexity of Algorithm~\ref{ag:fasttau} is $O(n\log n)$. As such, $\tau_w$ can be computed with $O(n\log n)$ floating-point operations, making our local colocalization analysis applicable to large scale images.

\begin{algorithm}[h!]
\caption{WeightMergeSort}
\label{ag:fasttau}
\begin{algorithmic}
\REQUIRE The data with weight $(Y_i,w_i)$, $1\le i\le n$.
\ENSURE Weighted concordance $s_w$ and ordered $(Y_i,w_i)$ with respect to $Y_i$

\STATE Split data $Z^{L}=\{(Y_i,w_i),1\le i\le n/2\}$ and $Z^{R}=\{(Y_i,w_i),n/2< i\le n\}$
\STATE Run WeightMergeSort($Z^L$) to get the weighted concordance $s_w^1$ and ordered data $Z^{Ls}$
\STATE Run WeightMergeSort($Z^R$) to get the weighted concordance $s_w^2$ and ordered data $Z^{Rs}$
\STATE Run CumSum($Z^{Ls}(w)$) to get the cumulative sum of weight $cw^{Ls}$ in $Z^{Ls}$
\STATE $i=1$, $j=1$ and $k=1$ 
\STATE $s_w=0$
\WHILE{$i\le n/2$ \AND $j\le n/2$}
\IF{$Z^{Ls}_i(Y)<Z^{Rs}_j(Y)$}
\STATE $Z^s_k=Z^{Rs}_j$ and $s_w=s_w+Z^{Rs}_j(w)*(cw^{Ls}_{n/2}-cw^{Ls}_{i-1})$
\STATE $j=j+1$ and $k=k+1$
\ELSE
\STATE $Z^s_k=Z^{Ls}_i$
\STATE $i=i+1$ and $k=k+1$
\ENDIF
\ENDWHILE
\STATE $s_w=s_w+s_w^1+s_w^2$
\RETURN $Z^s$ and $s_w$
\end{algorithmic}
\end{algorithm}

\begin{algorithm}[h!]
\caption{CumSum}
\label{ag:cumsum}
\begin{algorithmic}
\REQUIRE The data $w_i$, $1\le i\le n$.
\ENSURE Cumulative sum of $w_i$

\STATE $cw_0=0$
\FOR{$i=1$ to $n$}
\STATE $cw_i=cw_{i-1}+w_i$
\ENDFOR
\RETURN $cw$
\end{algorithmic}
\end{algorithm}

\subsection{Choices of Parameters}
\label{sc:paramter}

To implement the SACA framework, we also need to assign appropriate values for several tuning parameters. In this section, we discuss the choice of these parameters. In particular, we suggest two types of neighborhoods $B(k;r)$: a ball with Euclidian distance $\l_2$ and distance $\l_\infty$, i.e.  
$$
B(k;r)=\{i\in \II|\ d(k,i)<r\},
$$
where $d(k,i)=\|k-i\|_{\l_2}\ {\rm or}\ \|k-i\|_{\l_\infty}$ because they are easily implementable. We use the $\l_\infty$ ball in Section~\ref{sc:numerical}. The radius $r_t$ in each step is actually the bandwidth of local kernel methods. To learn the edge of regions with different colocalization levels, we suggest the initial neighborhood $r_0$ shall be chosen to be as small as possible, e.g. $r_0=1$. A large $r_0$ will cause an over-smoothing problem as the procedure without adaptive kernel, which is also discussed in detail in Section~\ref{sc:numerical}. The $r_t$ can be chosen as a geometric sequence, i.e. $r_t=r_0c_r^t$, where $c_r$ controls the growth speed of $r_t$. If a large value for $c_r$ is chosen, a large number of new points are brought into that new neighborhood so that the estimation is not stable and robust. The choice of $c_r=1.15$ and $r_0=1$ is supported by our experience and will be used in all numerical experiments in Section~\ref{sc:numerical}. There are a number of choices for kernel $\bK_l$, as the weighted model is widely used in nonparametric statistics \citep[see][]{gyorfi2006}. Since the $r_t$ increases slowly as suggested above, we choose $\bK_l$, not placing too much penalty on the points near the edge of neighborhood $B(k,r)$; in particular, we choose $\bK_l(x)=\max(1-x,0)$.

Since the adaptive weights are crucial to separate the regions with different colocalization levels, $\bK_s$ and $D_n$ need to be chosen carefully. If we put too little penalty on large $D_i(k;r_t)$, then the algorithm performs very similarly as the one without adaptive weights. On the other hand, the estimation is forced to be piecewise constant if too much penalty is placed on $D_i(k;r_t)$. As our theoretical developments, more specifically Theorems~\ref{thm:globalhomo}, \ref{thm:localhomo}, and \ref{thm:separate}, suggest, a good choice is $D_n=\sqrt{\log n}$ and $\bK_s(x)=\max((1-x/2)^2,0)$, and we shall use these values in all numerical experiments of this paper. In this way, the pixel $i$ plays no role in estimating $\tau(k,r_t)$ if $D_i(k;r_t)>2\sqrt{\log n}$. Based on our experience in simulations, the performance of SACA is sensitive to the choices of $\bK_s$ and $D_n$.

Three additional parameters $T^L$, $\Lambda$, and $T^U$ are used in our stopping criteria. As the estimation at step $T_L$ is regarded as a benchmark for subsequent estimation, $T_L$ cannot be too small. We assume $T^L\ge 8$ because the estimation at the first few steps is typically not reliable. Large $T_U$ incurs heavy computational cost and potential over-smoothing. We suggest to choose $T^U\le 15$. As the theoretical properties suggest, $\Lambda=\eta\sqrt{\log n}$ and $\eta=1$ are reasonable choices and will be used in our implementation. In addition, we also need to estimate the thresholds for $t_X$ and $t_Y$ in kernel $\bK_b$ and recommend to do so using the Otsu method \citep[see][]{sezgin2004} for each channel. 

\section{Numerical Experiments}
\label{sc:numerical}

We now conduct numerical experiments to further demonstrate the practical merits of our SACA framework in this section.

\subsection{Simulation}

\begin{figure}[h!]
    \centering
     \begin{subfigure}[b]{0.24\textwidth}
     \centering
        \begin{tikzpicture}[scale=1]
  \node[anchor=south west,inner sep=0] at (0,0) {\includegraphics[width=0.85\textwidth]{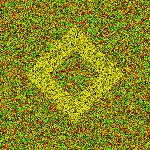}};
    \end{tikzpicture}
        \caption{Simulated example $(\theta=6)$}
        \label{fg:simsample}
    \end{subfigure}
    \begin{subfigure}[b]{0.24\textwidth}
     \centering
        \begin{tikzpicture}[scale=1]
  \node[anchor=south west,inner sep=0] at (0,0) {\includegraphics[width=0.85\textwidth]{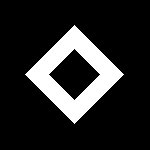}};
    \end{tikzpicture}
        \caption{True region distribution}
        \label{fg:simmask}
    \end{subfigure}
    \begin{subfigure}[b]{0.24\textwidth}
    \centering
        \begin{tikzpicture}[scale=1]
  \node[anchor=south west,inner sep=0] at (0,0) {\includegraphics[width=0.97\textwidth]{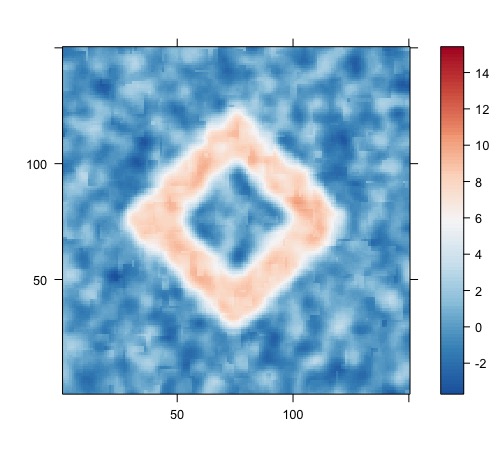}};
    \end{tikzpicture}
        \caption{LCA $(r=6)$}
        \label{fg:noadptd6}
    \end{subfigure}
    \begin{subfigure}[b]{0.24\textwidth}
    \centering
        \begin{tikzpicture}[scale=1]
  \node[anchor=south west,inner sep=0] at (0,0) {\includegraphics[width=0.97\textwidth]{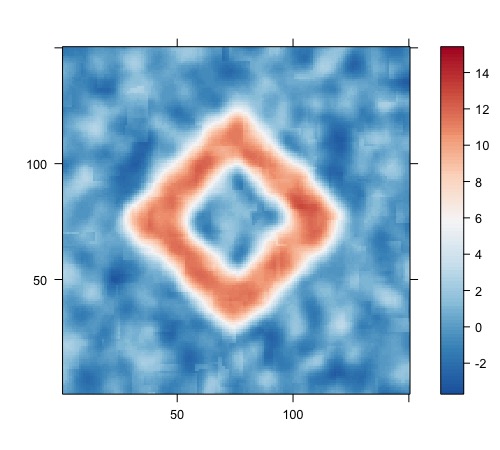}}; 
    \end{tikzpicture}
        \caption{LCA $(r=8)$}
        \label{fg:noadptd8}
    \end{subfigure}
    \begin{subfigure}[b]{0.24\textwidth}
    \centering
        \begin{tikzpicture}[scale=1]
  \node[anchor=south west,inner sep=0] at (0,0) {\includegraphics[width=0.97\textwidth]{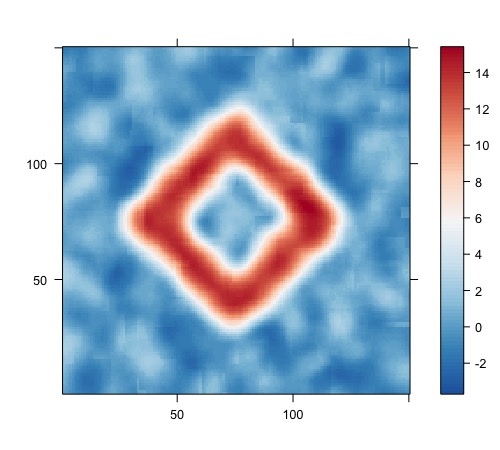}};
    \end{tikzpicture}
        \caption{LCA $(r=10)$}
        \label{fg:noadptd10}
    \end{subfigure}
    \begin{subfigure}[b]{0.24\textwidth}
    \centering
        \begin{tikzpicture}[scale=1]
  \node[anchor=south west,inner sep=0] at (0,0) {\includegraphics[width=0.97\textwidth]{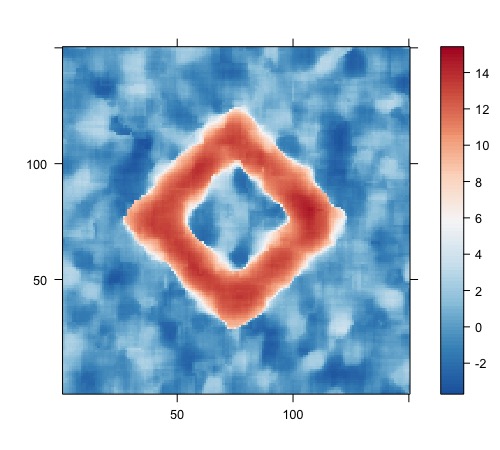}}; 
    \end{tikzpicture}
        \caption{SACA}
        \label{fg:adp}
    \end{subfigure}
    \caption{A typical simulated example and corresponding analysis results when comparing LCA with fixed radii to SACA.}\label{fg:smootheffect}
\end{figure}

The first simulation experiment we consider here is to assess the effect of adaptive weights of SACA in equation (\ref{eq:adpwt}) of Section~\ref{sc:adpwt}. To this end, we simulated an image on a $150\times 150$ squared lattice. The pixels on simulated images can be divided into two groups: background (null hypothesis $H_0$) and colocalized pixels (alternative hypothesis $H_1$), as shown in black and white regions of Figure~\ref{fg:simmask}. In the background, the data $(X_i,Y_i)$ is simulated from independent uniform distributions between $0$ and $1$; the data $(X_i,Y_i)$ in the colocalized signal region is generated from the distribution $F_\theta(x,y)=C_\theta(x^2,y^2)$ for Clayton copula $C_\theta(u,v)$ \citep[see][]{nelsen2006}
$$
C_\theta(u,v)=\left[\max\left(u^{-\theta}+v^{-\theta}-1,0\right)\right]^{-1/\theta},
$$
so that the marginal distribution of $X_i$ and $Y_i$ are $x^2$ and $y^2$, and the population Kendall tau of $F_\theta(x,y)$ is $\tau_\theta=\theta/(\theta+2)$. Clearly, $(X_i,Y_i)$ is colocalized if $\theta>0$ and non-colocalized otherwise. A typical example under this simulation setting is shown in Figure~\ref{fg:simsample} when $\theta=6$. 

In this simulation experiment, we compare local colocalization analysis (LCA) with a fixed radius and the spatially adaptive approach SACA. We chose $r=6, 8,$ and $10$ and $t_X=t_Y=0.3$ in LCA and applied them to the data example in Figure~\ref{fg:simsample}, obtaining pixel-wise $z$-scores $Z(k,r)$ plotted in Figure \ref{fg:noadptd6}, \ref{fg:noadptd8}, and \ref{fg:noadptd10}. When $r$ is small, e.g. $r=6$, $Z(k;r)$s do not have enough power to detect colocalization, as shown in Figure \ref{fg:noadptd6}. The $Z(k;r)$s with large $r$, e.g. $r=10$, blur the edges of regions with different colocalization levels as shown in Figure~\ref{fg:noadptd10}. Therefore, an appropriate way to determine the radius $r$ in LCA is crucial for best performance. We also applied SACA to the example in Figure~\ref{fg:simsample}, and the result is shown in Figure~\ref{fg:adp}. Figure~\ref{fg:adp} suggests SACA can learn the sharp edge of colocalized regions and has more detection power for subtle levels of colocalization.
 
To investigate the effect of colocalization levels, equivalently $\theta$ in the model above, we also conducted the simulation by varying $\theta$ from $1$ to $7$. After $z$-scores were calculated by SACA, we transformed them to pixel-wise $p$-values and corrected multiple comparisons by the Bonferroni method at level $5\%$. To assess the performance of LCA and SACA in each experiment, we recorded the false discovery rate $\alpha$ and true positive rate $\beta$
$$
\alpha={\#{\rm False\ Positive}\over \#{\rm Prediction\ Positive}}\ {\rm and}\ \beta={\#{\rm True\ Positive} \over \#{\rm Condition\ Positive}}.
$$
The experiment was repeated 500 times for each combination of methods and colocalization level $\theta$.  The results are summarized in Table~\ref{tb:smootheffect}. These results show that SACA, with adaptive weights, has a better ability to control false discovery rates and more detection power than LCA.

\begingroup
\renewcommand{\arraystretch}{1.6}
\begin{table*}[h!]
\begin{center}
\begin{tabular}{c c c c c c c c c c c c c}
\hline
\hline
 & & \multicolumn{2}{c}{LCA $(r=6)$} & &  \multicolumn{2}{c}{LCA $(r=8)$}& &\multicolumn{2}{c}{LCA $(r=10)$}& &\multicolumn{2}{c}{SACA} \\
 \cline{3-4}  \cline{6-7}  \cline{9-10} \cline{12-13}
 & & $\alpha$ & $\beta$ & & $\alpha$ & $\beta$ & & $\alpha$ & $\beta$ & & $\alpha$ & $\beta$ \\
\hline
$\theta=1$ & &  $ 0.008 $ & $ 0.01 $ & &  $ 0.009 $ & $ 0.075 $ & &  $ 0.016 $ & $ 0.237 $ & &  $ 0.026 $ & $ 0.223 $ \\ 
$\theta=2$ & &  $ 0.003 $ & $ 0.342 $ & &  $ 0.018 $ & $ 0.766 $ & &  $ 0.066 $ & $ 0.922 $ & &  $ 0.062 $ & $ 0.903 $ \\ 
$\theta=3$ & &  $ 0.007 $ & $ 0.765 $ & &  $ 0.049 $ & $ 0.944 $ & &  $ 0.136 $ & $ 0.989 $ & &  $ 0.094 $ & $ 0.981 $ \\ 
$\theta=4$ & &  $ 0.014 $ & $ 0.887 $ & &  $ 0.080 $ & $ 0.981 $ & &  $ 0.176 $ & $ 0.996 $ & &  $ 0.103 $ & $ 0.991 $ \\ 
$\theta=5$ & &  $ 0.021 $ & $ 0.929 $ & &  $ 0.102 $ & $ 0.991 $ & &  $ 0.197 $ & $ 0.998 $ & &  $ 0.103 $ & $ 0.993 $ \\ 
$\theta=6$ & &  $ 0.028 $ & $ 0.951 $ & &  $ 0.116 $ & $ 0.994 $ & &  $ 0.211 $ & $ 0.999 $ & &  $ 0.099 $ & $ 0.994 $ \\ 
$\theta=7$ & &  $ 0.034 $ & $ 0.963 $ & &  $ 0.127 $ & $ 0.996 $ & &  $ 0.22 $ & $ 0.999 $ & &  $ 0.095 $ & $ 0.994 $ \\
\hline
\hline
\end{tabular}
\end{center}
\caption{Comparison between LCA $(r=6,8,10)$ and SACA.}
\label{tb:smootheffect}
\end{table*}%
\endgroup

\begin{figure}[h!]
\begin{center}
\includegraphics[width=0.4\textwidth]{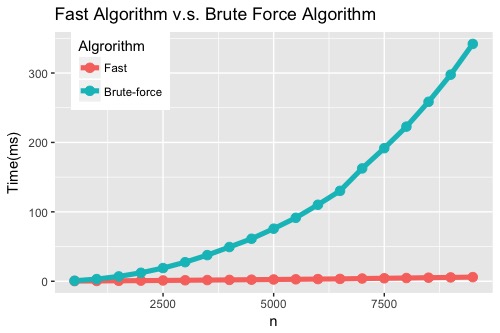}
\caption{Comparison between the naive brute force algorithm and fast algorithm.}
\label{fg:compalgm}
\end{center}
\end{figure}

Our next set of simulations was designed to compare two ways to compute Kendall tau correlation coefficients: the naive brute force algorithm and the fast algorithm proposed in Section~\ref{sc:fastalgm}. To this end, we simulated $X_i$, $Y_i$, and $w_i$, $i=1,\ldots n$, from independent uniform distribution between $0$ and $1$.  The computing times of both algorithms are reported in Figure~\ref{fg:compalgm}, which are also based on $1000$ runs for each $n$. It is clear from Figure~\ref{fg:compalgm} that the fast algorithm in Section~\ref{sc:fastalgm} is much more efficient than the brute force algorithm, which was expected.

\subsection{Real Data Examples}

\begin{figure}[h!]
    \centering
    \begin{tikzpicture}[scale=1]
  \node[inner sep=0] at (0,0) {\includegraphics[width=0.23\textwidth]{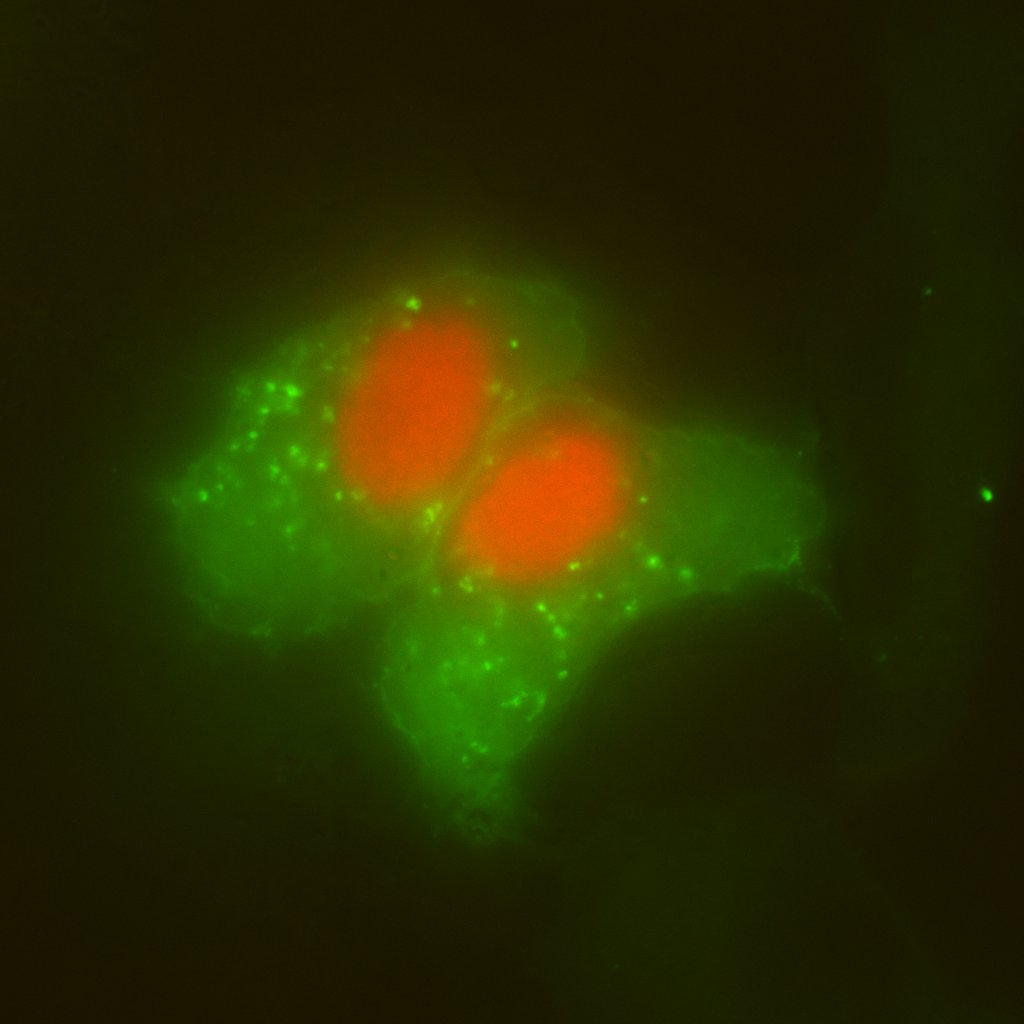}};
  \node[inner sep=0] at (0,-4.4) {\includegraphics[width=0.23\textwidth]{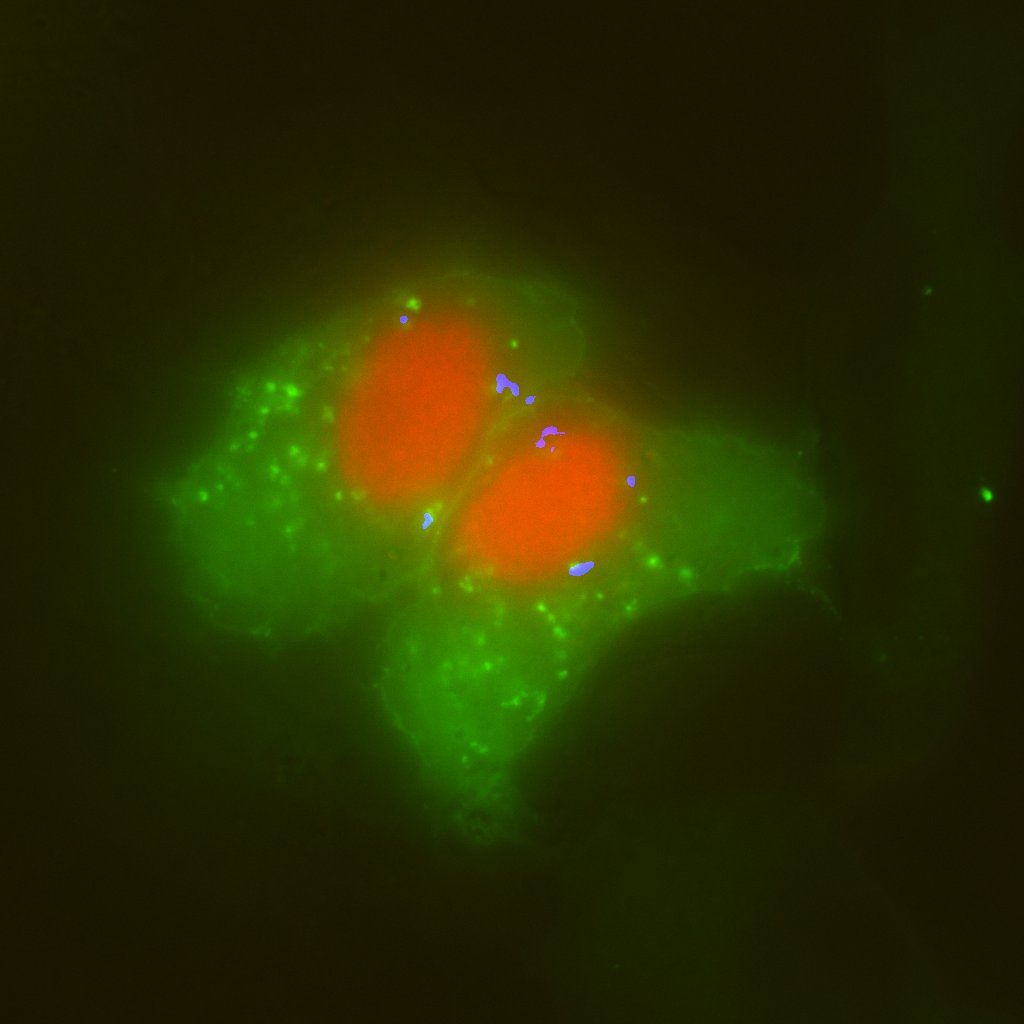}}; 
  \node[inner sep=0] at (0,-8) {\includegraphics[width=0.24\textwidth]{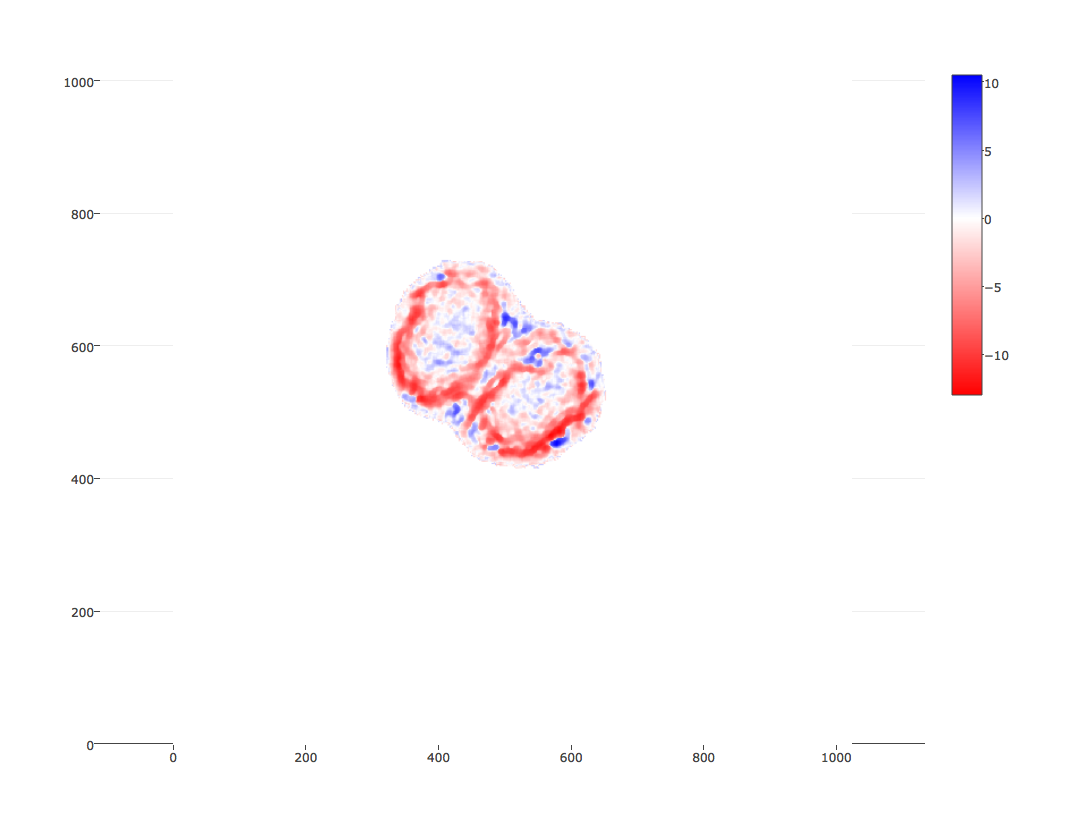}}; 
  \node[inner sep=0] at (4.4,0) {\includegraphics[width=0.23\textwidth]{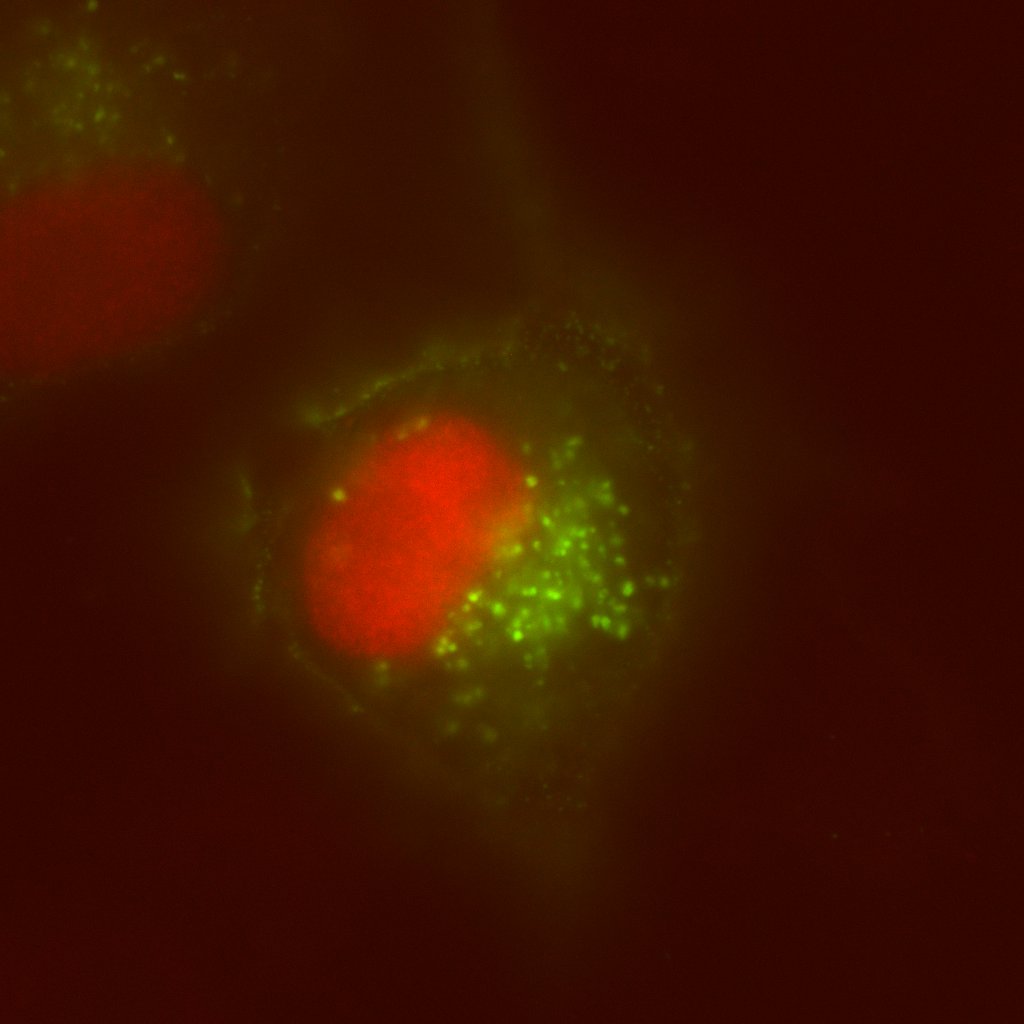}};
  \node[inner sep=0] at (4.4,-4.4) {\includegraphics[width=0.23\textwidth]{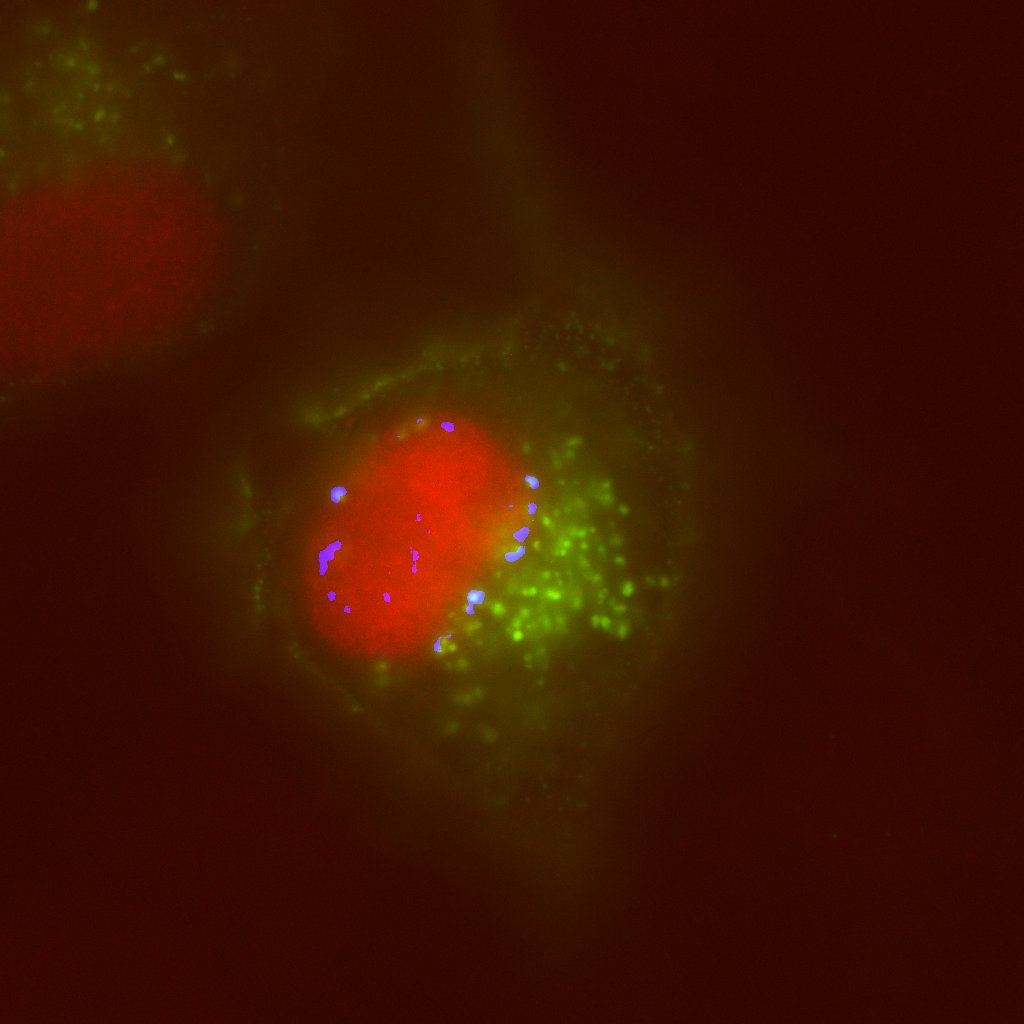}};
  \node[inner sep=0] at (4.4,-8) {\includegraphics[width=0.24\textwidth]{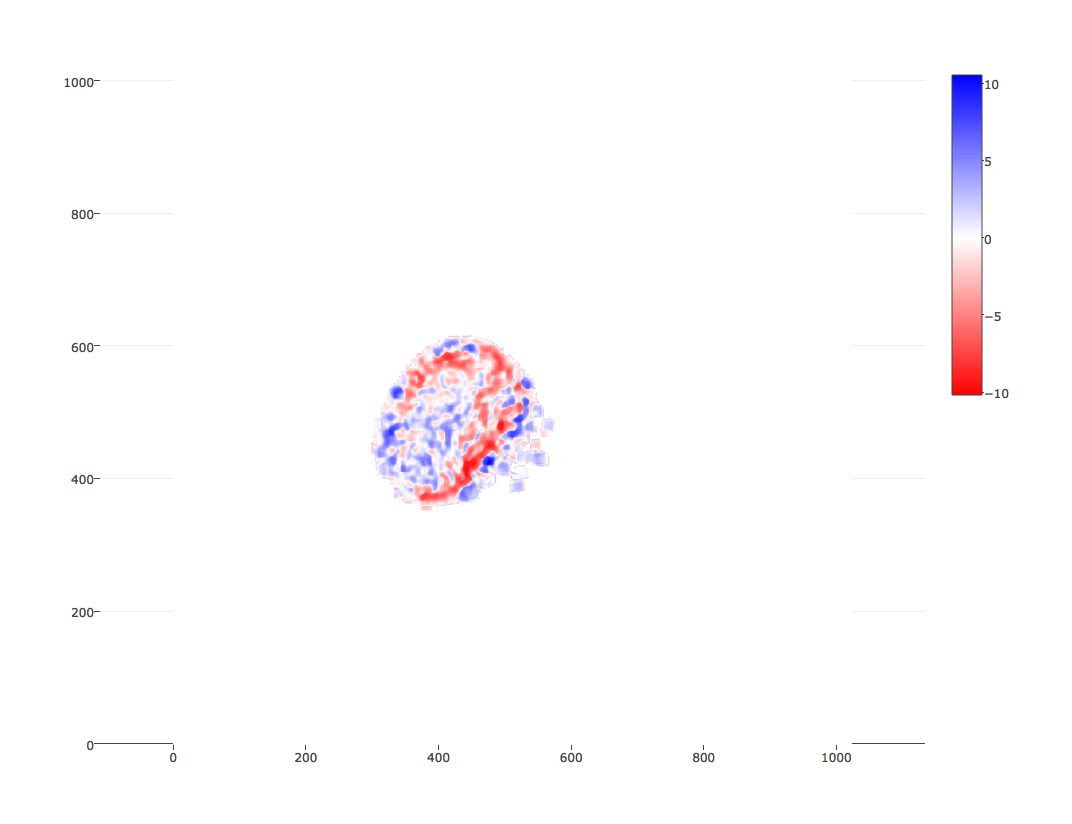}};
     \draw[line width=2pt,black!80!white] (-2.2,2.2) rectangle (6.6,-9.65);
   \draw[line width=2pt,black!80!white] (2.2,2.25) to (2.2,-9.65);
    \end{tikzpicture}
    \caption{No colocalization is expected between MS2-YFP (red channel) and Gag-CFP (green channel). Original overlay image (upper); colocalized region labelled in blue (middle); and heat map of $z$-scores (lower). }\label{fg:realdatanocol}
\end{figure}

\begin{figure}[h!]
    \centering
    \begin{tikzpicture}[scale=1]
  \node[inner sep=0] at (0,0) {\includegraphics[width=0.23\textwidth]{figures/Pos1Merge}};
  \node[inner sep=0] at (0,-4.4) {\includegraphics[width=0.23\textwidth]{figures/Pos1Loc}}; 
  \node[inner sep=0] at (0,-8) {\includegraphics[width=0.24\textwidth]{figures/Pos1HM}}; 
  \node[inner sep=0] at (4.4,0) {\includegraphics[width=0.23\textwidth]{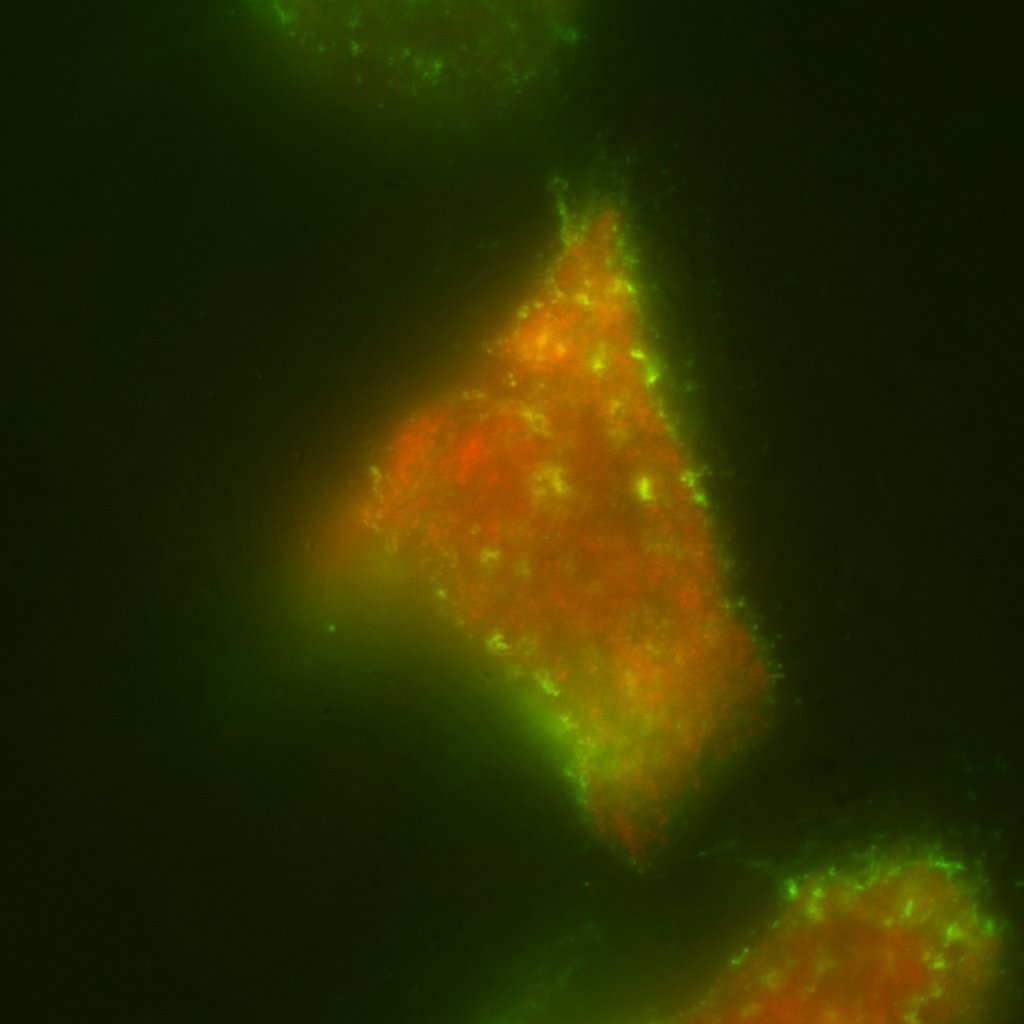}};
  \node[inner sep=0] at (4.4,-4.4) {\includegraphics[width=0.23\textwidth]{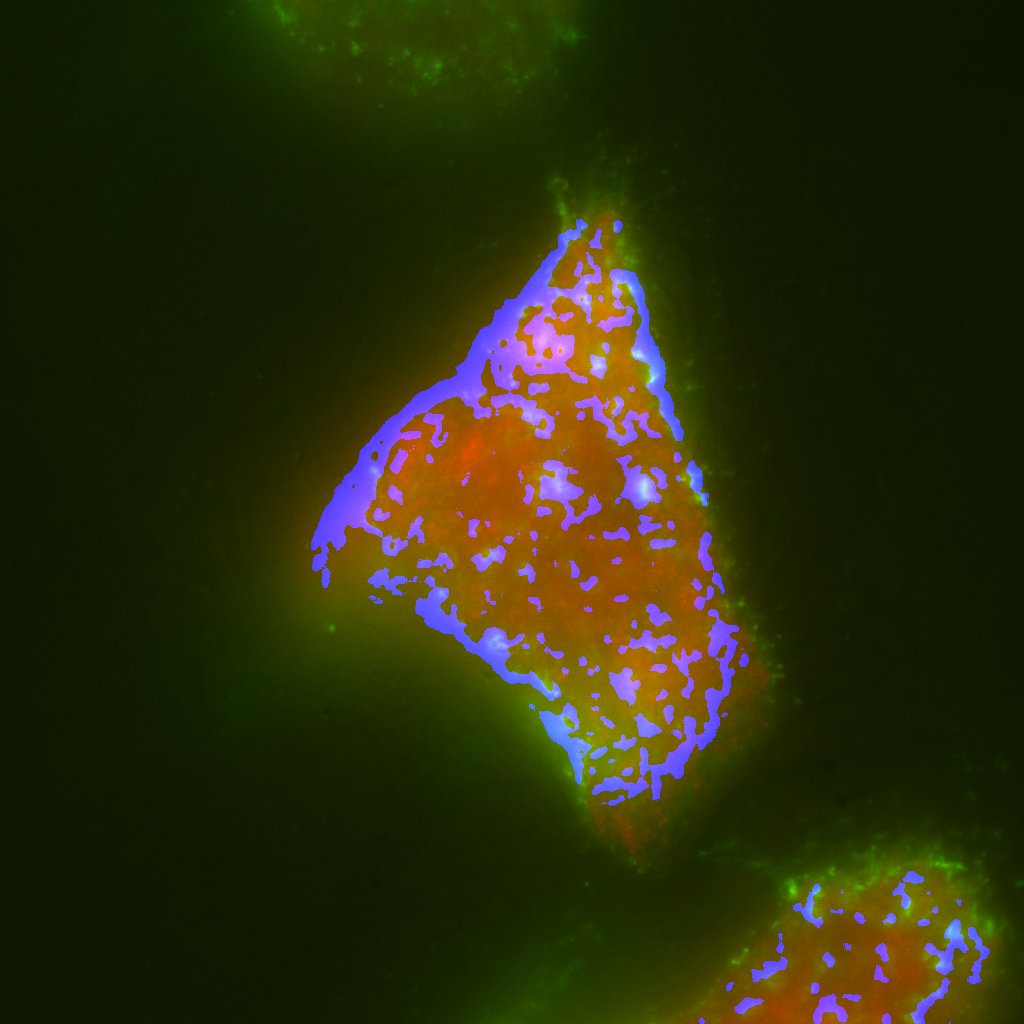}};
  \node[inner sep=0] at (4.4,-8) {\includegraphics[width=0.24\textwidth]{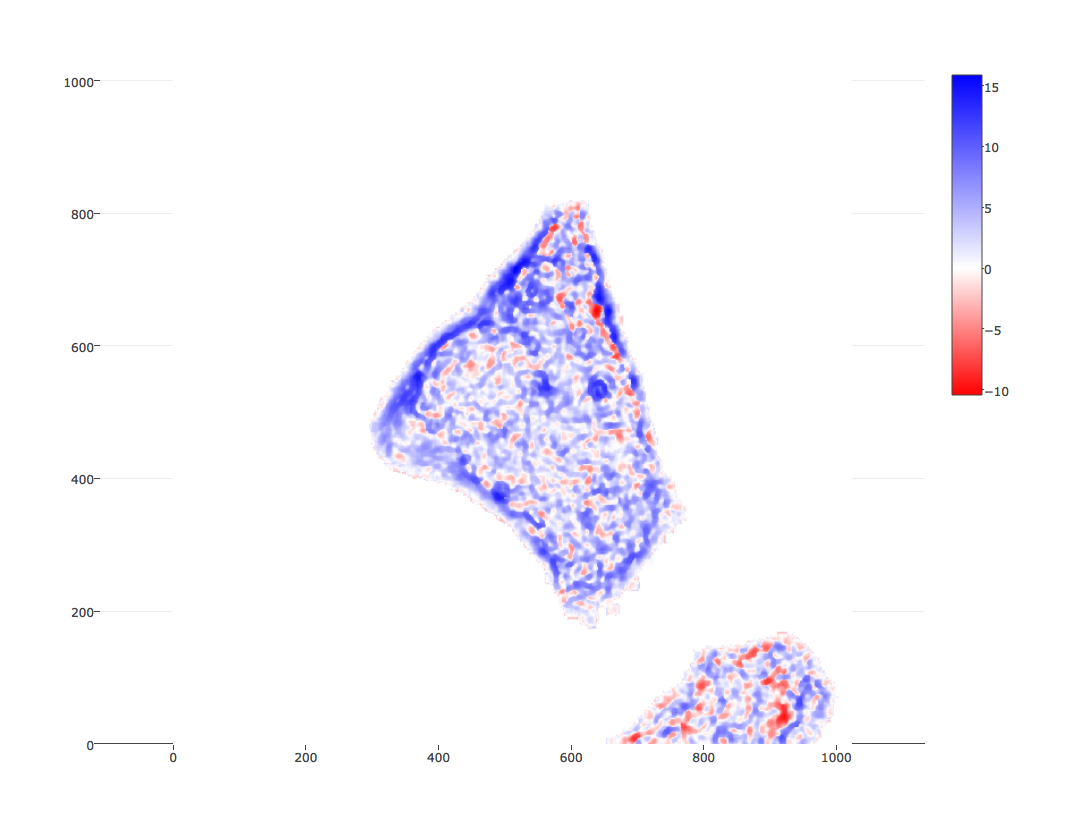}};
  \draw[line width=2pt,black!80!white] (-2.2,2.2) rectangle (6.6,-9.65);
   \draw[line width=2pt,black!80!white] (2.2,2.25) to (2.2,-9.65);
    \end{tikzpicture}
    \caption{Partial colocalization at the edge of cell is expected between MS2-YFP (red channel) and Gag-CFP (green channel). Original overlay image (upper); colocalized region labelled in blue (middle); and heat map of $z$-scores (lower).}\label{fg:realdatapacol}
\end{figure}

\begin{figure}[h!]
    \centering
    \begin{tikzpicture}[scale=1]
  \node[inner sep=0] at (0,0) {\includegraphics[width=0.23\textwidth]{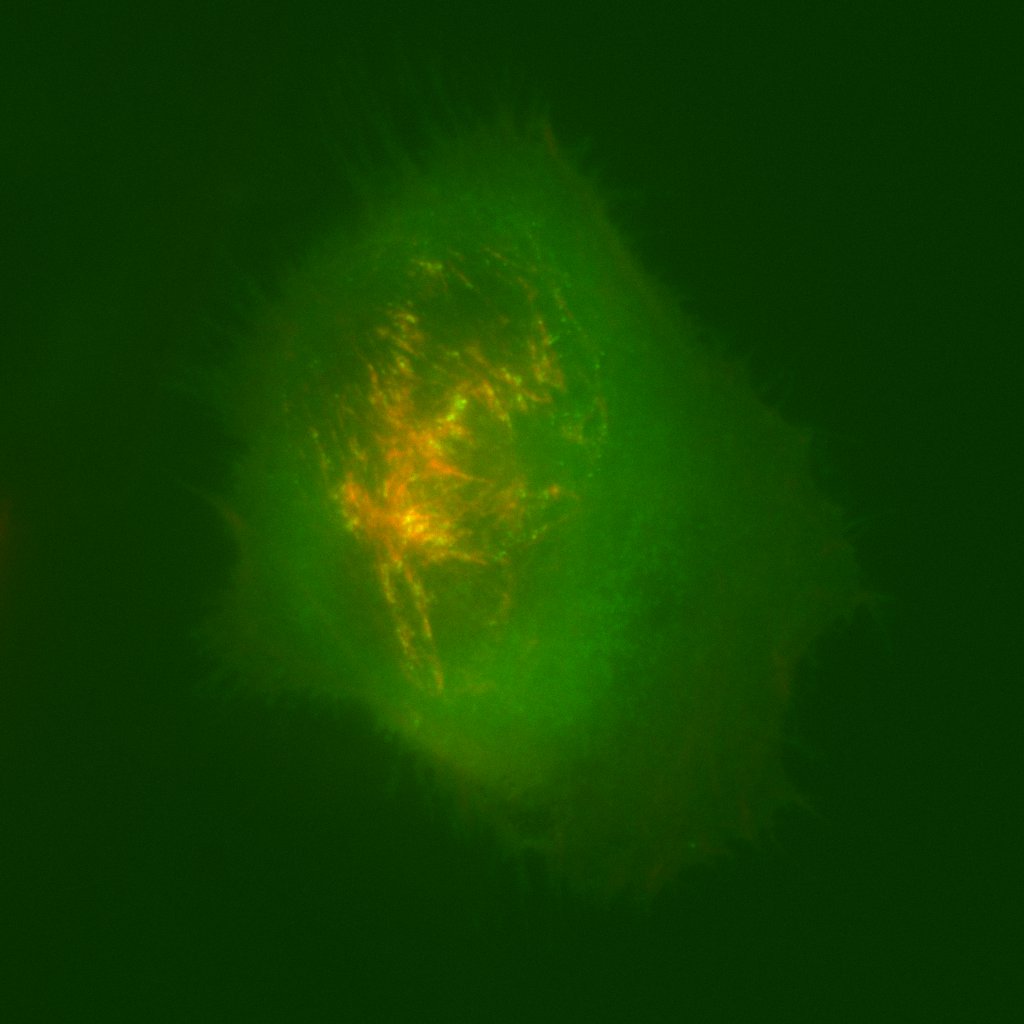}};
  \node[inner sep=0] at (0,-4.4) {\includegraphics[width=0.23\textwidth]{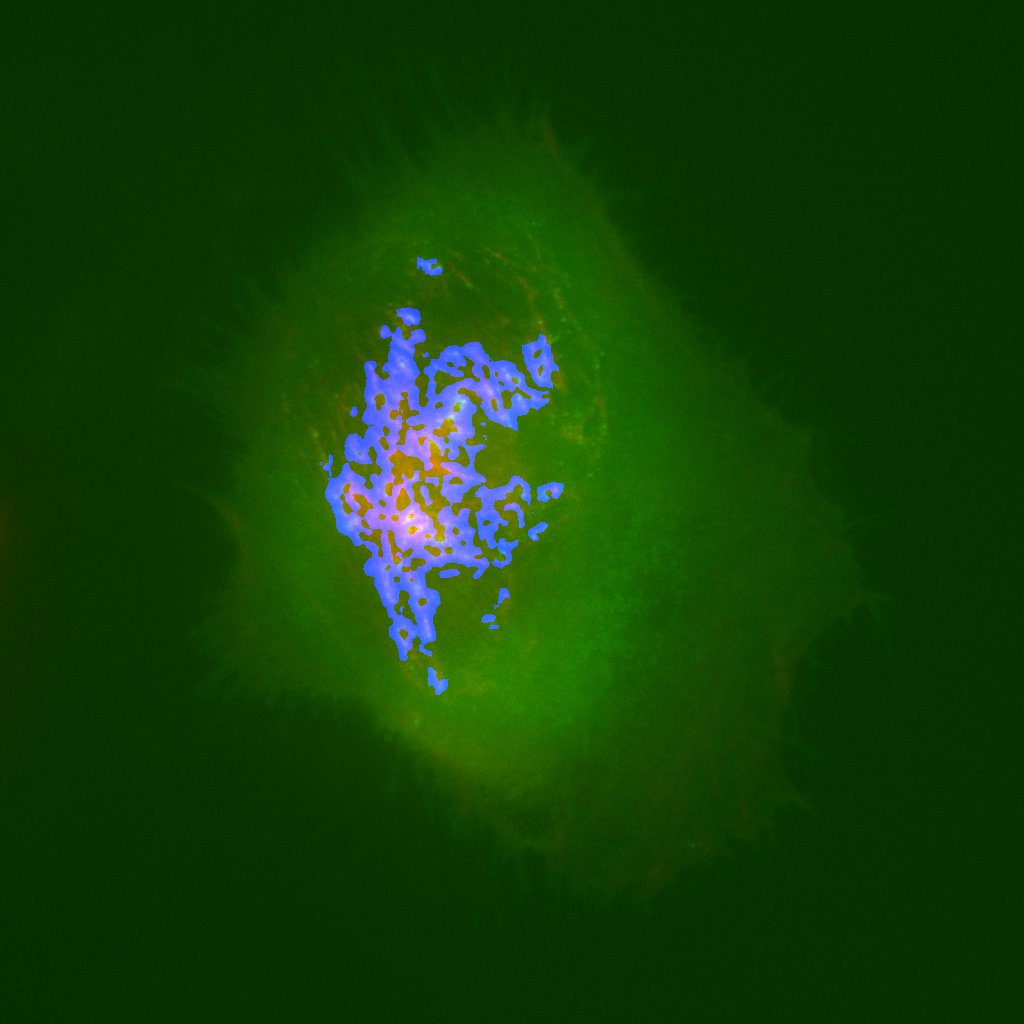}}; 
  \node[inner sep=0] at (0,-8) {\includegraphics[width=0.24\textwidth]{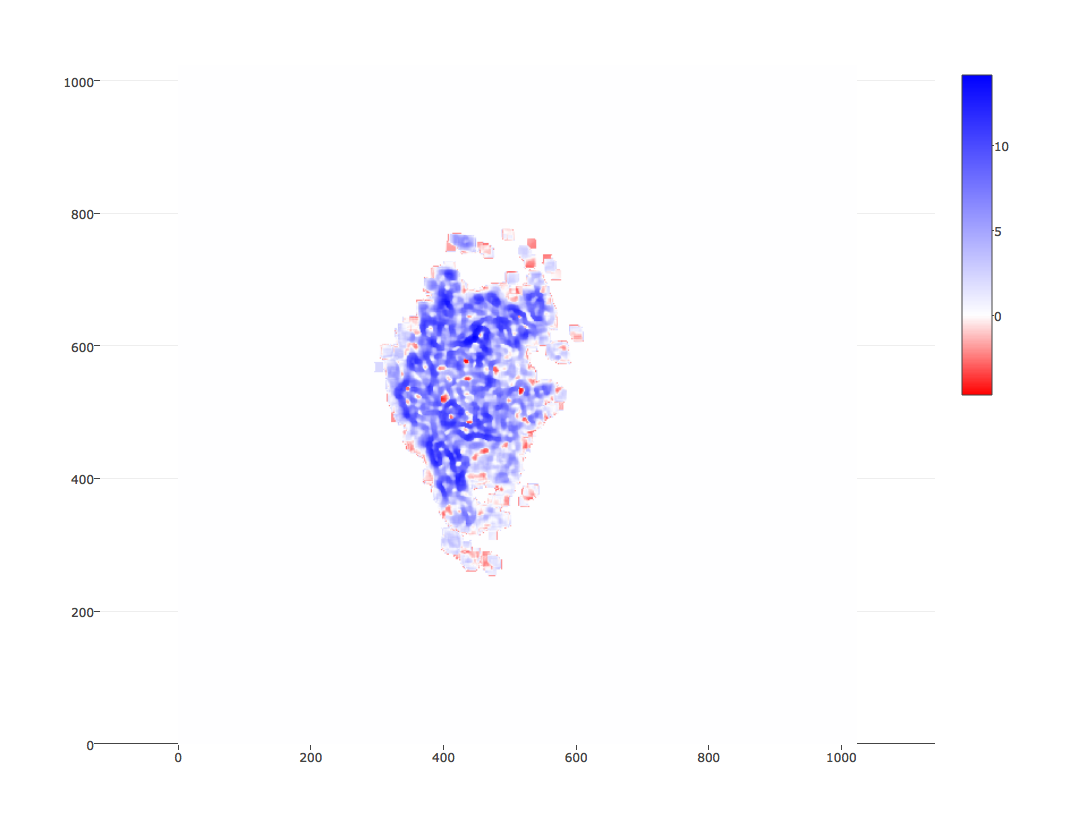}}; 
  \node[inner sep=0] at (4.4,0) {\includegraphics[width=0.23\textwidth]{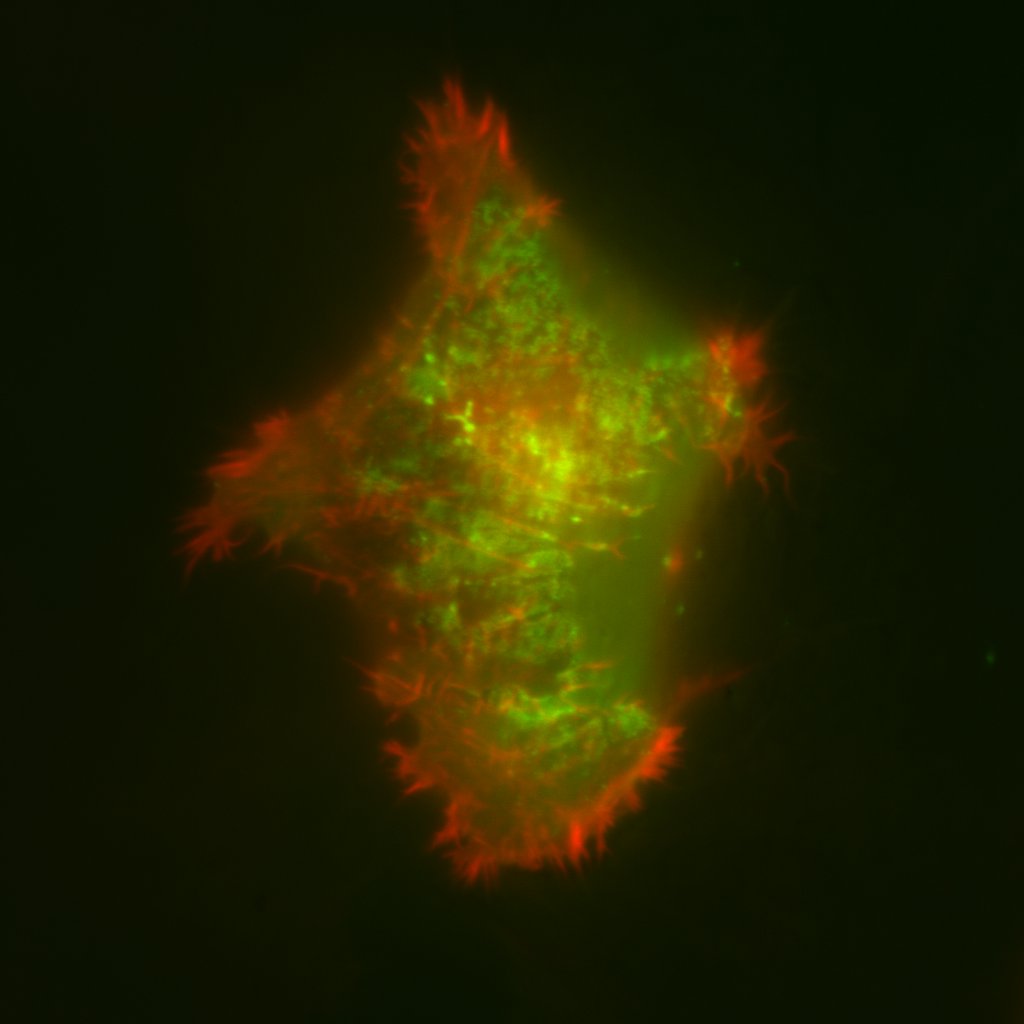}};
  \node[inner sep=0] at (4.4,-4.4) {\includegraphics[width=0.23\textwidth]{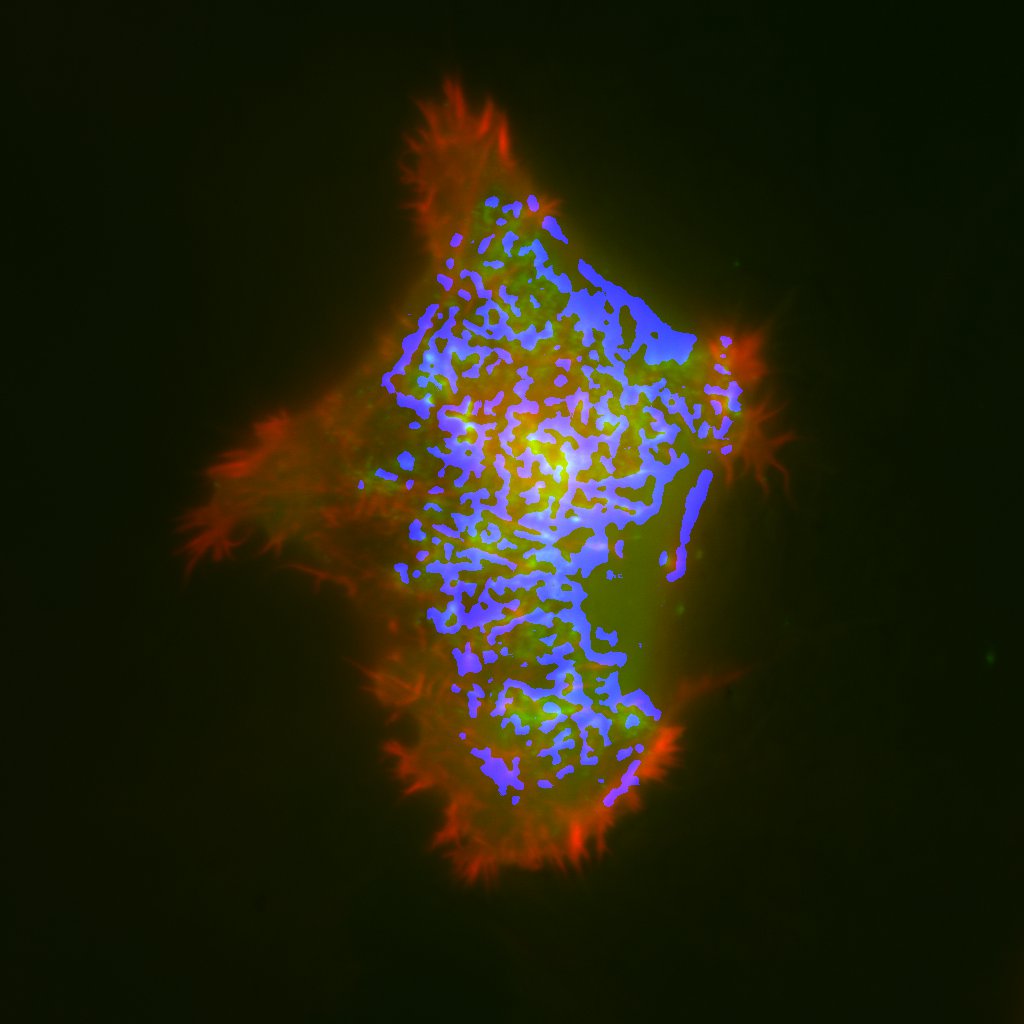}};
  \node[inner sep=0] at (4.4,-8) {\includegraphics[width=0.24\textwidth]{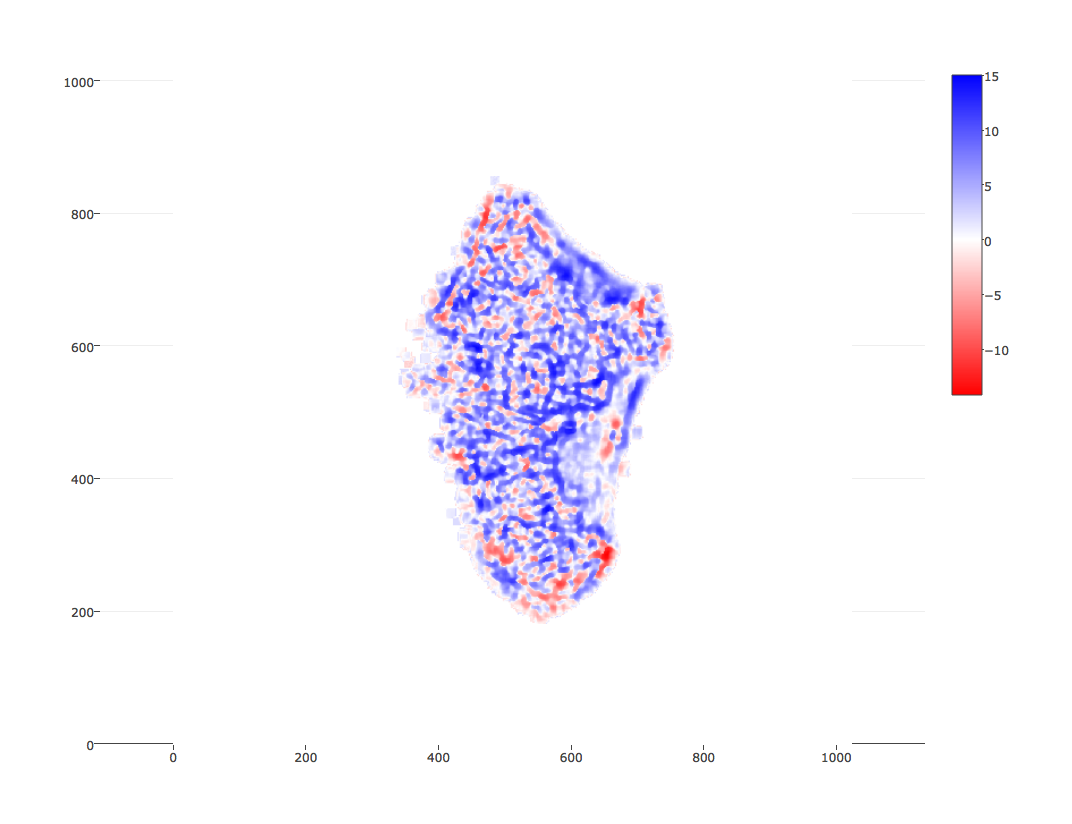}};
  \draw[line width=2pt,black!80!white] (-2.2,2.2) rectangle (6.6,-9.65);
   \draw[line width=2pt,black!80!white] (2.2,2.25) to (2.2,-9.65);
    \end{tikzpicture}
    \caption{Partial colocalization in actin fiber is expected between Lifeact-MS2-YFP (red channel) and Gag-CFP (green channel). Original overlay image (upper); colocalized region labelled in blue (middle); and heat map of $z$-scores (lower).}\label{fg:realdataactin}
\end{figure}

\begin{figure}[h!]
    \centering
    \begin{tikzpicture}[scale=1]
  \node[inner sep=0] at (0,0) {\includegraphics[width=0.23\textwidth]{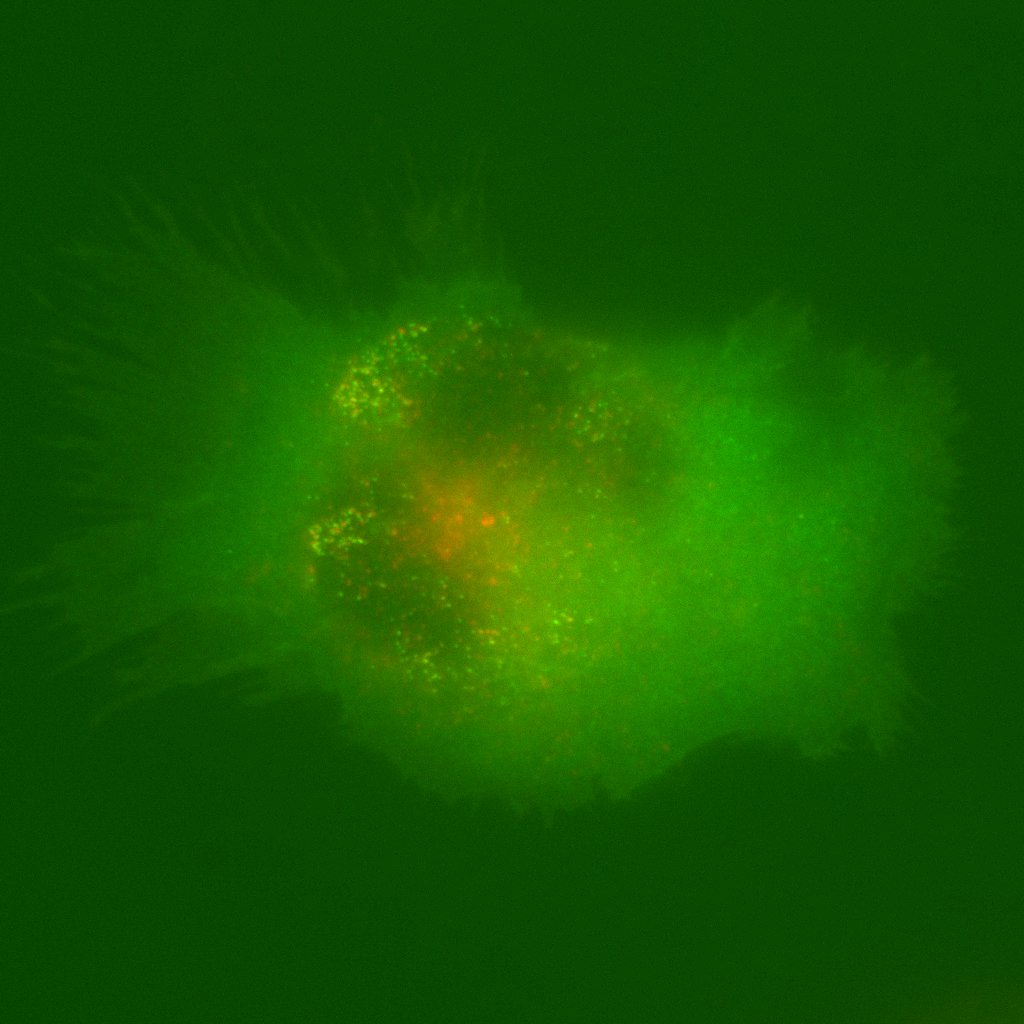}};
  \node[inner sep=0] at (0,-4.4) {\includegraphics[width=0.23\textwidth]{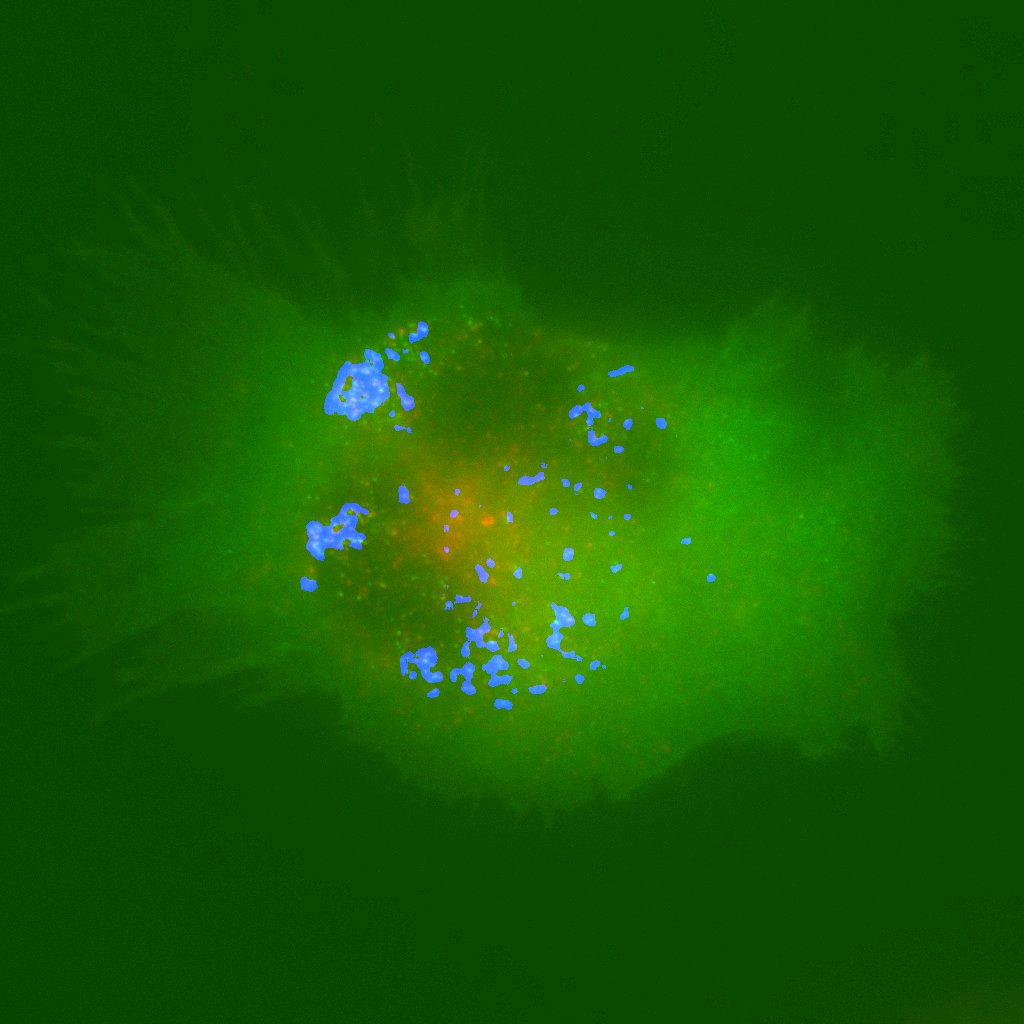}}; 
  \node[inner sep=0] at (0,-8) {\includegraphics[width=0.24\textwidth]{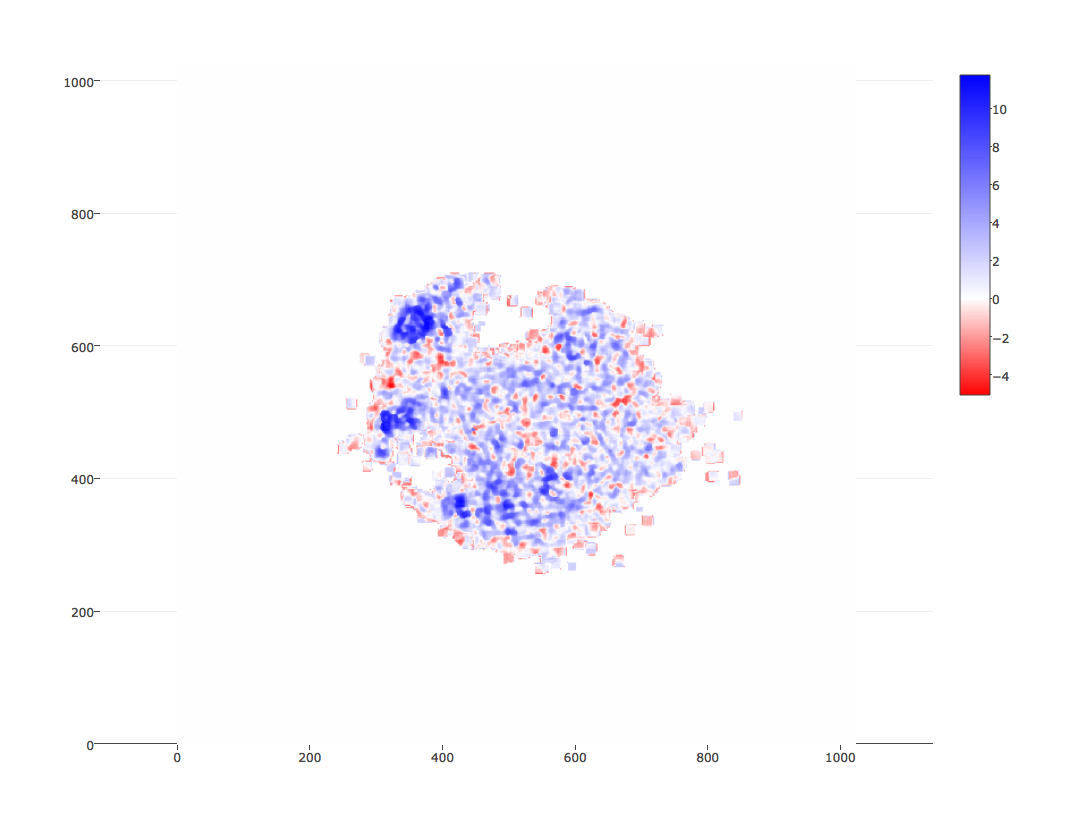}}; 
  \node[inner sep=0] at (4.4,0) {\includegraphics[width=0.23\textwidth]{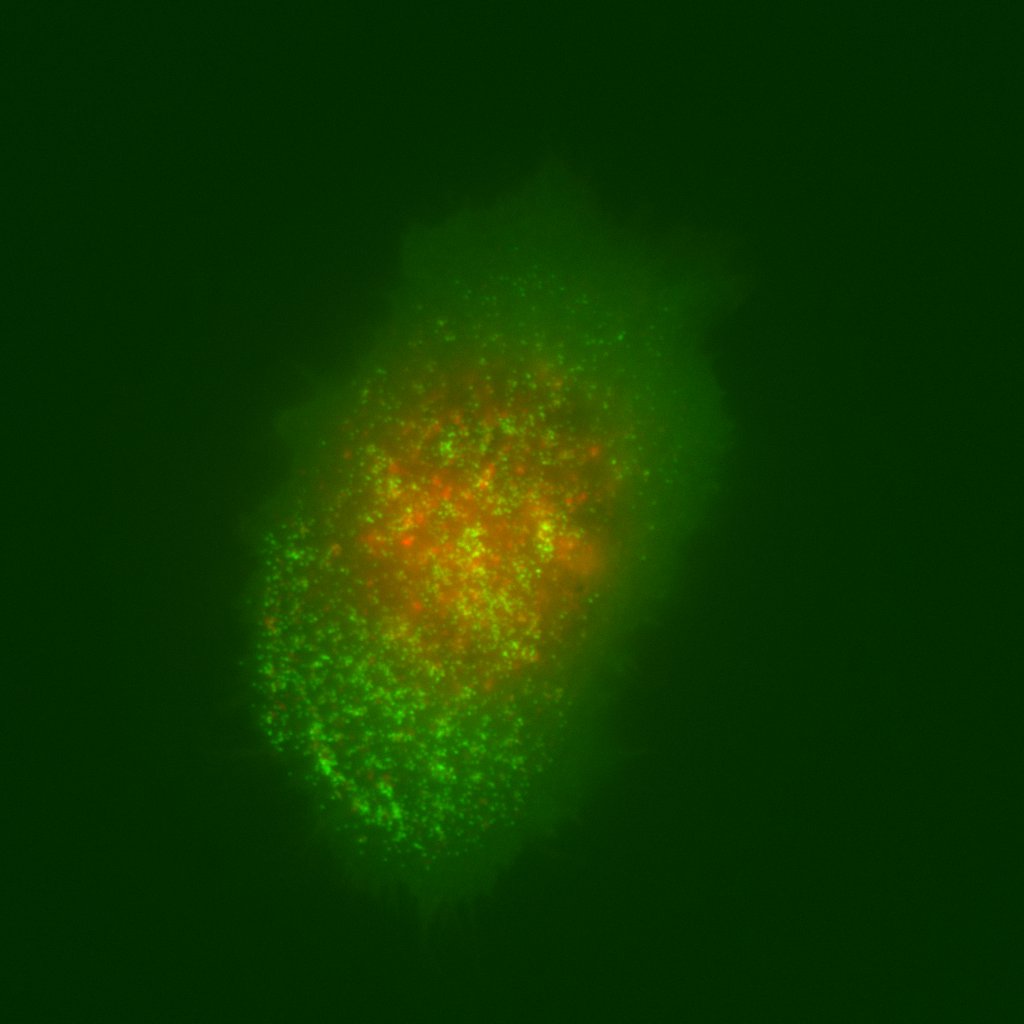}};
  \node[inner sep=0] at (4.4,-4.4) {\includegraphics[width=0.23\textwidth]{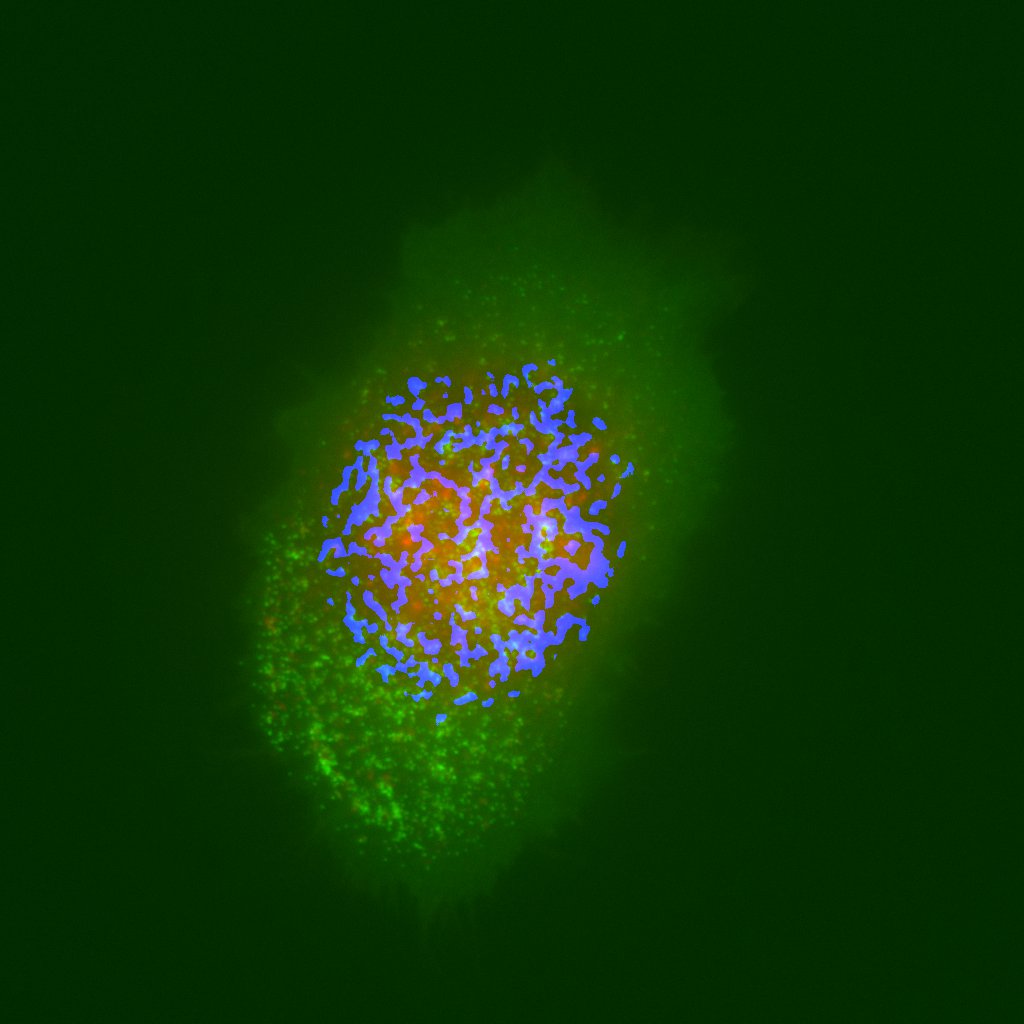}};
  \node[inner sep=0] at (4.4,-8) {\includegraphics[width=0.24\textwidth]{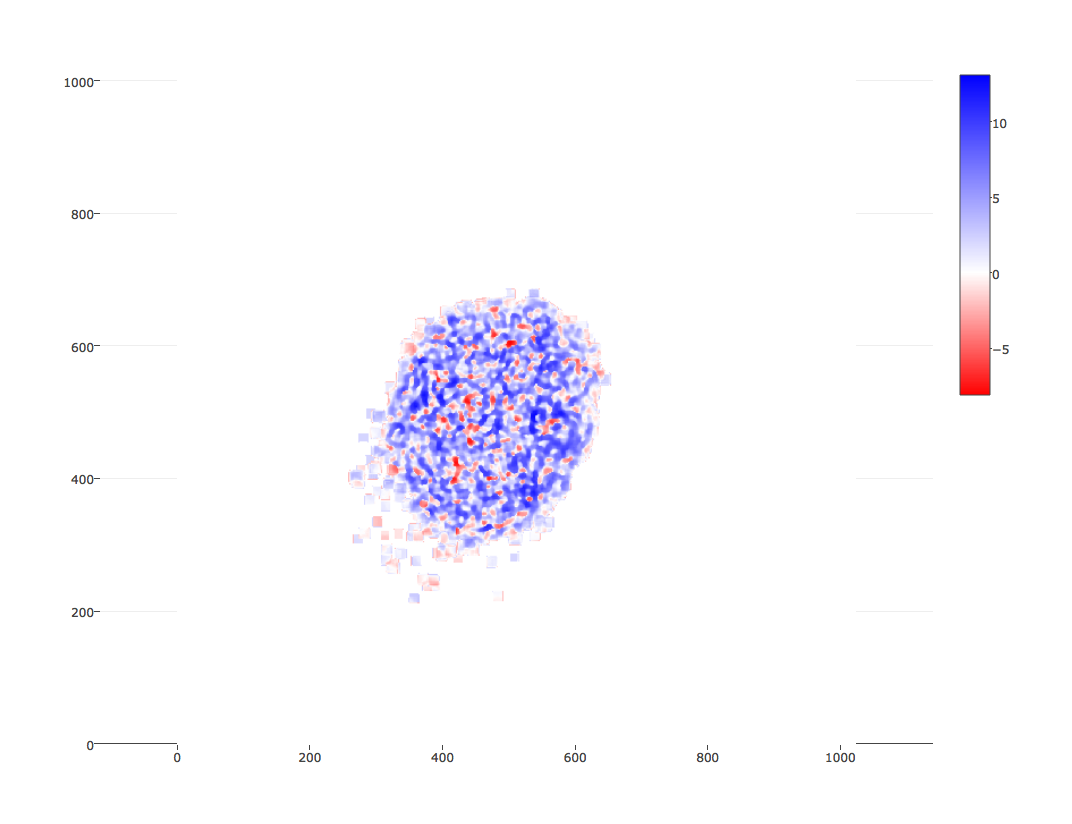}};
  \draw[line width=2pt,black!80!white] (-2.2,2.2) rectangle (6.6,-9.65);
   \draw[line width=2pt,black!80!white] (2.2,2.25) to (2.2,-9.65);
    \end{tikzpicture}
    \caption{Partial colocalization in intracellular vesicles is expected between Src-MS2-YFP (red channel) and Gag-CFP (green channel). Original overlay image (upper); colocalized region labelled in blue (middle); and heat map of $z$-scores (lower).}\label{fg:realdatamem}
\end{figure}

\begin{figure}[h!]
    \centering
    \begin{tikzpicture}[scale=1]
  \node[inner sep=0] at (0,0) {\includegraphics[width=0.23\textwidth]{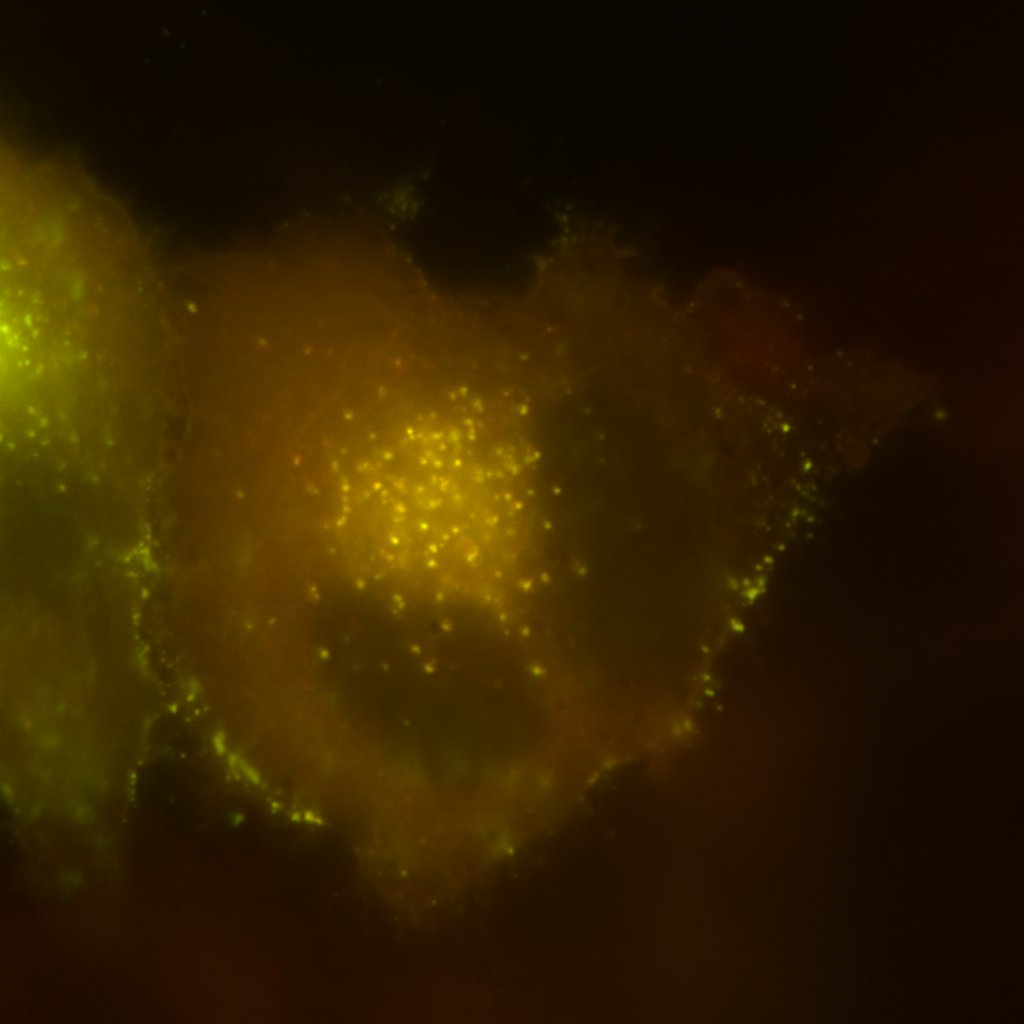}};
  \node[inner sep=0] at (0,-4.4) {\includegraphics[width=0.23\textwidth]{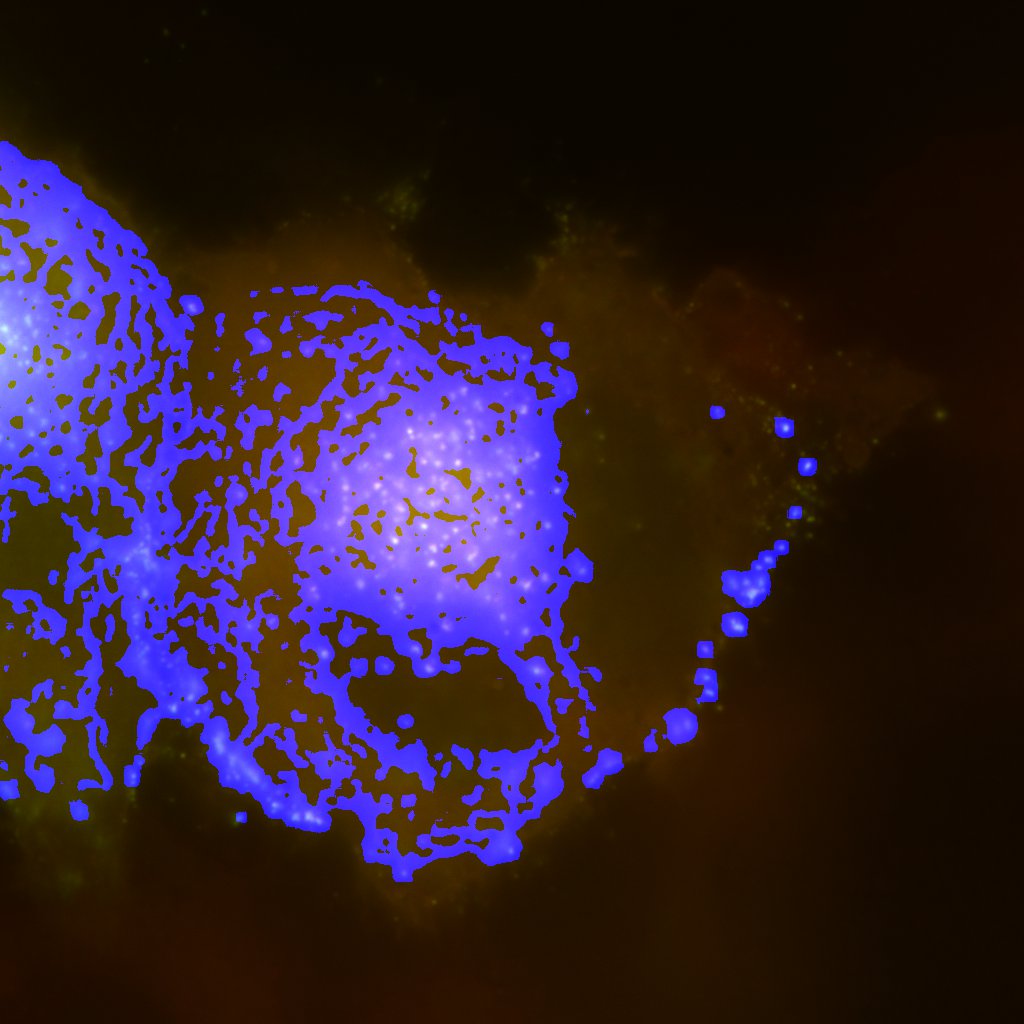}}; 
  \node[inner sep=0] at (0,-8) {\includegraphics[width=0.24\textwidth]{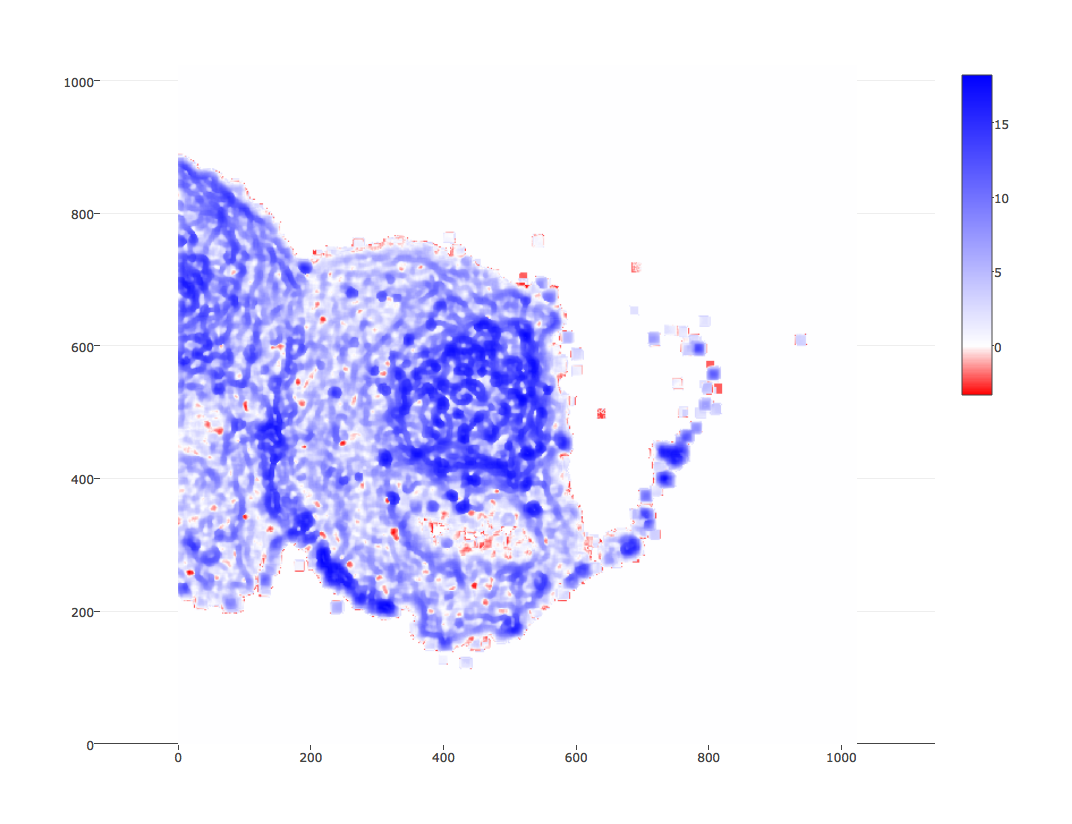}}; 
  \node[inner sep=0] at (4.4,0) {\includegraphics[width=0.23\textwidth]{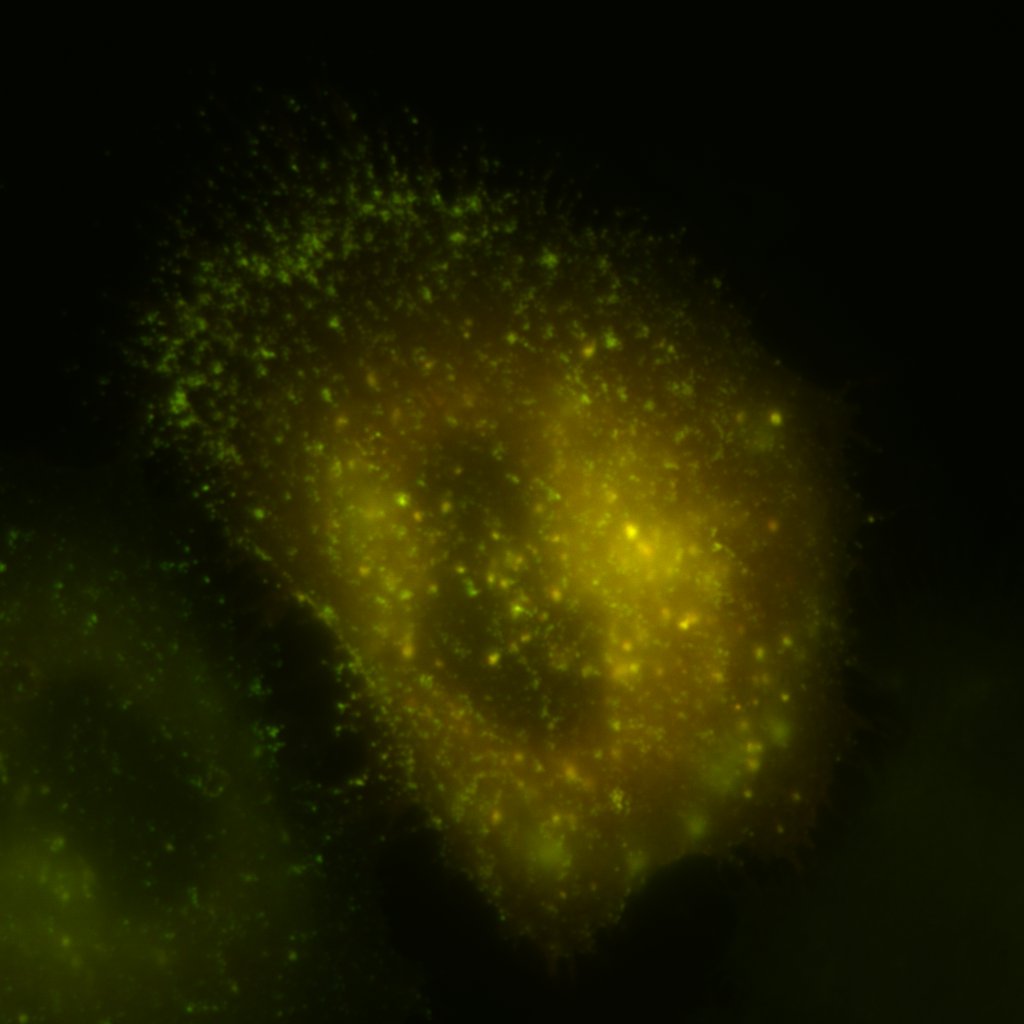}};
  \node[inner sep=0] at (4.4,-4.4) {\includegraphics[width=0.23\textwidth]{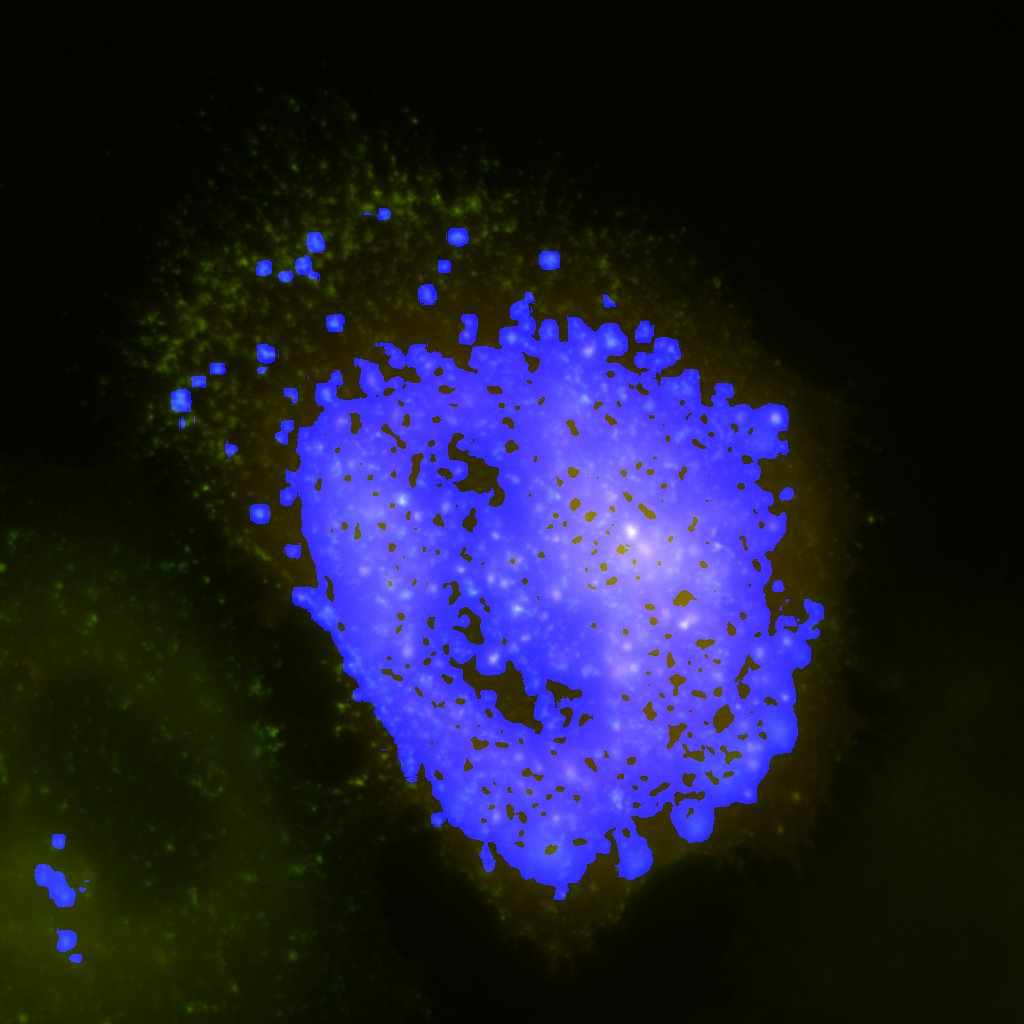}};
  \node[inner sep=0] at (4.4,-8) {\includegraphics[width=0.24\textwidth]{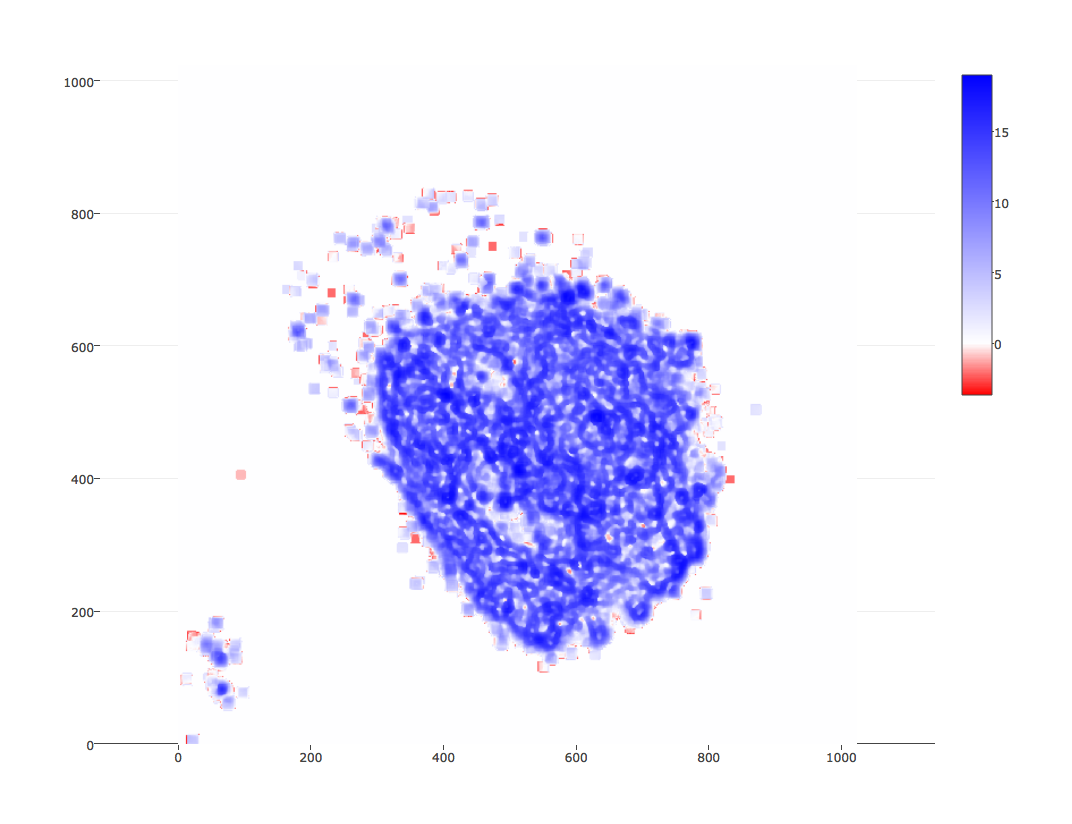}};
  \draw[line width=2pt,black!80!white] (-2.2,2.2) rectangle (6.6,-9.65);
   \draw[line width=2pt,black!80!white] (2.2,2.25) to (2.2,-9.65);
    \end{tikzpicture}
    \caption{Strong colocalization is expected between Gag-CFP (red channel) and Gag-YFP (green channel). Original overlay image (upper); colocalized region labelled in blue (middle); and heat map of $z$-scores (lower).}\label{fg:realdatacpcol}
\end{figure}

In this section, we applied SACA to two real biological datasets. The first dataset is microscopic images (image size: $1024\times 1024$) of HeLa cells expressing the human immunodeficiency virus type 1 (HIV-1) capsid protein, Gag; HIV-1 virus particles assemble at the plasma membrane, composed of $\sim$2000 Gag molecules and incorporating two copies of the viral RNA genome \citep[see][]{sundquist2012,freed2015}. Here we asked if we could detect subcellular sites of HIV-1 genome packaging based on the colocalization of differentially tagged Gag and viral genome molecules \cite[see][]{Jouvenet2009,Chen2009,Pocock2016}. In these experiments, $t_X$ and $t_Y$ were calculated by Otsu's method for each channel, and multiple comparison was corrected by the Bonferroni method at level $5\%$ (equivalently, it is significant when $z$-score $Z(k;r_T)$ is larger than 5.335 when the image size is $1024\times 1024$). All other parameters were chosen according to guidance in Section~\ref{sc:paramter}.

Specifically, we considered five different conditions of HeLa cells with corresponding images (Figure~\ref{fg:realdatanocol}, \ref{fg:realdatapacol}, and \ref{fg:realdatacpcol} are from \cite{wang2017}). In the first four conditions (Figure~\ref{fg:realdatanocol}--\ref{fg:realdatamem}), HIV-1-Gag (green channel) was fused to cyan fluorescence protein (CFP) and MS2 protein (red channel), which tracks the viral RNA genome, was fused to yellow fluorescent protein (YFP). In the first condition (Figure~\ref{fg:realdatanocol}), only coincidental colocalization between Gag-CFP and MS2-YFP was expected, as the MS2 protein in this case was designed to remain in the nucleus in the absence of the RNA genome, resulting in a negative control. In the second condition (Figure~\ref{fg:realdatapacol}), partial colocalization was expected at the edge of the cell, because Gag-CFP was expressed from an mRNA (and packageable genome) engineered to contain multiple copies of an RNA stem loop that binds MS2-YFP with high specificity \citep[see][]{becker2017}. Again, partial colocalization was expected in the third and fourth conditions (Figure~\ref{fg:realdataactin} and \ref{fg:realdatamem}); however, the location of colocalization should vary, being detected at actin fibers and intracellular vesicles, respectively, due to molecular retargeting of Gag and viral RNA genomes to these subcellular locations.  For the former (Figure~\ref{fg:realdataactin}), Lifeact-MS2-YFP and Gag colocalized to peripheral actin fibers \citep[see][]{becker2017} due to Lifeact's preferential binding to and targeting of F-actin bundles with high specificity \citep[see][]{riedl2008}; for the latter (Figure~\ref{fg:realdatamem}), a protein myristoylation signal (MGSSKSKPKD) derived from the proto-oncogene Src kinase was fused to MS2-YFP (Src-MS2-YFP) to target the gRNAs to cellular membranes, and the majority of particles were located in intracellular vesicles \citep[see][]{becker2017}. In the final condition (Figure~\ref{fg:realdatacpcol}), strong colocalization was expected for two constructs expressing synthetic Gags fused to CFP and YFP, respectively; Gag self-assembles, therefore we expected and found multi-colored particles with the highest levels of colocalization of all conditions.

We summarized heat maps of $z$ scores $Z(k;r_T)$ and reported colocalized regions in Figure~\ref{fg:realdatanocol}--\ref{fg:realdatacpcol}. In addition, several summarized statistics of heat maps are shown in Table~\ref{tb:heatmapstat}, including the proportion of colocalized pixels $r_p$, the mean of  $z$ scores $\bar{Z}$, the maximum value of $z$ scores $M^\ast$, and  the mean of  $z$ scores within colocalized regions $\bar{Z}_s$. The results show almost no colocalized regions were discovered in Figure~\ref{fg:realdatanocol}, as only $0.1\%-0.2\%$ pixels are categorized as colocalized; colocalized regions concentrated at the plasma membrane in Figure~\ref{fg:realdatapacol}, along actin fibers of cells in Figure~\ref{fg:realdataactin}, and in intracellular vesicles of cells in Figure~\ref{fg:realdatamem}; a very strong, significant level of colocalization in cells was discovered in Figure~\ref{fg:realdatacpcol}. Having not only the statistical information regarding colocalization levels, but also the spatial information on where the colocalization is occurring, is incredibly powerful.  SACA is able to confirm the biological outcomes expected in this particular dataset regarding HIV-1 particle assembly and provides a robust tool to aid scientists in their investigations into macromoleular dynamics and regional associations.

\begin{figure}[h!]
    \centering
    \begin{tikzpicture}[scale=1]
  \node[inner sep=0] at (0,0) {\includegraphics[width=0.23\textwidth]{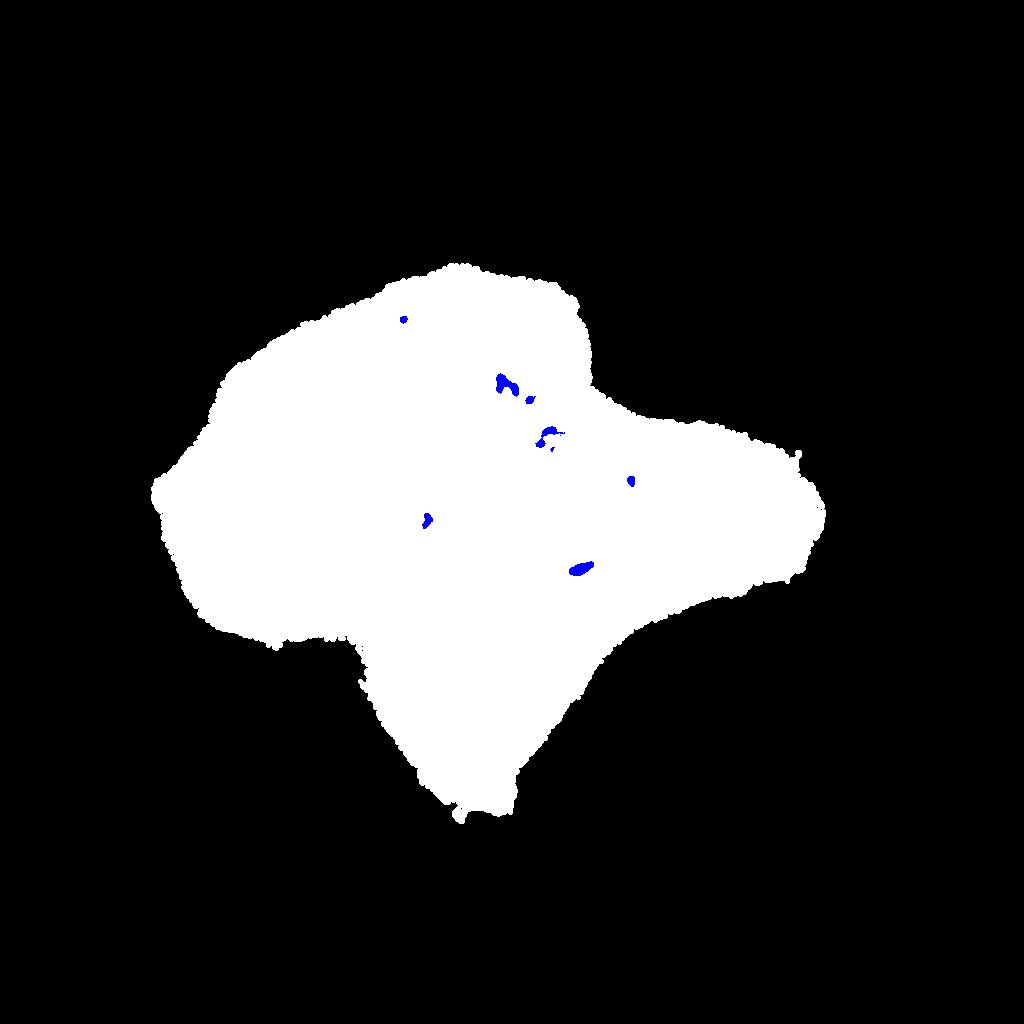}};
  \node[inner sep=0] at (4.4,0) {\includegraphics[width=0.23\textwidth]{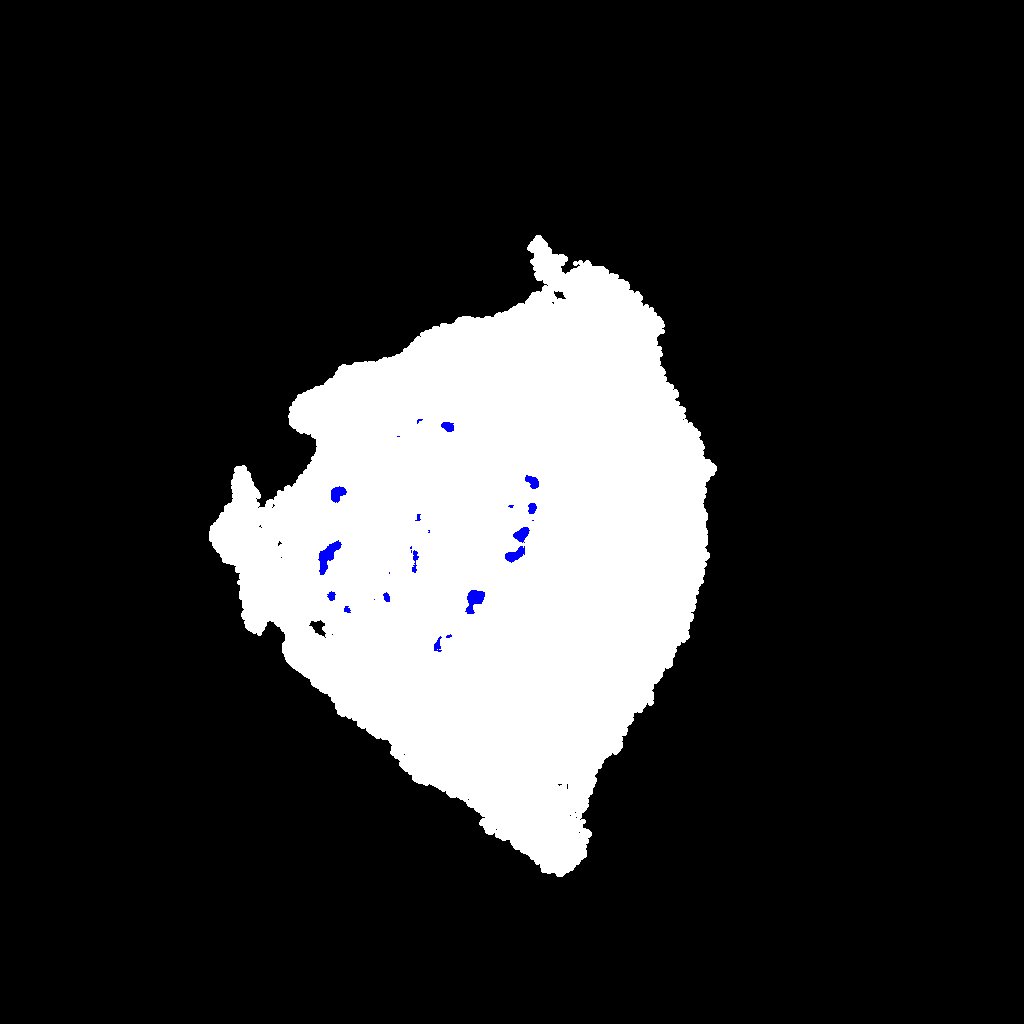}};
  \draw[line width=2pt,black!80!white] (-2.2,2.2) rectangle (6.6,-2.2);
   \draw[line width=2pt,black!80!white] (2.2,2.2) to (2.2,-2.2);
    \end{tikzpicture}
    \caption{The cell region (white) by segmentation and colocalized region (blue) in Figure~\ref{fg:realdatanocol}. $0.50\%$ (left) and $0.98\%$ (right) area of cell regions are reported as colocalized.}\label{fg:realdatanocolvalid}
\end{figure}

\begin{figure}[h!]
    \centering
    \begin{tikzpicture}[scale=1]
  \node[inner sep=0] at (0,0) {\includegraphics[width=0.23\textwidth]{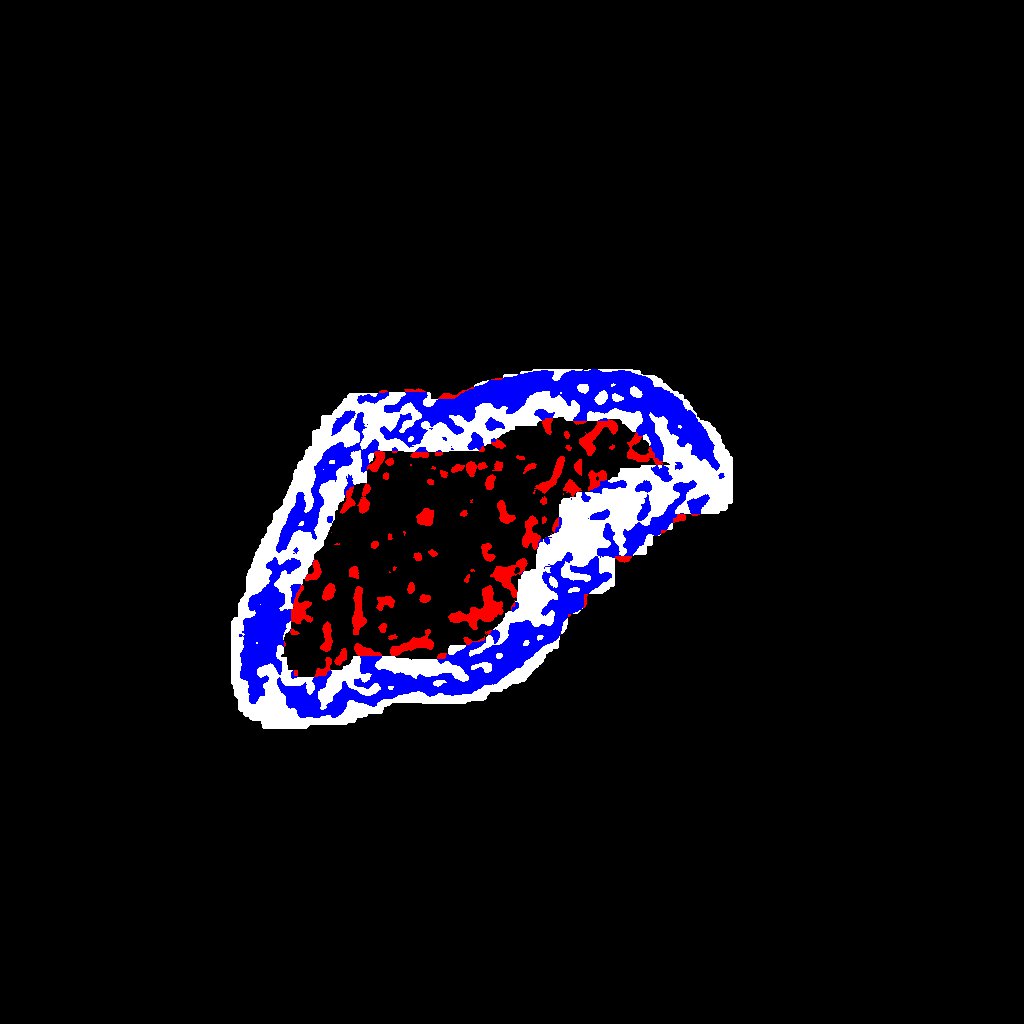}};
  \node[inner sep=0] at (4.4,0) {\includegraphics[width=0.23\textwidth]{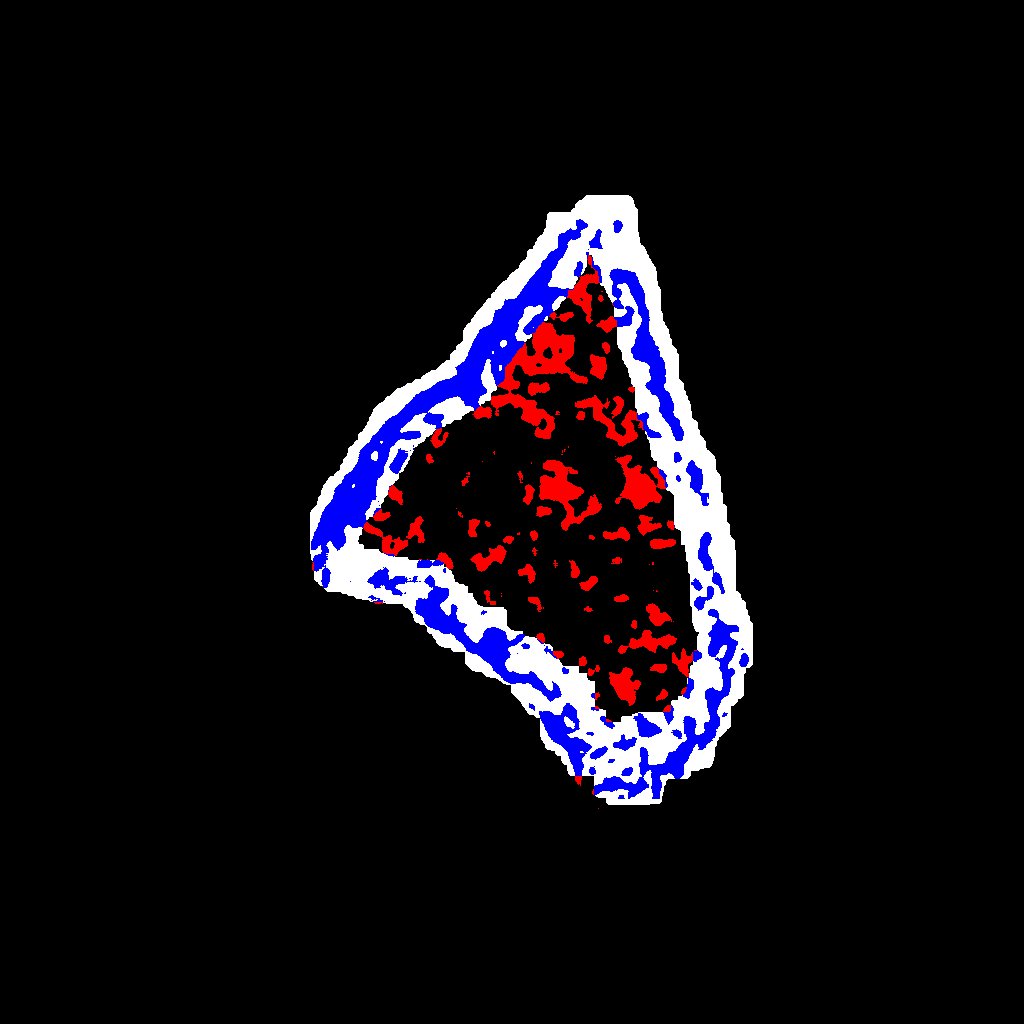}};
    \node[inner sep=0] at (0,-3.3) {\includegraphics[width=0.23\textwidth]{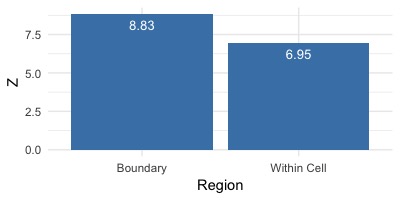}};
  \node[inner sep=0] at (4.4,-3.3) {\includegraphics[width=0.23\textwidth]{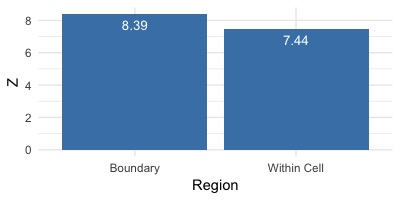}};
  \draw[line width=2pt,black!80!white] (-2.2,2.2) rectangle (6.6,-4.4);
   \draw[line width=2pt,black!80!white] (2.2,2.2) to (2.2,-4.4);
    \end{tikzpicture}
    \caption{Boundary region of cells (white) and colocalized region (blue in boundary region and red otherwise) in Figure~\ref{fg:realdatapacol} (upper figure), and mean of $z$-scores in blue and red regions (lower figure). $75.7\%$ (left) and $63.4\%$ (right) area of colocalized regions belong to boundary region of cells.}\label{fg:realdatapacolvalid}
\end{figure}

\begin{figure}[h!]
    \centering
    \begin{tikzpicture}[scale=1]
  \node[inner sep=0] at (0,0) {\includegraphics[width=0.23\textwidth]{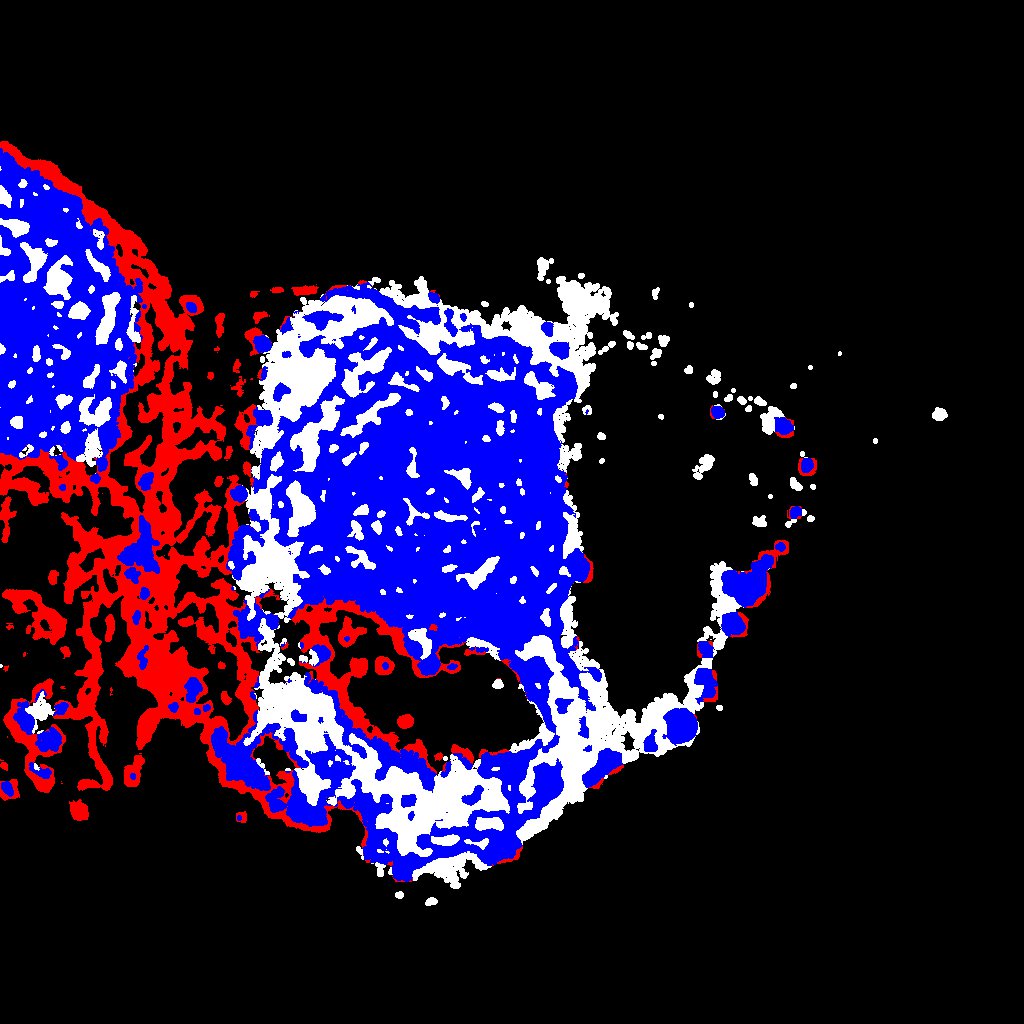}};
  \node[inner sep=0] at (4.4,0) {\includegraphics[width=0.23\textwidth]{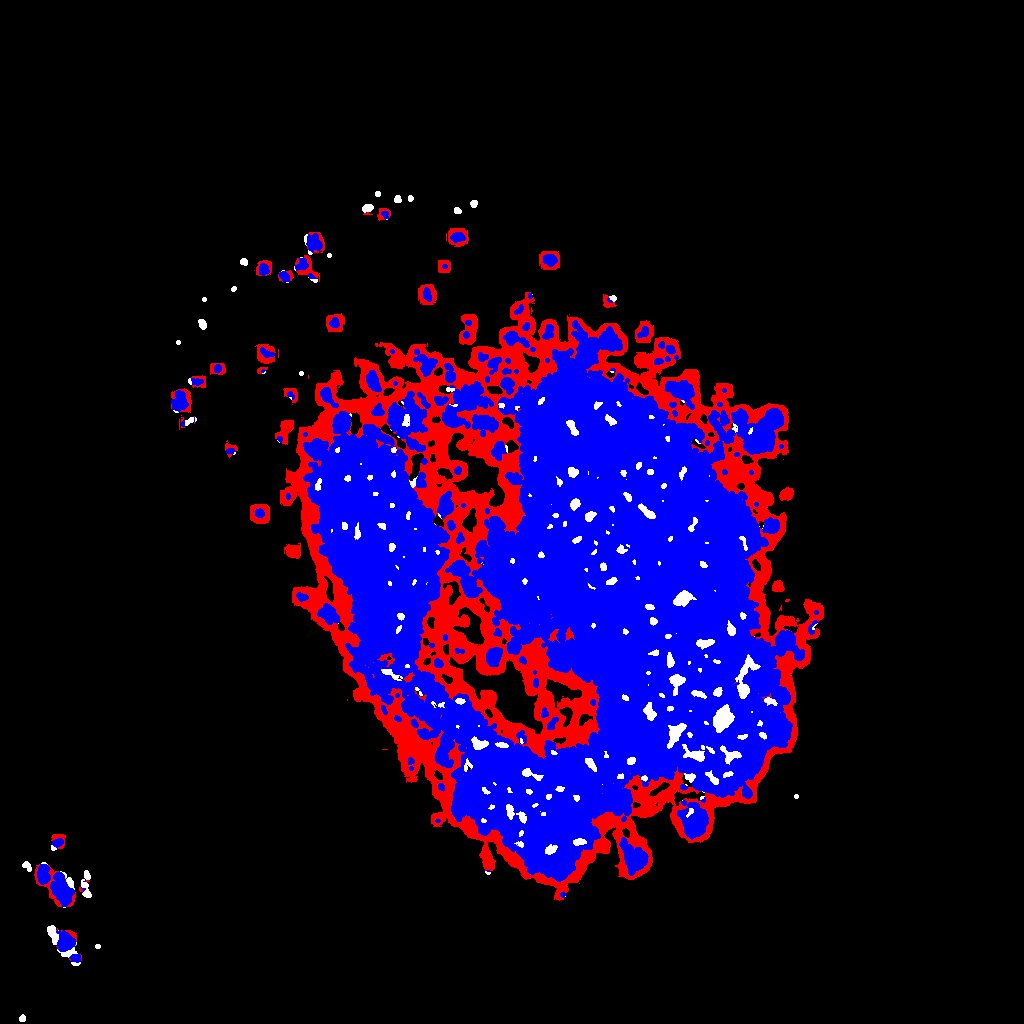}};
  \draw[line width=2pt,black!80!white] (-2.2,2.2) rectangle (6.6,-2.2);
   \draw[line width=2pt,black!80!white] (2.2,2.2) to (2.2,-2.2);
    \end{tikzpicture}
    \caption{``True" colocalized region (white) and colocalized region reported by SACA (blue in ``true" region and red otherwise) in Figure~\ref{fg:realdatacpcol}. $64.1\%$ (left) and $93.6\%$ (right) area of ``true" regions are reported, and $29\%$ (left) and $30.3\%$ (right) area of reported regions are false discovery.}\label{fg:realdatacpcolvalid}
\end{figure}

To further validate SACA's performance on this biological dataset, we quantitively compare the colocalized regions identified by SACA with ``ground truth" in Figures~\ref{fg:realdatanocol}, \ref{fg:realdatapacol}, and \ref{fg:realdatacpcol}. In Figure~\ref{fg:realdatanocol}, the ``ground truth" is that no colocalization is expected within cell. We applied a cell segmentation algorithm from EBImage package \citep[see][]{pau2010ebimage} to distinguish the cells from the background. The cell segmentation results and colocalized regions are shown in Figure~\ref{fg:realdatanocolvalid}. SACA reports almost no colocalized regions, as only $0.50\%$ and $0.98\%$ of cell regions are identified as colocalized. In Figure~\ref{fg:realdatapacol}, most colocalized regions are expected at the edge of cells. In Figure~\ref{fg:realdatapacolvalid}, we identified the boundary region of cells by cell segmentation and boundary detection algorithms from EBImage package and imager package (see \url{https://dahtah.github.io/imager/}). It shows that most colocalized regions concentrate at the boundary of cells ($75.7\%$ and $63.4\%$, respectively), and  $z$-scores are larger in boundary regions than ones within cells. In Figure~\ref{fg:realdatacpcol}, colocalization is expected at pixels where signals appear in both channels. Thus, we apply adaptive thresholding in EBImage package to identify signals from both channels,  and regard pixels with signals in both channels as the ``ground truth" for the colocalized region. Through comparison in Figure~\ref{fg:realdatacpcolvalid}, most of the ``true" colocalized region is recovered, and there are few false discoveries due to spatial autocorrelation among $Z(k;r_T)$ of adjacent pixels. 

\begingroup
\renewcommand{\arraystretch}{1.5}
\begin{table*}[!h]
\centering
\begin{tabular}{c c c c c c c c c c c c c c c c c c c c}
\hline
\hline
& & \multicolumn{2}{c}{Fig.~\ref{fg:realdatanocol}} & & \multicolumn{2}{c}{Fig.~\ref{fg:realdatapacol}} & &  \multicolumn{2}{c}{Fig.~\ref{fg:realdataactin}}& &\multicolumn{2}{c}{Fig.~\ref{fg:realdatamem}}& &\multicolumn{2}{c}{Fig.~\ref{fg:realdatacpcol}} & & \multicolumn{3}{c}{Fig.~\ref{fg:realdatabill}}  \\
 \cline{3-4}  \cline{6-7}  \cline{9-10}  \cline{12-13} \cline{15-16} \cline{18-20}
& & L & R & & L & R & & L & R & & L & R & & L & R & & L & M & R \\
\hline
$r_p$ & & $0.10\%$ &  $0.17\%$ & &  $4.0\%$ & $4.9\%$ & &  $3.0\%$ & $5.5\%$ & &  $1.2\%$ & $2.9\%$ & &  $18.0\%$ & $19.4\%$ & & $1.6\%$ & $3.4\%$ & $3.6\%$ \\ 
$\bar{Z}$ & & $-0.16$ &  $-0.02$ & &  $0.39$ & $0.55$ & &  $0.33$ & $0.48$ & &  $0.24$ & $0.31$ & &  $2.13$ & $2.38$ & & $-0.06$ & $0.38$ & $0.36$\\ 
$M^\ast$ & & $10.5$ &  $10.5$ & &  $15.1$ & $15.9$ & &  $14.1$ & $15.0$ & &  $11.8$ & $13.1$ & &  $18.2$ & $19.0$& & $13.0$ & $16.4$ & $14.8$ \\
$\bar{Z}_s$ & & $6.83$ &  $6.59$ & &  $8.37$ & $7.93$ & &  $8.10$ & $8.34$ & &  $7.26$ & $7.52$ & &  $9.66$ & $11.44$& & $7.38$ & $10.59$ & $8.97$ \\
\hline
\hline
\end{tabular}
\caption{Summarized statistics of heat maps in Figure~\ref{fg:realdatanocol}--\ref{fg:realdatabill}: proportion of colocalized pixels $r_p$, the mean of  $z$ scores $\bar{Z}$, the maximum value of $z$ scores $M^\ast$ and  the mean of  $z$ scores within colocalized regions $\bar{Z}_s$. (L=left, M=middle, R=right)}\label{tb:heatmapstat}
\end{table*}
\endgroup

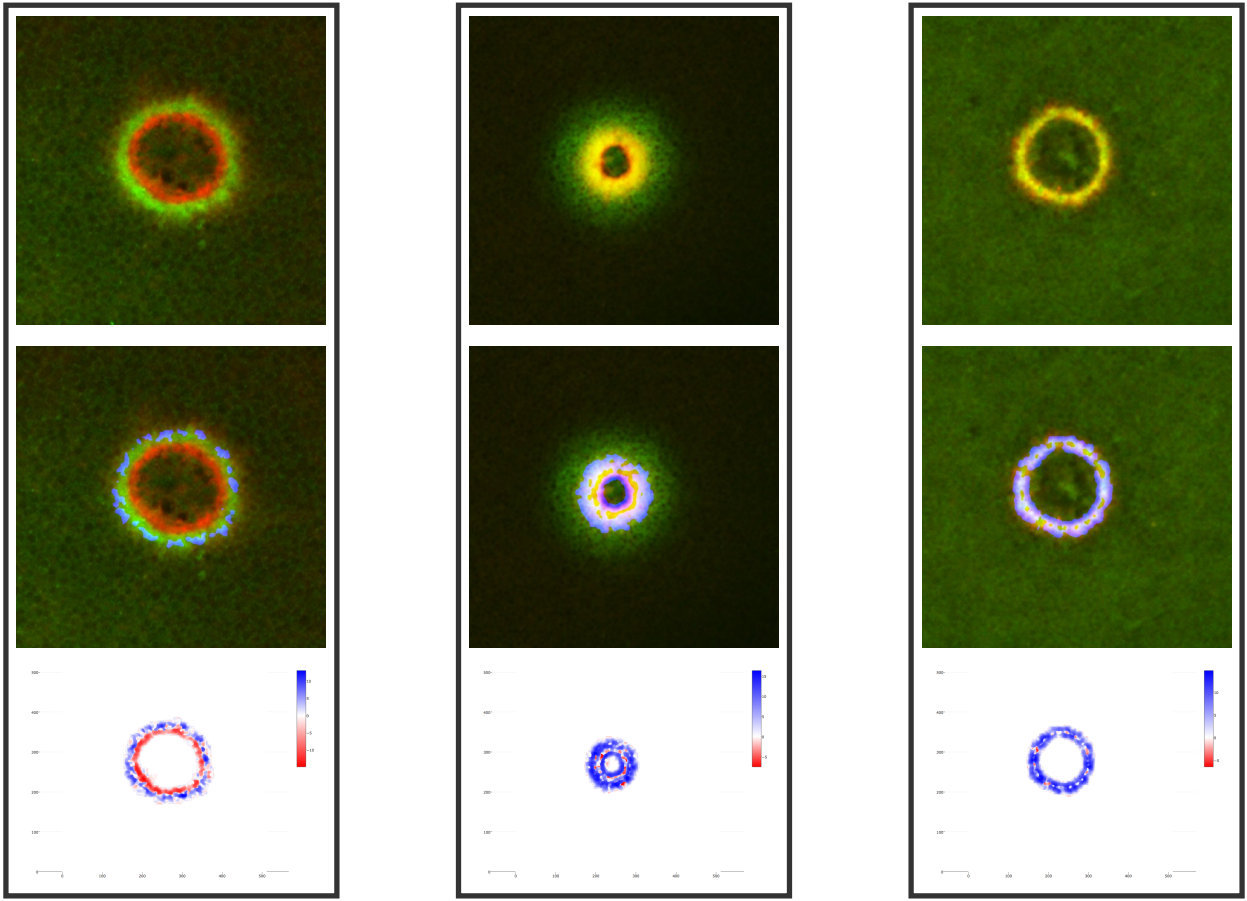
\begin{figure*}[h!]
    \centering
    \begin{subfigure}[b]{0.325\textwidth}
    \centering
        \begin{tikzpicture}[scale=1]
    \node[inner sep=0] at (0,0) {\includegraphics[width=0.697\textwidth]{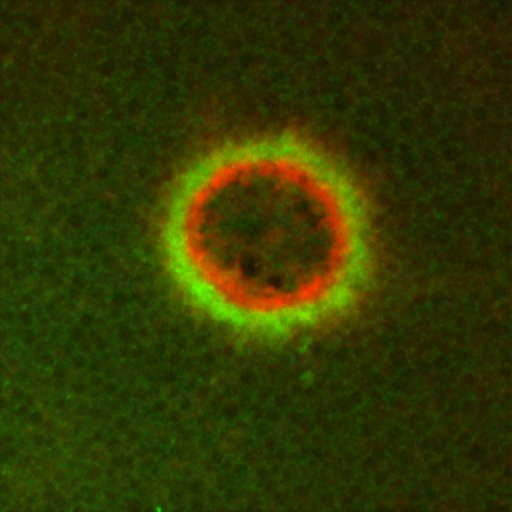}};
  \node[inner sep=0] at (0,-4.4) {\includegraphics[width=0.697\textwidth]{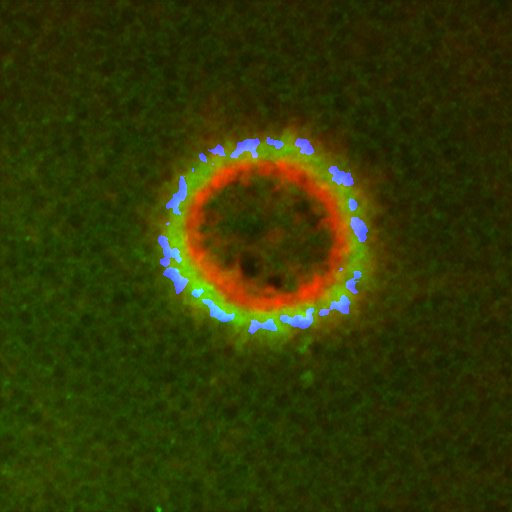}}; 
  \node[inner sep=0] at (0,-8) {\includegraphics[width=0.727\textwidth]{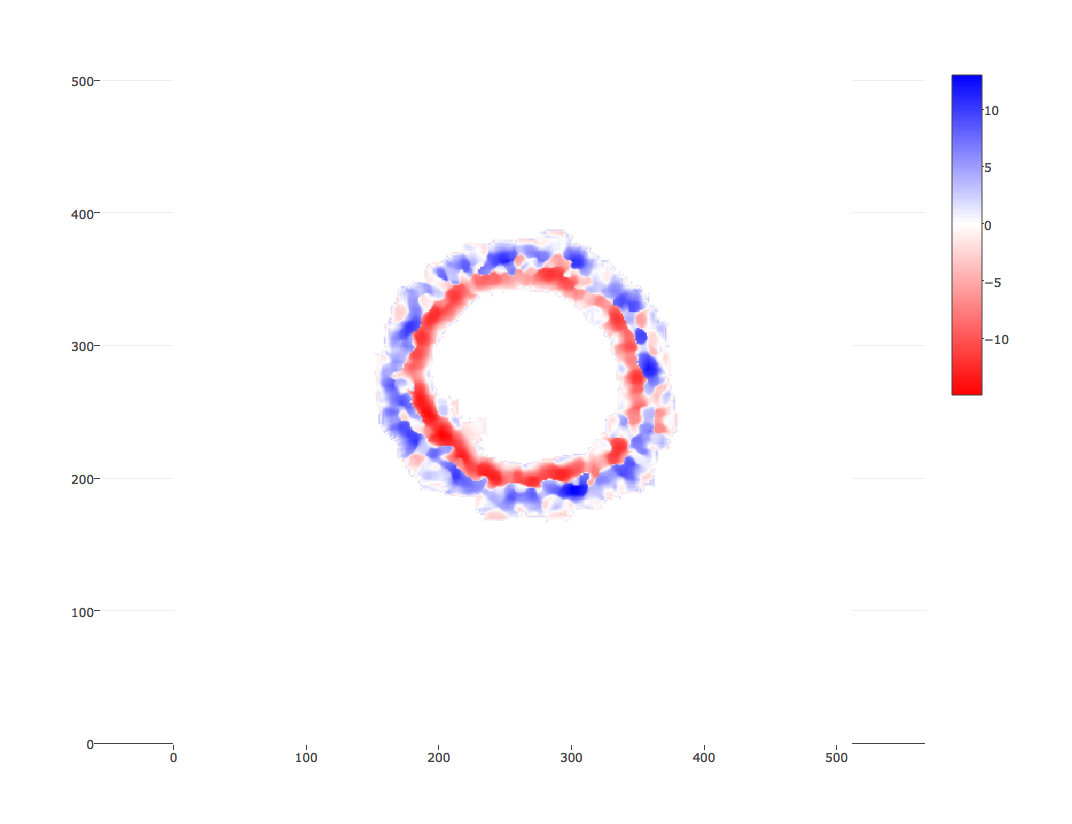}}; 
  \draw[line width=2pt,black!80!white] (-2.2,2.2) rectangle (2.2,-9.65);
    \end{tikzpicture}
        \caption{Low levels of colocalization: Rho (red) and Cdc42 (green)}
        \label{fg:bill1}
    \end{subfigure}
    \begin{subfigure}[b]{0.325\textwidth}
    \centering
    \begin{tikzpicture}[scale=1]
  \node[inner sep=0] at (0,0) {\includegraphics[width=0.697\textwidth]{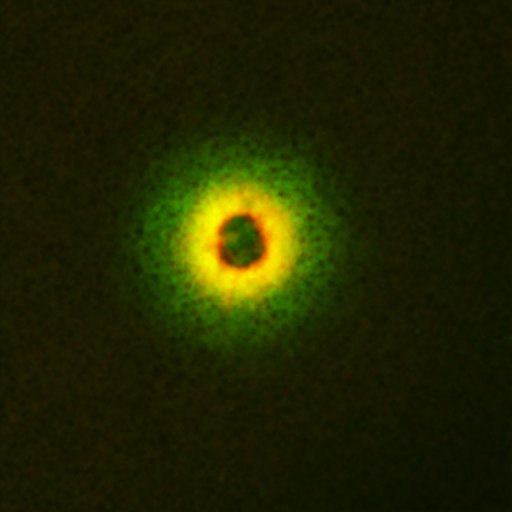}};
  \node[inner sep=0] at (0,-4.4) {\includegraphics[width=0.697\textwidth]{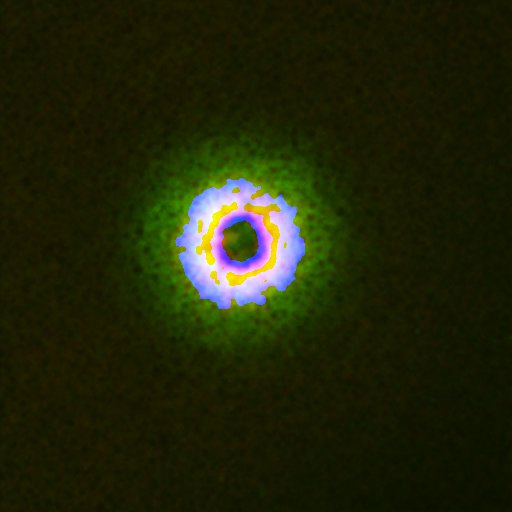}}; 
  \node[inner sep=0] at (0,-8) {\includegraphics[width=0.727\textwidth]{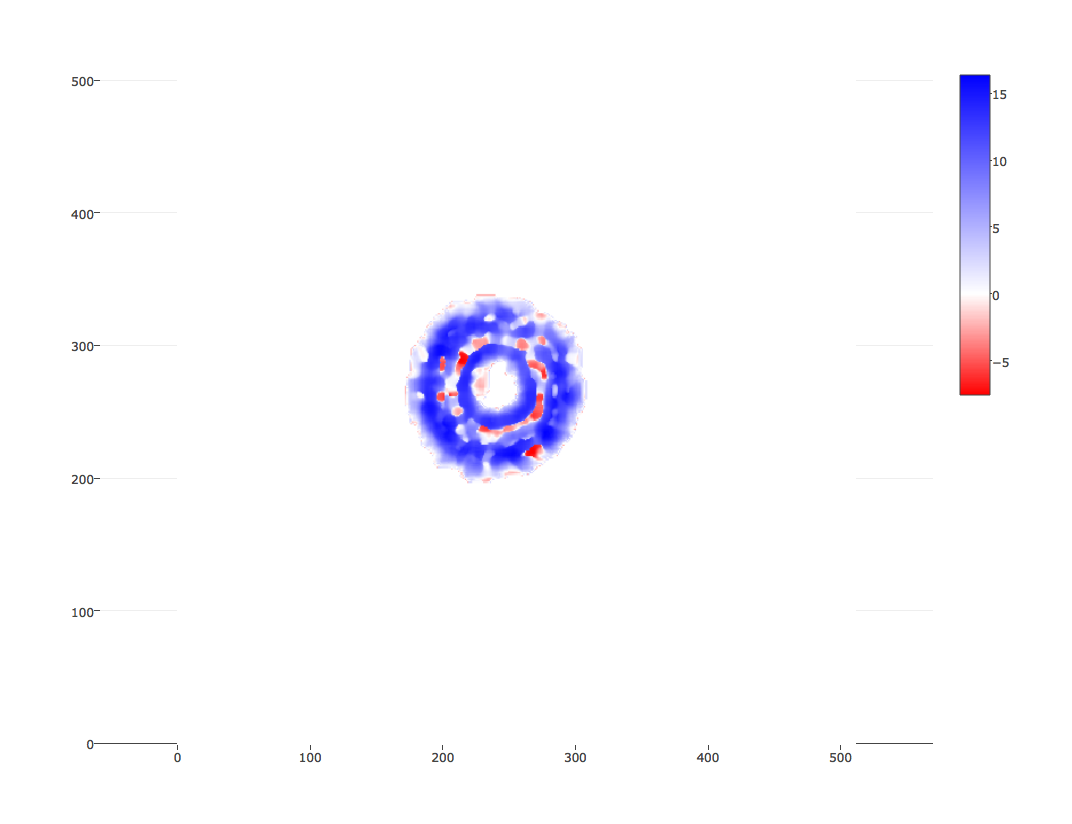}}; 
  \draw[line width=2pt,black!80!white] (-2.2,2.2) rectangle (2.2,-9.65);
    \end{tikzpicture}
        \caption{High levels of colocalization: PKC$\beta$ (red) and calcium (green)}
        \label{fg:bill2}
    \end{subfigure}
    \begin{subfigure}[b]{0.325\textwidth}
    \centering
    \begin{tikzpicture}[scale=1]
  \node[inner sep=0] at (0,0) {\includegraphics[width=0.697\textwidth]{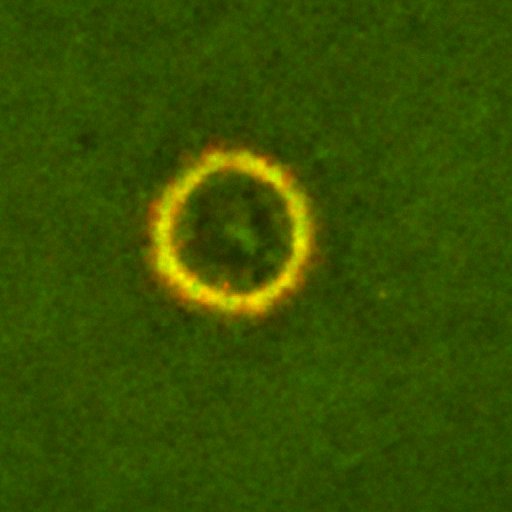}};
  \node[inner sep=0] at (0,-4.4) {\includegraphics[width=0.697\textwidth]{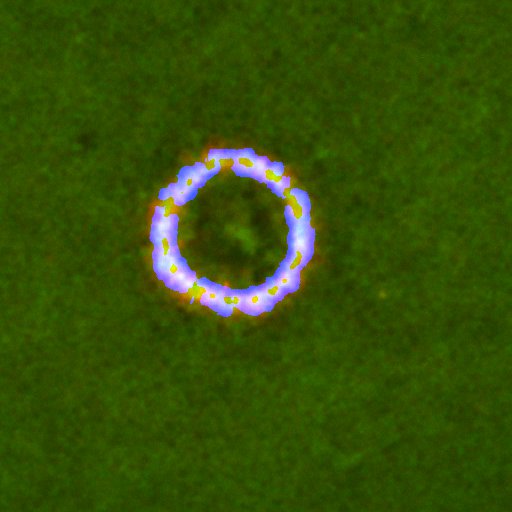}}; 
  \node[inner sep=0] at (0,-8) {\includegraphics[width=0.727\textwidth]{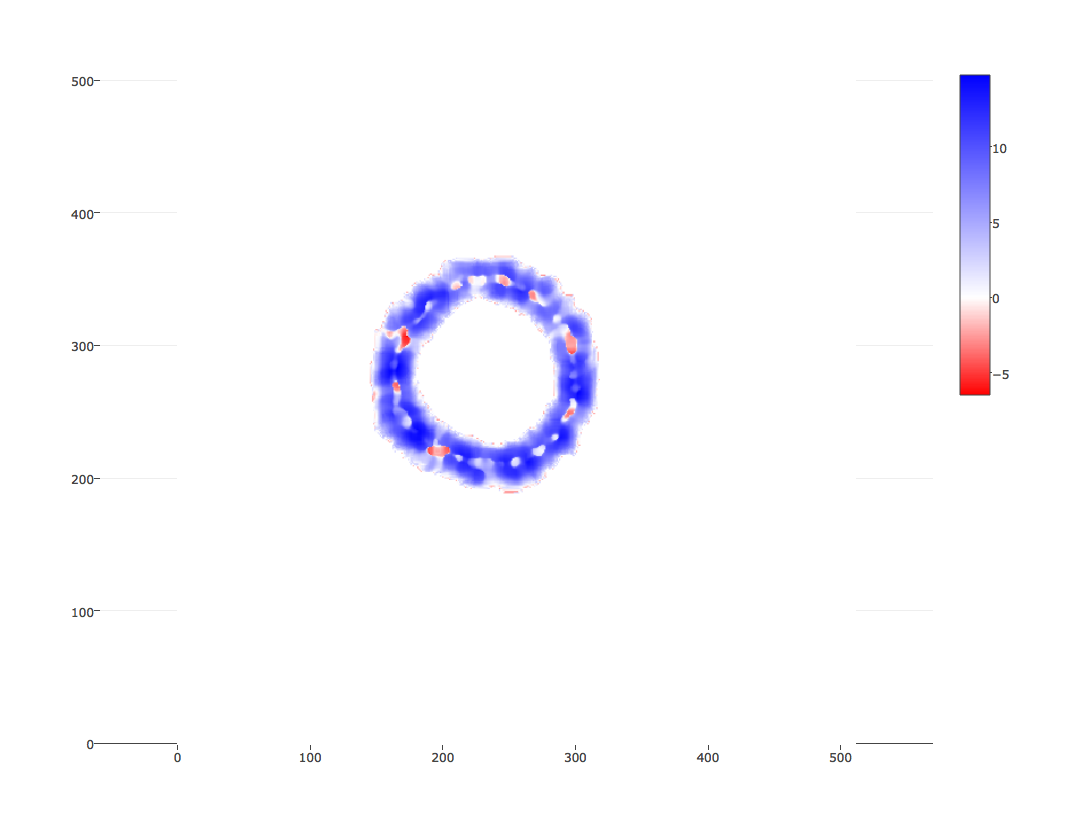}}; 
  \draw[line width=2pt,black!80!white] (-2.2,2.2) rectangle (2.2,-9.65);
    \end{tikzpicture}
        \caption{High levels of colocalization: Cdc42 (red) and cortactin (green)}
        \label{fg:bill3}
    \end{subfigure}
    \caption{Xenopus wounding model: denoised overlay image (upper), colocalized region labelled in blue color (middle) and heat map of $z$-scores (lower).}\label{fg:realdatabill}
\end{figure*}

We also applied our new colocalization method to another biological dataset, this time to microscopic images (image size: $512\times 512$) of a  Xenopus developmental model used to elucidate signal responses during cellular wounding and the subsequent repair process \citep[see][]{wang2017}. Before we applied SACA, we denoised these images using the ImageJ plugin, PureDenoise (see \url{http://bigwww.epfl.ch/algorithms/denoise/}) because these particular biological images were heavily corrupted by noise.  We then applied SACA on these images with the same settings used for the  previous real data examples. Rho GTPases, including Rho and Cdc42, play a role during Xenopus oocyte wound repair \citep[see][]{simon2013}; saying that, they do not overlap during the wound repair process, as is shown in Figure~\ref{fg:bill1}. PKC$\beta$ participates in Rho and Cdc42 activation and is also recruited to cell wounds \citep[see][]{vaughan2014}; calcium defines a broad region within which PKC$\beta$ can be found, and therefore, some level of colocalization was expected between the two; this region is shown in Figure~\ref{fg:bill2}. Finally, cortical cytoskeleton repair is important in wound healing, so the actin regulatory protein, cortactin, largely overlaps with Cdc42, for example, during the wound healing process. Figure~\ref{fg:bill3} represents this final scenario with the highest levels of colocalization between Cdc42 and cortactin.  The heat maps of $z$ score $Z(k;r_T)$ and colocalized regions are summarized in Figure~\ref{fg:realdatabill}. Once again, all discoveries found by SACA demonstrate this method's robustness within complex, biological contexts. All together, these results in Figure~\ref{fg:realdatanocol}--\ref{fg:realdatabill} demonstrate that SACA is able to reveal the the degree of colocalization at each pixel location, which provides an extremely powerful tool to researchers interested in colocalization.

\subsection{Comparisons with Previous Methods}

\begin{figure}[h!]
    \centering
    \begin{subfigure}[b]{0.23\textwidth}
    \centering
        \begin{tikzpicture}[scale=1]
    \node[inner sep=0] at (0,0) {\includegraphics[width=\textwidth]{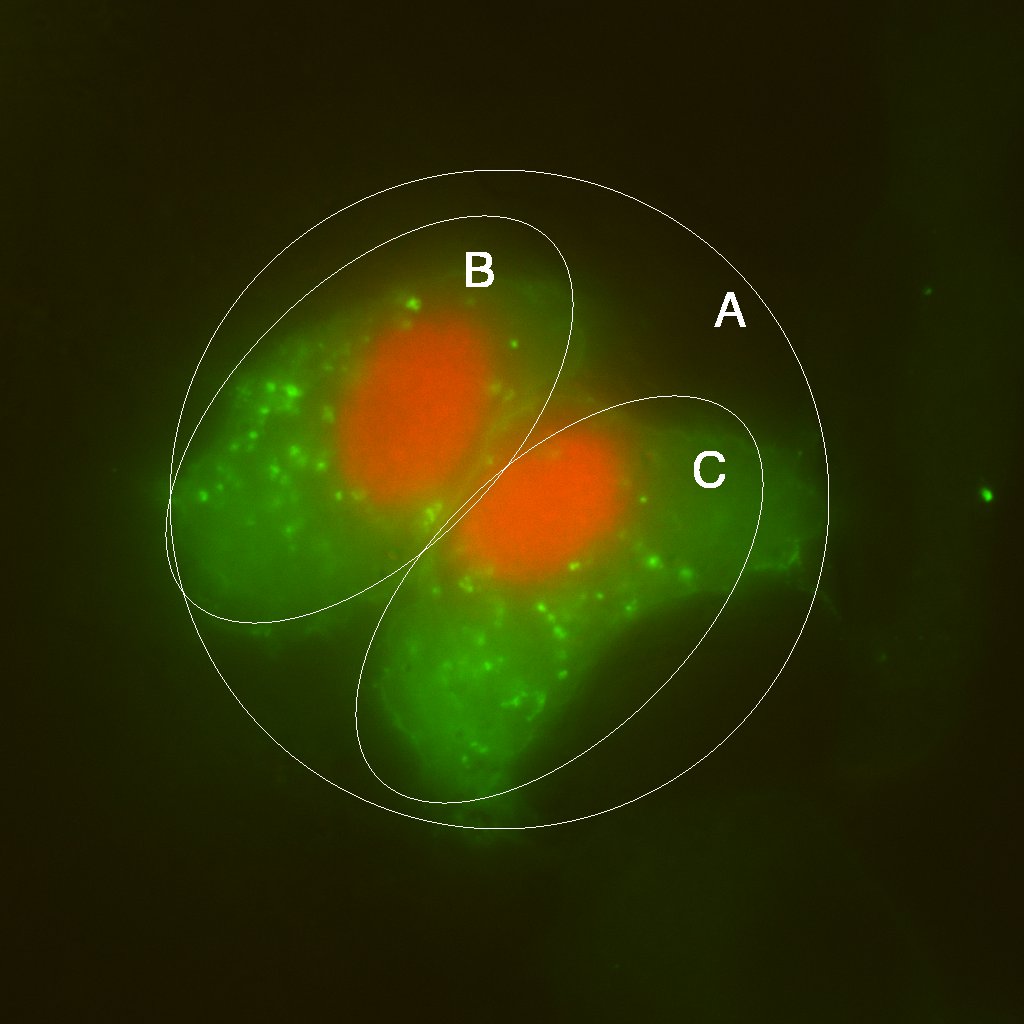}};
    \end{tikzpicture}
        \caption{Testing ROIs in Figure~\ref{fg:realdatanocol}}
        \label{fg:compare1}
    \end{subfigure}
    \begin{subfigure}[b]{0.23\textwidth}
    \centering
    \begin{tikzpicture}[scale=1]
  \node[inner sep=0] at (0,0) {\includegraphics[width=\textwidth]{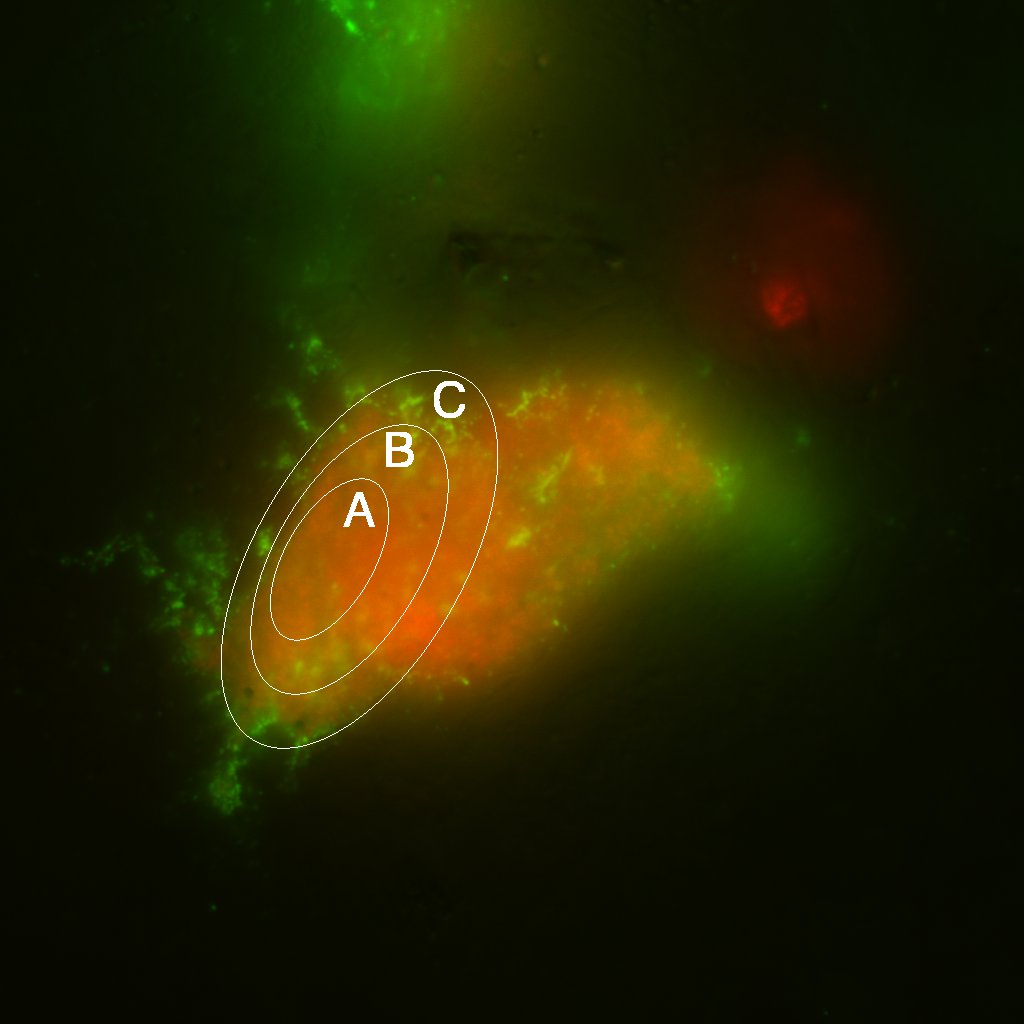}};
    \end{tikzpicture}
        \caption{Testing ROIs in Figure~\ref{fg:realdatapacol}}
        \label{fg:compare2}
    \end{subfigure}
    \begin{subfigure}[b]{0.23\textwidth}
    \centering
    \begin{tikzpicture}[scale=1]
  \node[inner sep=0] at (0,0) {\includegraphics[width=\textwidth]{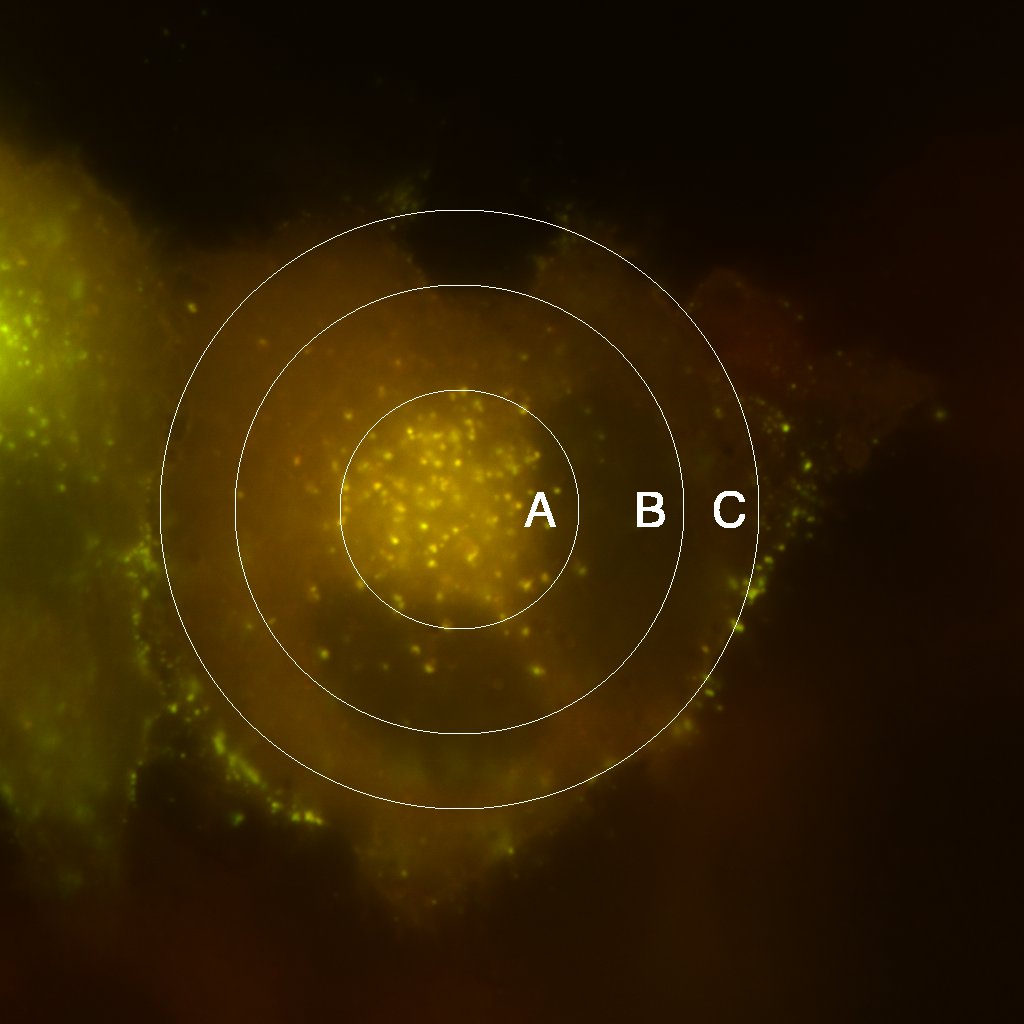}};
    \end{tikzpicture}
        \caption{Testing ROIs in Figure~\ref{fg:realdatacpcol}}
        \label{fg:compare3}
    \end{subfigure}
        \begin{subfigure}[b]{0.23\textwidth}
    \centering
    \begin{tikzpicture}[scale=1]
  \node[inner sep=0] at (0,0) {\includegraphics[width=\textwidth]{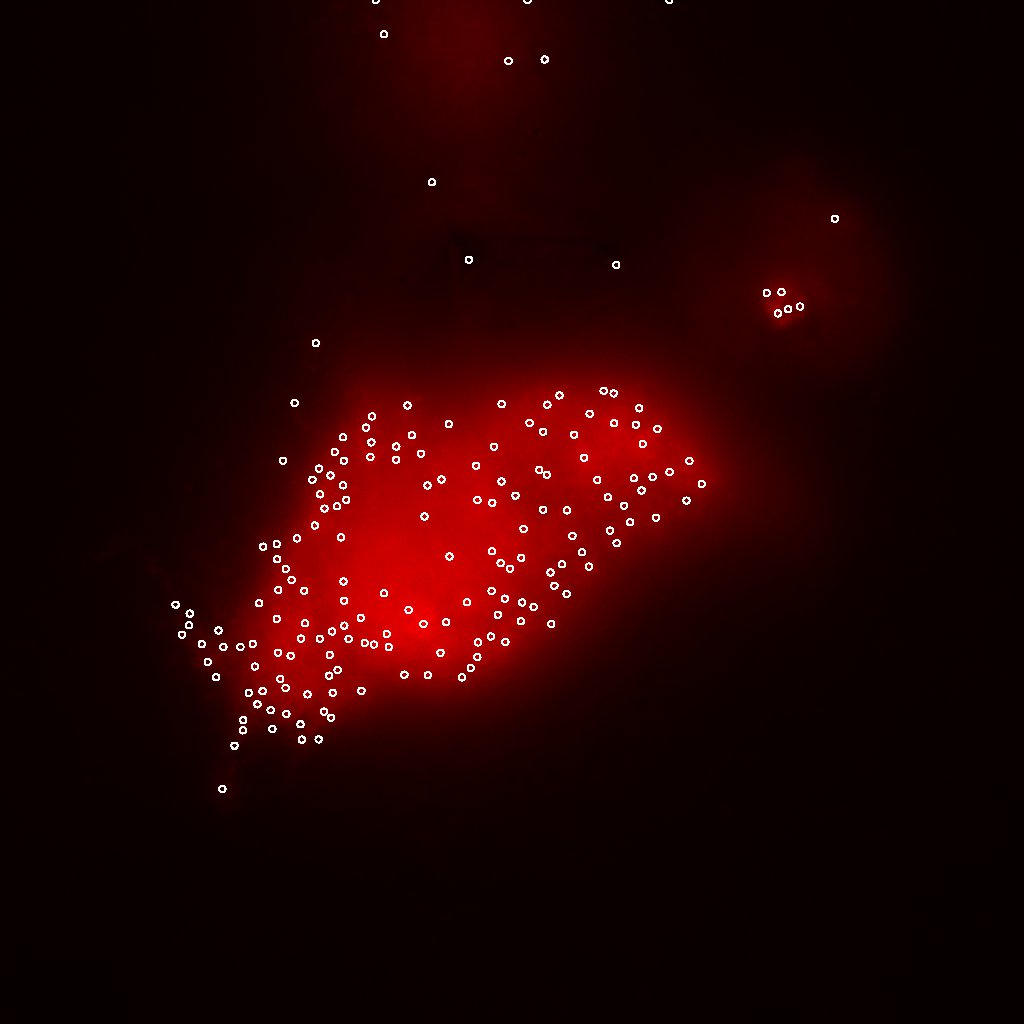}};
    \end{tikzpicture}
        \caption{Spots detection results in \ref{fg:compare2}}
        \label{fg:spots}
    \end{subfigure}
    \caption{ROIs for colocalization methods comparison and spot detection results.}\label{fg:compareregion}
\end{figure}

In this section, we compare other related colocalization methods with SACA. Three different regions of interest (see Figure~\ref{fg:compareregion}) in the left most images of Figures~\ref{fg:realdatanocol}, \ref{fg:realdatapacol} and \ref{fg:realdatacpcol} were chosen to evaluate the performance of previous ROI-based methods. Following the 3-step procedure, we applied Pearson correlation coefficient \cite{manders1992dynamics}, Manders' split coefficients $(M_1,M_2)$ \cite{manders1993measurement}, Intensity correlation quotient (ICQ) \cite{Li04}, and maximum truncated Kendall tau correlation coefficient $\tau^\ast$ (MTKT) \cite{wang2017}. For $M_1$, $M_2$, and $\tau^\ast$, the thresholds or lower bound of thresholds were determined by Otsu's method. To obtain a $p$-value, the images were permuted pixel-wise $1000$ times within each ROI. The values of these colocalization measures and corresponding $p$-values calculated by the permutation tests are summarized in Table~\ref{tb:comparison}. The results in Table~\ref{tb:comparison} suggest that colocalization measures and $p$-values are very sensitive to the choices of ROI. For example, most colocalization measures in Figure~\ref{fg:compare1} show no evidence of colocalization in region B and C, but strong evidence of colocalization in region A, which contradicts the expected biological outcomes of no colocalization \citep[see more discussion on Simpson's paradox][]{simpson1951interpretation}. In contrast, SACA reports almost no colocalized regions. Moreover, these methods in Table~\ref{tb:comparison} can only provide the colocalization degree, i.e. how strong on average the degree of colocalization is within a particular ROI, while SACA is able to reflect the spatial change of colocalization levels within any region. For instance, almost all colocalization measures in Table~\ref{tb:comparison} suggest colocalization occurs in regions A, B, and C of Figure~\ref{fg:compare2}. Besides the reported existence of colocalization, SACA also shows that such levels primarily locate near the edge of cells. Through these comparisons, we can conclude that SACA can provide more robust and accurate colocalization quantification, as it does not call for pre-determined ROIs and can provide pixel-wise colocalization measures. 

\begingroup
\renewcommand{\arraystretch}{1.5}
\begin{table*}[h!]
\centering
\begin{tabular}{c c c c c c c c c c c c c c}
\hline
\hline
\multirow{2}{*}{Methods} & & & \multicolumn{3}{c}{Fig.~\ref{fg:compare1}} & & \multicolumn{3}{c}{Fig.~\ref{fg:compare2}} & &  \multicolumn{3}{c}{Fig.~\ref{fg:compare3}}\\
 \cline{4-6}   \cline{8-10}  \cline{12-14} 
& & & A & B & C & & A & B & C & & A & B & C \\
\hline
\multirow{2}{*}{Pearson } &  Index &      & $0.351$ &  $-0.036$ & $0.052$ & & $0.135$ & $0.175$ & $0.453$ & & $0.983$ & $0.984$ & $0.984$\\ 
& $p$-value & & $<0.1\%$ & $100\%$ & $<0.1\%$ & & $<0.1\%$ & $<0.1\%$ & $<0.1\%$ & & $<0.1\%$ & $<0.1\%$  & $<0.1\%$  \\
\hline
\multirow{2}{*}{Manders $M_1$ } &  Index &      & $0.998$ &  $0.996$ & $1.000$ & & $1.000$ & $1.000$ & $1.000$ & & $0.993$ & $0.898$ & $0.884$\\ 
& $p$-value & & $<0.1\%$ & $<0.1\%$ & $<0.1\%$ & & $<0.1\%$ & $<0.1\%$ & $<0.1\%$ & & $<0.1\%$ & $<0.1\%$  & $<0.1\%$  \\
\hline
\multirow{2}{*}{Manders $M_2$ } &  Index &      & $0.302$ &  $0.369$ & $0.274$ & & $1.000$ & $0.994$ & $0.858$ & & $0.999$ & $0.997$ & $0.997$\\ 
& $p$-value & & $<0.1\%$ & $<0.1\%$ & $<0.1\%$ & & $<0.1\%$ & $<0.1\%$ & $<0.1\%$ & & $<0.1\%$ & $<0.1\%$  & $<0.1\%$  \\
\hline
\multirow{2}{*}{ICQ } &  Index &      & $0.196$ &  $-0.037$ & $-0.018$ & & $0.029$ & $0.014$ & $0.170$ & & $0.448$ & $0.432$ & $0.433$\\ 
& $p$-value & & $<0.1\%$ & $100\%$ & $100\%$ & & $<0.1\%$ & $<0.1\%$ & $<0.1\%$ & & $<0.1\%$ & $<0.1\%$  & $<0.1\%$  \\
\hline
\multirow{2}{*}{MTKT $\tau^\ast$  } &  Index &      & $5.936$ &  $1.842$ & $1.543$ & & $3.640$ & $4.573$ & $2.674$ & & $148.5$ & $334.5$ & $408.9$\\ 
& $p$-value & & $<0.1\%$ & $97.0\%$ & $98.3\%$ & & $0.80\%$ & $0.30\%$ & $46.6\%$ & & $<0.1\%$ & $<0.1\%$  & $<0.1\%$  \\ 
\hline
\hline
\end{tabular}
\caption{The colocalization measure values and corresponding $p$-values obtained by Pearson's correlation coefficient, Manders' split coefficients, ICQ, and MTKT on different ROIs in Figure~\ref{fg:compareregion}.}\label{tb:comparison}
\end{table*}
\endgroup

\begin{figure}[h!]
    \centering
    \begin{tikzpicture}[scale=1]
  \node[inner sep=0] at (0,0) {\includegraphics[width=0.24\textwidth]{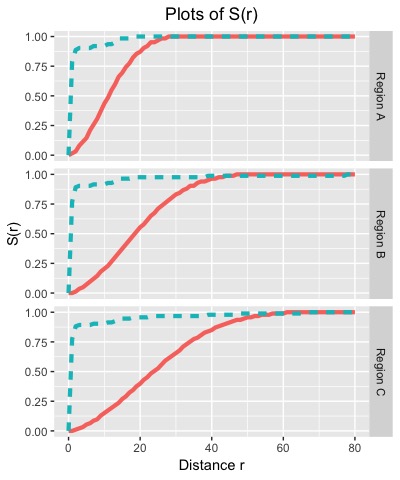}};
  \node[inner sep=0] at (4.5,0) {\includegraphics[width=0.24\textwidth]{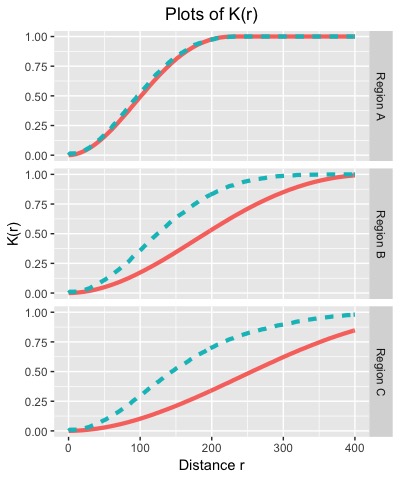}};
    \end{tikzpicture}
    \caption{Results of object based colocalization methods on Figure~\ref{fg:compare3}. The mean distance to the nearest neighbor $S$ are $1.81 (p<0.1\%)$, $2.89 (p<0.1\%)$, and $3.26 (p<0.1\%)$ on regions A, B, and C. The plots of $S(r)$ (left) and $K(r)$ (right) along $r$ are shown above: the green dashed line is the statistics calculated from data itself and the red line is $5\%$ upper quantile of statistics on random data.}\label{fg:objectcompare}
\end{figure}

We also compared SACA with object based colocalization methods on the same dataset. We first applied a spot detection algorithm on each channel. The spot detection algorithm we applied is from the imager package and is used to search for local maximum based on hessian (see more details in \url{https://dahtah.github.io/imager/}). The results of spot detection are not satisfied on the red channel of Figure~\ref{fg:compare1} and \ref{fg:compare2} as the spots are too dense with respect to resolution to be distinguished from each other. For instance, the spot detection results on Figure~\ref{fg:compare2} is shown in Figure~\ref{fg:spots}. Thus, we only apply object based spatial analysis on Figure~\ref{fg:compare3}. In particular, we applied mean distance to the nearest neighbors $S={1\over n_1}\sum_{i=1}^{n_1}d_i$ \cite{clark1954distance,Lagache2015}, the nearest neighbor function $S(r)={1\over n_1}\sum_{i=1}^{n_1}\mathbf{I}_{(d_i<r)}$ \cite{Helmuth2010,torquato1990nearest,Lagache2015} and Ripley's K function $K(r)$ \cite{lagache2013statistical,lagache2018mapping,Lagache2015}. To access the statistical significance, the spots were randomly drawn from each ROI $1000$ times. The results are summarized in Figure~\ref{fg:objectcompare}. Results in Figure~\ref{fg:objectcompare} suggest that object based colocalization methods have similar problems in choices of ROI as pixel-based colocalization methods (shown in Table~\ref{tb:comparison}). Once again, SACA demonstrates its unique advantages in robust colocalization quantification and precise colocalized region identification through comparisons with previous methods.

\section{Theoretical Properties}
\label{sc:theprop}

To complement the numerical studies, we now provide some theoretical justifications for the SACA framework. Our development builds upon earlier work \cite{Polzehl2000,Polzehl2006,becker2013}. The detailed proofs are included in the supplemental materials for completeness.

We begin by providing several basic properties of weighted Kendall's tau coefficient when the weights are non-stochastic.
\begin{proposition}
\label{pr:basickendall}
Suppose the weights $w_i$ in (\ref{eq:wtkendalltau}) are deterministic, then
$$
\PP\left(\left|\tau_w-\EE(\tau_w)\right|>z\right)\le 2\exp\left( -{(\sum_{i\ne j}w_iw_j)^2\over 2\sum_i w_i^2(\sum_{j\neq i}w_j)^2}z^2\right).
$$
 Moreover, if we assume $\sum_i w_i^2\le c(\sum_i w_i)^2$ for a small constant $c>0$, then
$$
\PP\left(\left|\tau_w-\EE(\tau_w)\right|>z\right)\le 2\exp\left(-{(1-c)^2Nz^2\over 2}\right),
$$
where $N$ is defined as
$$
N=\left(\sum_{i}w_i\right)^2/\sum_i w_i^2.
$$
\end{proposition}

If all $(X_i,Y_i)$ come from the same distribution $F$, then $\EE(\tau_w)=\EE(\sign(X_i-X_j)(Y_i-Y_j))$. Here $N$ can be seen as ``effective sample size" in the weight Kendall's tau correlation coefficient and can be used in standardization (in $D_i(k;r_t)$ and $\Delta \tau_k^{(t)}$).

Next, we show that the adaptive estimation can be extended freely in a homogeneous situation, and it can separate two homogeneous regions with significant difference. Without loss of generality, we assume $\bK_b(x,y)=\mathbf{I}_{(x>t_X,y>t_Y)}$  for some known $t_X$ and $t_Y$, and the quantity of interest is $Q(F)$ in (\ref{eq:colocaldef}) throughout the rest of this section. More general cases with arbitrary $\bK_b$ can be treated similarly.

Denoted by
$$
\SS=\left\{i\in\II:X_i>t_X,Y_i>t_Y\right\}
$$
for some known constant $t_X$ and $t_Y$. As $\bK_b(X_i,Y_i)$ is expected to be $0$ for $i\notin \SS$, the expected sample size of weight Kendall's tau correlation coefficient at step $t$ is given by
$$
N^{(t)}_k={\left(\sum_{i\in B(k,r_t)\cap \SS}\bK_l\left(d(i,k)\over r_t\right)\right)^2 \over \sum_{i\in B(k,r_t)\cap \SS}\bK_l^2\left(d(i,k)\over r_t\right)}.
$$
$N^{(t)}_k$ can be seen as the expectation of $\tilde{N}^{(t)}_k$ in homogeneous regions and thus is determined. We assume $N^{(t)}_k$s do not change greatly within a neighborhood, i.e.
\begin{equation}
\label{eq:nbsize}
\sup_{i\in B(k,r_{t+1})} \sqrt{N^{(t)}_k/N^{(t)}_i}\le S_b,
\end{equation}
and the weights in $N^{(t)}_k$ do not concentrate on few of them
\begin{equation}
\label{eq:evensize}
\sum_{i\in \tilde{\SS}}\bK_l^2\left(d(i,k)\over r_t\right)\le S_c\left(\sum_{i\in \tilde{\SS}}\bK_l\left(d(i,k)\over r_t\right)\right)^2,
\end{equation}
where $\tilde{\SS}=B(k,r_t)\cap \SS$.
Conditions (\ref{eq:nbsize}) and (\ref{eq:evensize}) reflect the pattern of $\SS$ and the weight $\bK_l$.

We also want to make assumptions for parameters and adaptive component $\bK_s$ of weights. We assume $\bK_s$ is non-increasing kernel on $[0,\infty)$ such that $\bK_s(0)=1$,
\begin{equation}
\label{eq:adkernel}
\bK_s\left( 2\left(1+2S_b\right) \over 1-2S_c\right)\ge {\sqrt{2}\over 2}\quad {\rm and}\quad \bK_s\left( z\right)=0\quad {\rm if}\ z\ge A
\end{equation}
for some constant $A$. The parameter $D_n$ and $\Lambda$ can be chosen in the following way
\begin{equation}
\label{eq:para}
D_n=\sqrt{\log n}\qquad {\rm and}\qquad \Lambda=\eta\sqrt{\log n}
\end{equation}
for some constant $\eta$. Here $n=|\II|$.

The last assumption we want to make is an independence assumption, which is to simplify the theoretical analysis. The main difficulty in analyzing the iterative procedure is that $\tau_w(k;r_t)$ and $w_i(k;r_t)$ depend on each other in each step $t$. To overcome this dependence problem, Polzehl and Spokoiny proposed an assumption that the adaptive weights are independent from observations \cite{Polzehl2006}. Later in 2013, Becker and Math{\'e} proposed a propagation condition to replace this independence condition, which is not easy to be applied in our setting. To conduct theoretical analysis, we also included this independence assumption into our analysis
\begin{equation}
\label{eq:indep}
w_i(k;r_t)\ \forall i,k\in\II, \forall\ t\ {\rm are\ independent\ from\ }(X_i,Y_i), i\in \II.
\end{equation}
We show that if the whole region $\II$ is homogeneous, the adaptive iterative procedure behaves in almost the same manner as the non-adaptive way. 

\begin{theorem}
\label{thm:globalhomo}
Suppose $F_k, k\in\SS$ are the same distribution $F$ and assumptions (\ref{eq:nbsize})-(\ref{eq:indep}) hold. 
Then the adaptive components in weights $w(k;r_t)$ of every step $t$ is larger than $\sqrt{2}/2$ with probability $1-2T^U/n$, i.e.
$$
\PP\left(\bK_s\left(D_i(k;r_{t}) \over D_n \right)\ge {\sqrt{2}\over 2},\forall i,k,t\right)\ge 1-{2T^U\over n}.
$$
\end{theorem}
In other words, Theorem~\ref{thm:globalhomo} shows that the stochastic adaptive components $\bK_s$ of weights have little effect on the weight $w(k;r_t)$ in the homogeneous region, and there is not much cost in estimating $\tau_w(k;r_t)$ and $Z(k;r_t)$ with adaptive procedure. Theorem~\ref{thm:globalhomo} can be generalized to a local homogeneity case with the same arguments. To this end, denoted by $B_g(k,r_t,s)$, $0\le s\le t$, the generalized neighborhood of $k$. When $s=0$, then $B_g(k,r_t,0)=B(k,r_t)$ and, for $s\ge 1$, $B(k,r_t,s)$ can be defined recursively
$$
B_g(k,r_t,s)=\bigcup_{i\in B_g(k,r_t,s-1)} B(i,r_{t-s}).
$$
Now, we generalize the free propagation result of Theorem~\ref{thm:globalhomo} to the local homogeneity case.
\begin{theorem}
\label{thm:localhomo}
Suppose $F_i=F$ for all $i\in B_g(k,r_t,t)$ and all conditions in Theorem~\ref{thm:globalhomo} are satisfied, then 
$$
\PP\left(\bK_s\left(D_i(k;r_{t'}) \over D_n \right)\ge {\sqrt{2}\over 2},\forall i\in B(k,r_t),t'<t\right)\ge 1-{2t\over n}.
$$
\end{theorem}

Now, we discuss the case of multiple homogeneous regions and separation properties of our adaptive procedure. To the end, we assume there is an $m$-partition of $\SS$, denoted by $\{\SS_1,\ldots,\SS_m\}$, such that
$$
\bigcup_{1\le l\le m}\SS_l=\SS\qquad{\rm and}\qquad \SS_l\bigcap\SS_h=\emptyset,\ \forall\ l\ne h.
$$
We assume $(X_k,Y_k)$ is homogeneous within each region $\SS_i$ as well, i.e.
$$
F_k=F^{(l)},\qquad k\in \SS_l
$$
for some distribution $F^{(1)},\ldots, F^{(m)}$ such that $Q(F^{(i)})$s are different. Denoted by $\SS_l^{o}$, the collection of points $k$ of which generalized neighborhood belongs to $\SS_l$,
$$
\SS_l^{o}=\{k:B_g(k,r_{T^L},T^L)\subset \SS_l\}.
$$
Now we are ready to show separation properties. For simplicity, we only discuss the case where $m=2$.
\begin{theorem}
\label{thm:separate}
Suppose there exists some number $N^{(T^L)}$ such that $\tilde{N}_k^{(t)}\ge N^{(T^L)}$ for any $t\ge T^L$ and $k\in \SS_1^{o}\cup \SS_2^{o}$, and $Q(F^{(1)})$ and $Q(F^{(2)})$ obey
$$
|Q(F^{(1)})-Q(F^{(2)})|>C_Q\sqrt{\log n\over N^{(T^L)}}
$$
for some constant $C_Q>A+\eta+2/(1-2S_c)$. In addition, conditions (\ref{eq:nbsize})-(\ref{eq:indep}) hold.
Then,
$$
\PP\left(\bK_s\left(D_i(k;r_{t}) \over D_n \right)=0;\forall\ i\in \SS_1^{o},k\in \SS_2^{o}, t\right)\ge 1-{2T^L\over n}.
$$
\end{theorem}
Theorem~\ref{thm:separate} suggests that as long as the colocalization levels between regions are sufficiently large, our adaptive procedure can separate regions completely with high probability.

\section{Concluding Remarks}

In this paper, we propose a novel, spatially adaptive colocalization analysis (SACA) procedure. Instead of evaluating the average degree of colocalization within the limits of a ROI, as in the 3-step procedure, our new SACA procedure quantifies the level of colocalization  pixel-by-pixel across the entire image by taking advantage of the spatial information within microscopic images. By doing so, the regions of high level colocalization can be identified at the pixel-level to locate associations between labelled macromolecules accurately. Due to incorporation of multiscale propagation and separation scheme, SACA expands the neighborhood adaptively, allowing more precise quantification and more detection power of colocalization. The colocalized regions discovered by SACA are reliable as multiple comparison correction is imposed in SACA to guard against false discoveries.

In this paper, we only discuss the SACA procedure within the context of weighted Kendall tau correlation coefficient $\tau_w$ because of its robustness \citep[see][]{wang2017}. However, SACA, as a local analysis framework, is readily generalizable to more colocalization quantification indices. For instance, we can construct SACA based on the Pearson correlation coefficient by replacing the weighted Kendall tau correlation coefficient $\tau_w$ with the weighted Pearson correlation coefficient
$$
r_w:={\sum_i w_i(X_i-\bar{X}_w)(Y_i-\bar{Y}_w)\over\sqrt{\sum_iw_i(X_i-\bar{X}_w)^2\sum_iw_i(Y_i-\bar{Y}_w)^2}},
$$
where $\bar{X}_w$ and $\bar{Y}_w$ are the weighted average intensities
$$
\bar{X}_w={\sum_i w_iX_i\over\sum_i w_i}\qquad{\rm and}\qquad \bar{Y}_w={\sum_i w_iY_i\over\sum_i w_i}.
$$

Although we mainly demonstrate the merit of SACA on 2D microscopic images for convenience of illustration, SACA can be applied to $s$ dimensional data for any arbitrary integer $s$ as long as the neighborhood $B(k,r)$ is well defined in $s$ dimensional space. The adaptive colocalization analysis framework, recommended choices of parameters, and theoretical properties are still valid in higher dimensional space. For example, SACA can be  also applied to 3D confocal microscopic images when the neighborhood $B(k,r)$ is a sphere in 3D space. 

We also developed a fast algorithm for calculating weighted Kendall's tau coefficient $\tau_w$ to speed up SACA, as well as visualization tools, such as the $z$-score heat map plotter and colocalized region marker displayed in Figures~\ref{fg:realdatanocol}--\ref{fg:realdatabill}; this allows display of the analysis results in a user-friendly way. Applying these to real biological datasets demonstrates that our new colocalization analysis method, SACA, is able to reveal the spatial colocalization information efficiently and robustly. All colocalization analysis algorithms, including LCA, SACA, and visualization tools used in the paper are readily available in the \texttt{R} package \texttt{RKColocal} (see \url{https://github.com/lakerwsl/RKColocal}). These colocalization analysis algorithms will also ultimately be implemented in ImageJ, which is a more broadly used and user-friendly package for biological researchers \citep[see][]{arena2016quantitating}. We hope this new colocalization analysis method will bring more insight into spatial associations between bio-molecules and greatly aid researchers in making more scientific discoveries.


\ifCLASSOPTIONcaptionsoff
  \newpage
\fi

\bibliographystyle{IEEEtran}  
\bibliography{FDRColocal} 

\end{document}